\documentclass[%
 reprint,
 amsmath,amssymb,
 aps,
]{revtex4-2}
\usepackage{graphicx}
\usepackage{amsmath,amssymb}
\usepackage{overpic}

\usepackage{xcolor}

\newcommand{\fr}[2]{\frac{\displaystyle #1}{\displaystyle #2}}

\newcommand{\pf}[2]{\frac{\displaystyle \partial #1}{\displaystyle \partial #2}}

\newcommand{\Real}{\mathrm{Re}\:}
\newcommand{\Imag}{\mathrm{Im}\:}

\newcommand{\qed}{\hfill \rule{2.3mm}{2.3mm}}

\newtheorem{proposition}{Proposition}
\newtheorem{remark}{Remark}

\newcommand{\be}{\begin{equation}}
\newcommand{\ee}{\end{equation}}

\begin{document}

\title{Mean-field approach to finite-size fluctuations in the Kuramoto-Sakaguchi model}

\author{Oleh E. Omel'chenko}
\email{omelchenko@uni-potsdam.de}
\address{University of Potsdam,
Institute of Physics and Astronomy,
Karl-Liebknecht-Str. 24/25,
14476 Potsdam,
Germany}

\author{Georg A. Gottwald}
\email{georg.gottwald@sydney.edu.au}
\address{School of Mathematical and Statistics, 
University of Sydney, NSW 2006, Australia}

\date{\today}

\begin{abstract}
We develop an ab initio approach to describe the statistical behavior of finite-size fluctuations in the deterministic Kuramoto-Sakaguchi model. We obtain explicit expressions for the covariance function of fluctuations of the complex order parameter and determine the variance of its magnitude entirely in terms of the equation parameters. Our results rely on an explicit complex-valued formula for the solution of the Adler equation. We present analytical results for both the sub- and the super-critical case. Moreover, our framework does not require any prior knowledge about the structure of the partially synchronized state. We corroborate our results with numerical simulations of the full Kuramoto-Sakaguchi model. The proposed methodology is sufficiently general such that it can be applied to other interacting particle systems. 
\end{abstract}

\maketitle

{\it Introduction.}
Synchronization is a generic phenomenon inherent in interacting oscillators
that determines their ability to exhibit a wide range of complex dynamical behavior
\cite{Kuramoto84, Str2000,PikovskyRosenblumKurths01, AceBVRS2005, ArenasEtAl08, DorflerBullo14, RodriguesEtAl16}.
It has been observed in a plethora of natural and engineered  systems,
including pace-maker cells of circadian rhythms \cite{YamaguchiETAl03},
networks of neurons \cite{BhowmikShanahan12}, chemical oscillators \cite{KissEtAl02,TaylorEtAl09}
and power grids \cite{FilatrellaEtAl08}.
The main features of synchronization are captured
by the paradigmatic Kuramoto-Sakaguchi (KS) model~\cite{SakK1986}
\begin{equation}
\dot{\psi}_i = \omega_i + \fr{K}{N} \sum\limits_{j=1}^N \sin( \psi_j - \psi_i - \lambda )
\label{Eq:KS:original}
\end{equation}
for the collective behavior of a system of $N$ all-to-all coupled phase oscillators.
In this model, $\omega_i$ are natural frequencies drawn randomly and independently
from a given distribution $g(\omega)$ and $K$ and $\lambda$ are real parameters
quantifying the coupling strength and the phase lag, respectively.
It is often convenient to rewrite Eq.~\eqref{Eq:KS:original} as
$$
\dot{\psi_i} = \omega_i + K\: \mathrm{Im}\left( Z(t) e^{-i \psi_i} e^{-i \lambda} \right),
$$
which reveals that each oscillator interacts
with the other oscillators only via the mean field
\begin{equation}
Z(t) = \fr{1}{N} \sum\limits_{j=1}^N e^{i \psi_j(t)},
\label{Def:Z}
\end{equation}
which is also coined the complex order parameter and quantifies the degree of synchronization. 

A fully synchronized state with $\theta_1 = \theta_2 = \dots = \theta_N$ is characterized by $r(t)=|Z(t)|=1$,
whereas a completely disordered (incoherent) state is characterized
by $r(t) = \mathcal{O}(1/\sqrt{N})$.
It is well known that for sufficiently large values of the coupling strength $K$ the system becomes fully synchronized, whereas for sufficiently small values of $K$ the oscillators behave independently and form an incoherent state. For intermediate values of the coupling strength, partially synchronized states are observed in which $r(t)$ fluctuates around a constant value between $0$ and $1$. On a microscopic level the oscillator population splits into two groups: coherent oscillators $\mathcal{C}$, which are phase locked, and rogue oscillators $\mathcal{R}$, which drift with respect to the mean field. 

In the thermodynamic limit of infinitely many oscillators each oscillator interacts with the other oscillators via a constant collective mean. This limit is well understood and described by mean-field theory~\cite{SakK1986,OmeW2012,OmeW2013,OttA2008}. In particular, the rotation frequency of the coherent oscillators $\Omega_\infty$ and the constant order parameter $r_\infty$ are determined by a self-consistency relation,
first derived in~\cite{SakK1986} and later formulated in a more convenient complex form in~\cite{OmeW2012},
\begin{equation}
\fr{1}{K} e^{i \lambda} = i \int_{-\infty}^\infty g(\Omega_\infty + K r_\infty s) h(s) ds,
\label{Eq:SC}
\end{equation}
where
\begin{equation}
h(s) = \left\{
\begin{array}{lcl}
( 1 - \sqrt{1 - s^{-2}} ) s & \mbox{ for } & |s| > 1 \\[1mm]
s - i \sqrt{1-s^2} & \mbox{ for } & |s|\le 1.
\end{array}
\right.
\label{Def:h}
\end{equation}

However, recently finite-size effects have gained much attention. Numerical experiments have led to a better understanding of finite-size effects and provided clear evidence for their significance~\cite{Dai1994,Dai1996,CraD1999,NisTHA2012,HonCTP2015,SnyderEtAl21,YueG2024,SumJ2024}. In particular, several finite-size phenomena were observed such as a stochastic drift of oscillators which disappears in the thermodynamic limit \cite{GiacominPoquet15,Lucon15,Got2015,Gottwald17} and emergent random chimera switching in a deterministic two-population KS model \cite{IrvG2025}. Capturing finite-size effects analytically is a notoriously hard problem. Relaxation rates and pair correlation functions were obtained using kinetic theory~\cite{HildebrandEtAl07,BuiceChow07,YoonEtAl15} and the finite-size scaling near criticality has been successfully quantified~\cite{HongEtAl07,HonCTP2015,HonOS2016,ParkPark24}.
Recently, substantial progress has been made
for stochastic systems of coupled oscillators \cite{TyuGKP2018,Bue2025}.
However, analytical methods for deterministic
systems still remain underdeveloped.
A general scheme for analyzing fluctuations in deterministic systems
of coupled phase oscillators was proposed in~\cite{KurN1987,Daido89,Dai1990}.
In particular, accurate approximations were obtained in the subcritical case of the KS model~\eqref{Eq:KS:original} with $\lambda=0$
and a Lorentzian frequency distribution~\cite{Dai1990}.
However, this approach relies on several approximations
which do not carry over to the case of more general frequency distributions
such as Gaussians as we will show (see Appendix~\ref{app:daido}).
In this Letter we will capture finite-size effects in a first principle way
and provide explicit expressions for the covariance function
and the variance of the order parameter. 

Our theory is based on the assumption that fluctuations in the deterministic KS model constitute a Gaussian process decaying with $1/\sqrt{N}$ and determined entirely by its mean and covariance function (see Appendix~\ref{app:CLT}
for numerical evidence for this assumption).
This assumption would naturally be satisfied if their dynamics were governed
by an underlying functional central limit theorem.
We will additionally adopt several technical assumptions on replacing finite-size quantities
by their thermodynamic mean-field limits as well as replacing sums
by integrals which requires a sufficiently large system size $N$.

Specifically, we consider the modified order parameter
$$
Z_\mathrm{mod}(t) = Z(t) \frac{\overline{Z_\mathrm{c}(t)}}{|Z_\mathrm{c}(t)|}
\quad\mbox{with}\quad
Z_\mathrm{c}(t) = \frac{1}{N} \sum\limits_{j\in\mathcal{C}} e^{i \psi_j(t)},
$$
where the bar denotes the complex-conjugate. Compared to the original order parameter \eqref{Def:Z}, the modified order parameter is represented in the frame of reference moving with the synchronized cluster. This ensures that, contrary to $Z(t)$, the modified order parameter does not exhibit phase diffusion~\cite{PikR1999,PikBI2026} (see Appendix~\ref{app:phasediff}).
Recently, it was numerically recognized that fluctuations 
\begin{equation}
\zeta(t) = \sqrt{N} ( Z_\mathrm{mod}(t) - \langle Z_\mathrm{mod}(t) \rangle )
\label{Def:zeta}
\end{equation}
around the thermodynamic mean $\langle Z_\mathrm{mod}(t) \rangle$,
where $\langle\cdot\rangle$ denotes time averaging, can be approximated by a Gaussian process \cite{YueG2024,IrvG2025}. 
Fluctuations in $Z_\mathrm{mod}(t)$,
with a variance that decays as $1/N$ and vanishes in the thermodynamic limit,
lead to fluctuations in other macroscopic variables,
including the order parameter with fluctuations
\begin{equation}
\delta(t) = \sqrt{N} ( |Z_\mathrm{mod}(t)| - \langle |Z_\mathrm{mod}(t)| \rangle ) = \sqrt{N} ( r(t) - \langle r(t) \rangle ).
\label{Def:delta}
\end{equation}
Gaussian processes are entirely determined by their mean, accessible via classical mean-field theory, and their covariance function
\begin{equation}
R(\tau) = \langle \zeta(t) \overline{\zeta(t+\tau)} \rangle.
\label{Def:R}
\end{equation}
Previous work~\cite{YueG2024,IrvG2025} has shown the efficacy of such an approach, however, these works relied on approximating the general Gaussian process by an Ornstein-Uhlenbeck process the parameters of which were estimated only numerically and needed to be recalculated for each set of equation parameters. Moreover, a precise a priori knowledge of which oscillators are synchronized and which are not was required.  

In this Letter, we find explicit analytical expressions for the covariance function \eqref{Def:R} and the variance of finite-size fluctuations of the order parameter $V = \langle \delta^2(t) \rangle$, given entirely in terms of equation parameters $K$, $\lambda$ and $g(\omega)$. We present expressions for both the partially synchronized and the completely incoherent state. Remarkably, our expressions do not require any a priori knowledge about which oscillators partake in the synchronized cluster and which do not.
This opens up the way to perform stochastic model reduction
for deterministic systems of coupled oscillators
where the drift and diffusion coefficients are given entirely
in terms of the microscopic equation parameters.
Such reductions have so far been restricted to multi-scale systems \cite{Givonetal04}.

\begin{figure}
\begin{center}
\includegraphics[height=0.48\columnwidth,angle=270]{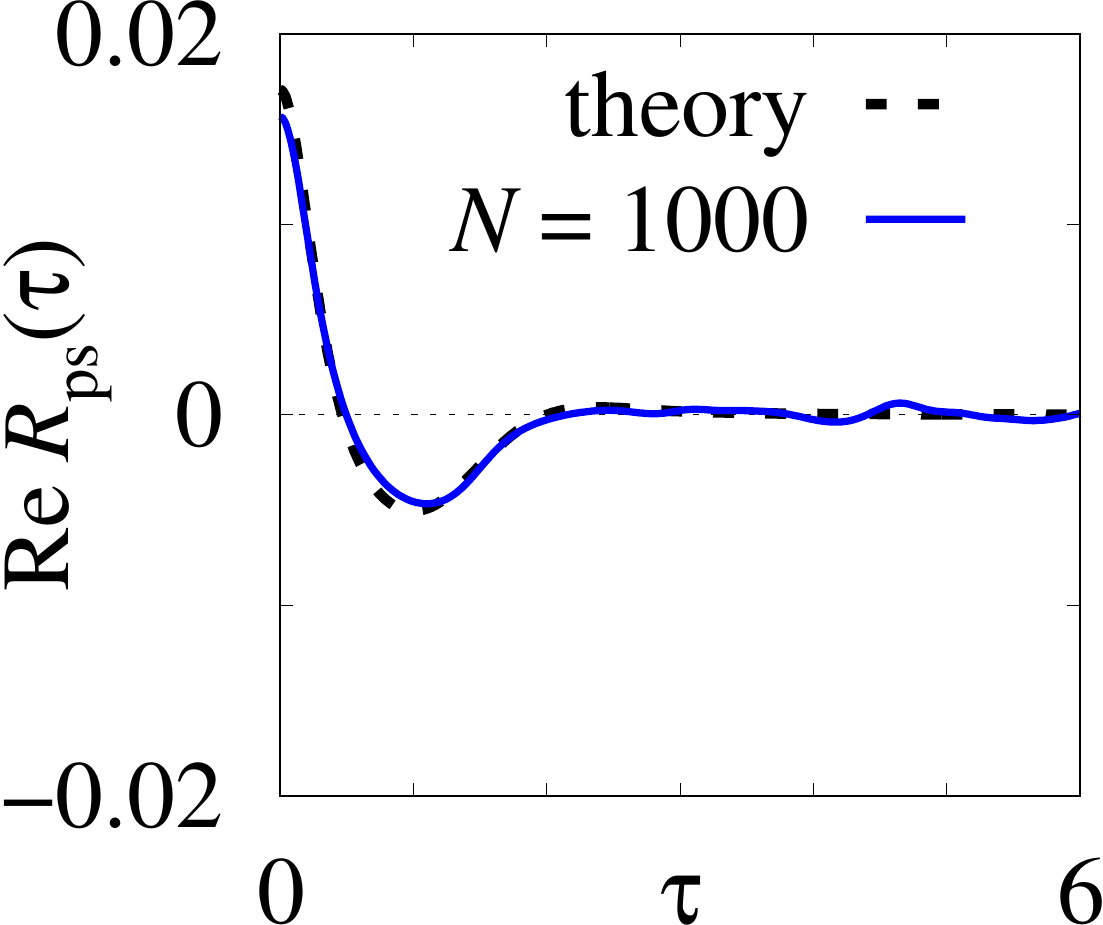}\hspace{0.02\columnwidth}
\includegraphics[height=0.48\columnwidth,angle=270]{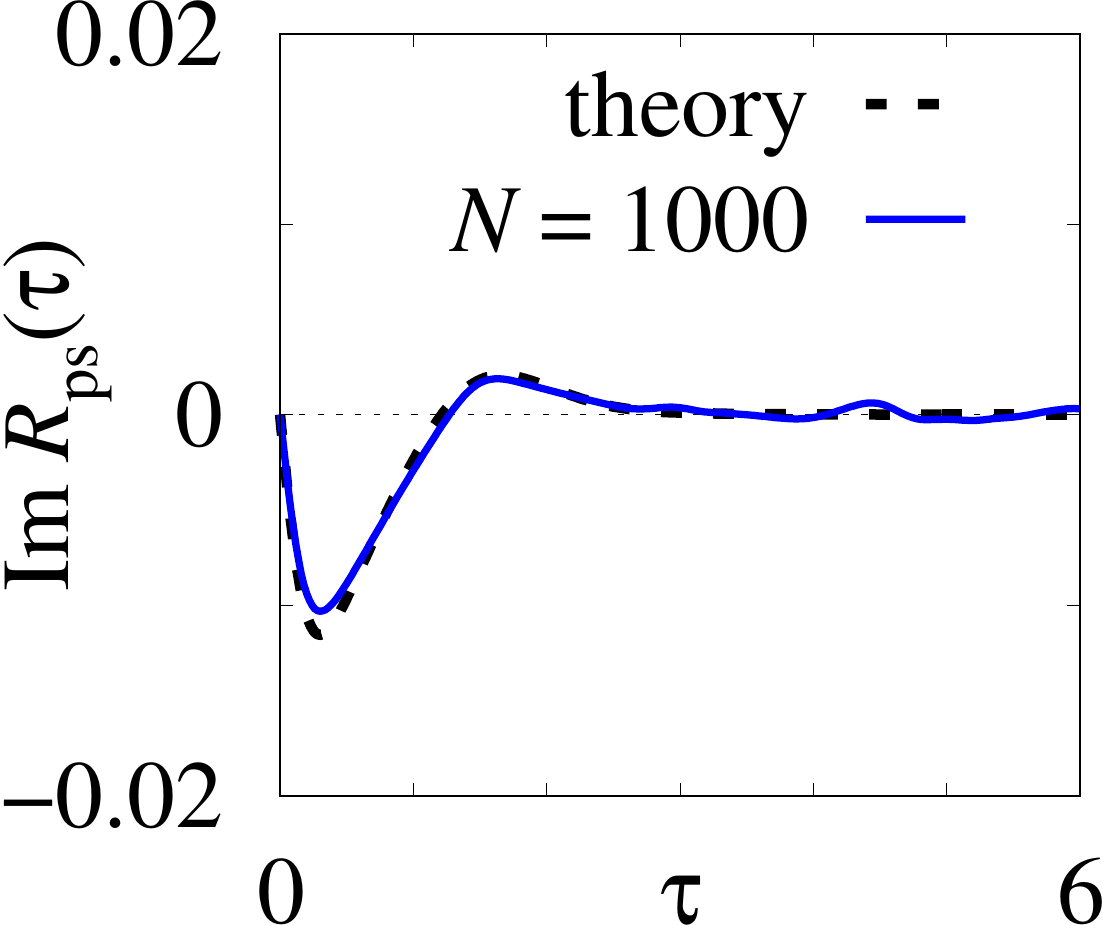}\\[1mm]
\includegraphics[height=0.48\columnwidth,angle=270]{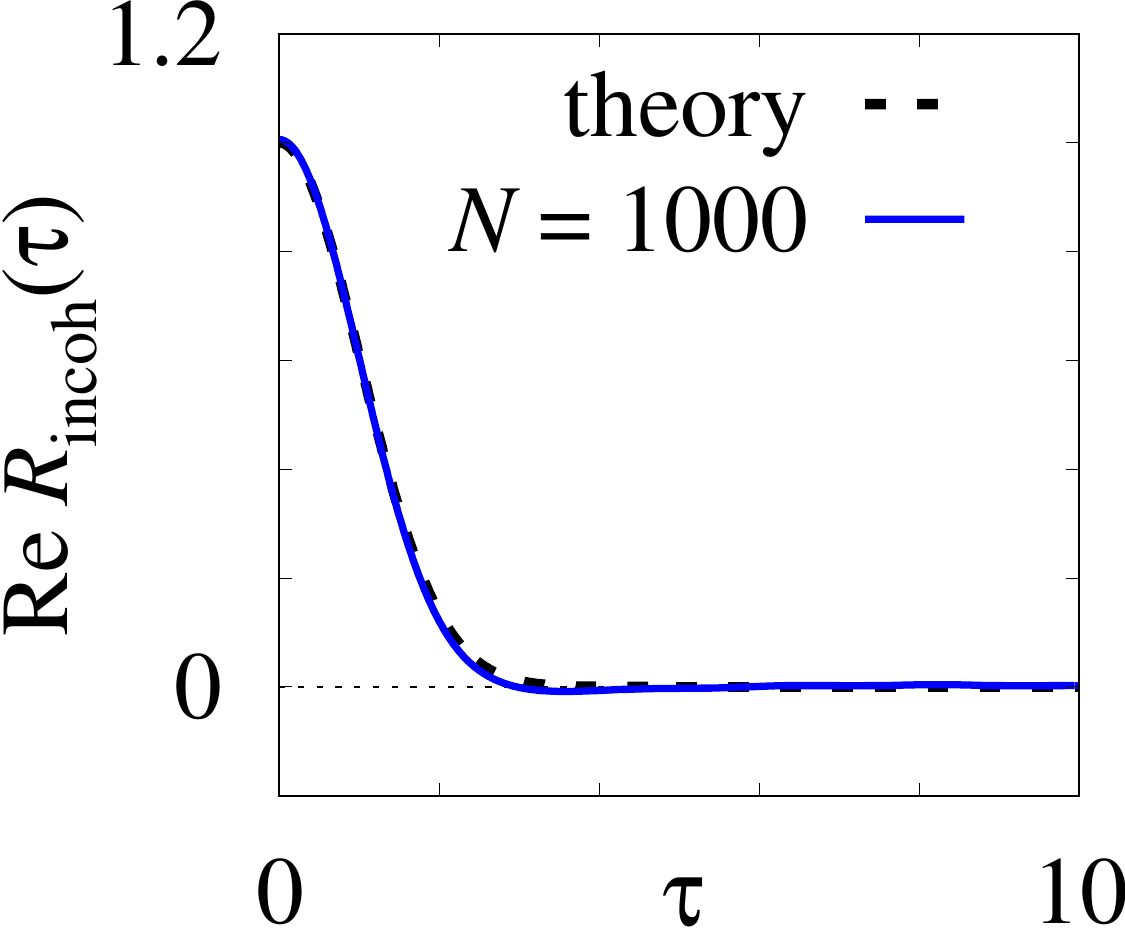}\hspace{0.02\columnwidth}
\includegraphics[height=0.48\columnwidth,angle=270]{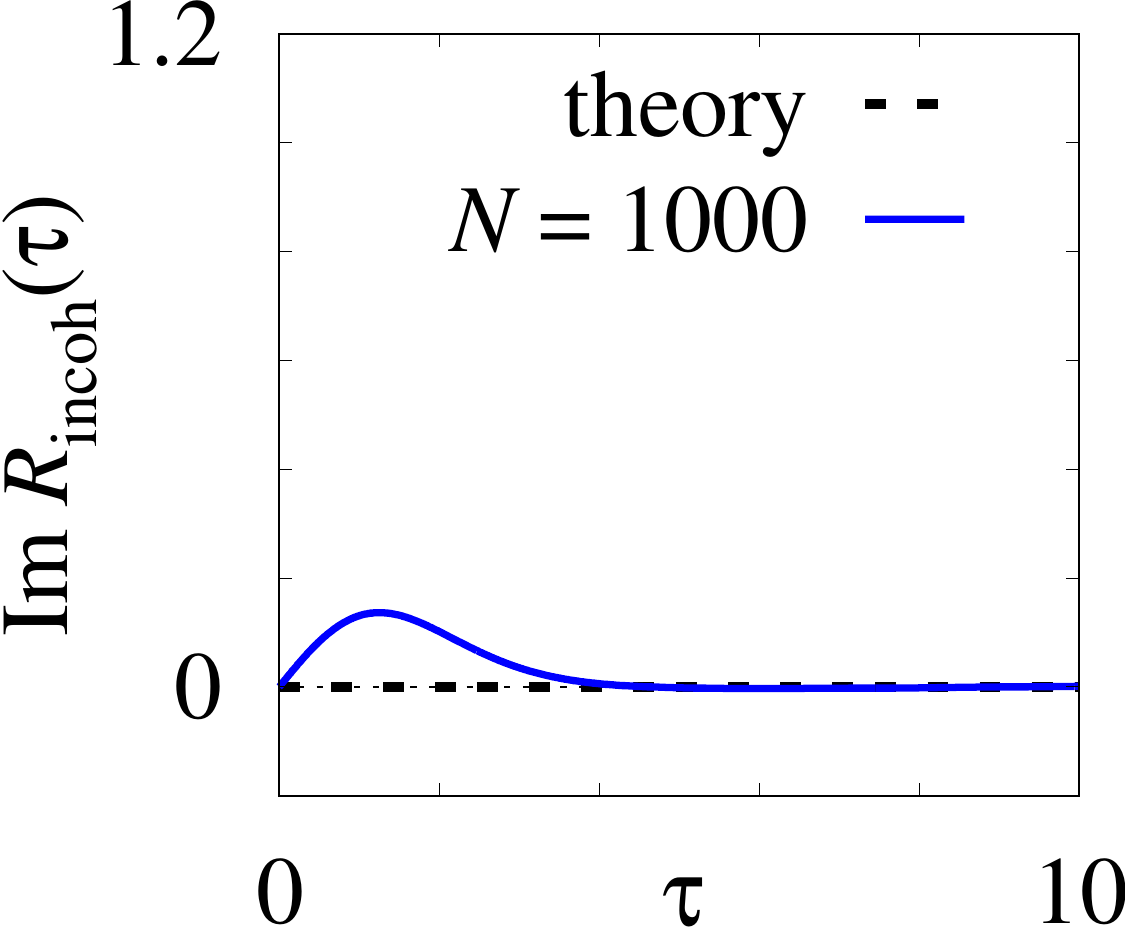}
\end{center}
\caption{Covariance $R(\tau)$ of fluctuations $\zeta(t)$ of the complex order parameter for the KS model \eqref{Eq:KS:original}. (Top row) Partially synchronized state at $K = 6$ and $\lambda = \pi/4$, with theoretical prediction \eqref{Formula:R:ps}. (Bottom row) Completely incoherent state at $K = 0.5$ and $\lambda = \pi/4$, with theoretical prediction \eqref{Formula:R:incoh}.}
\label{Fig:Covariances}
\end{figure}

\begin{figure}
\begin{center}
\includegraphics[height=0.48\columnwidth,angle=270]{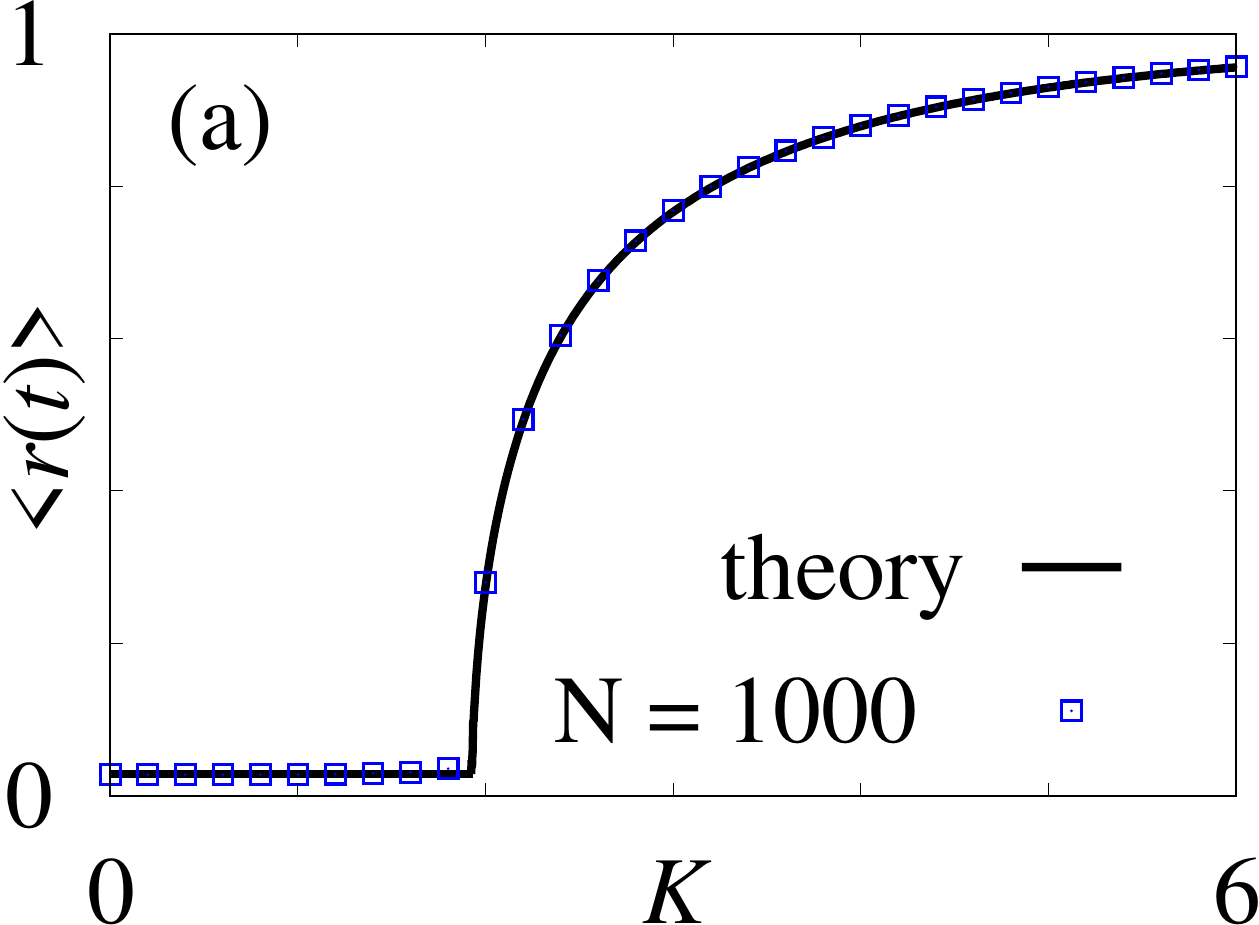}\hspace{0.02\columnwidth}
\includegraphics[height=0.48\columnwidth,angle=270]{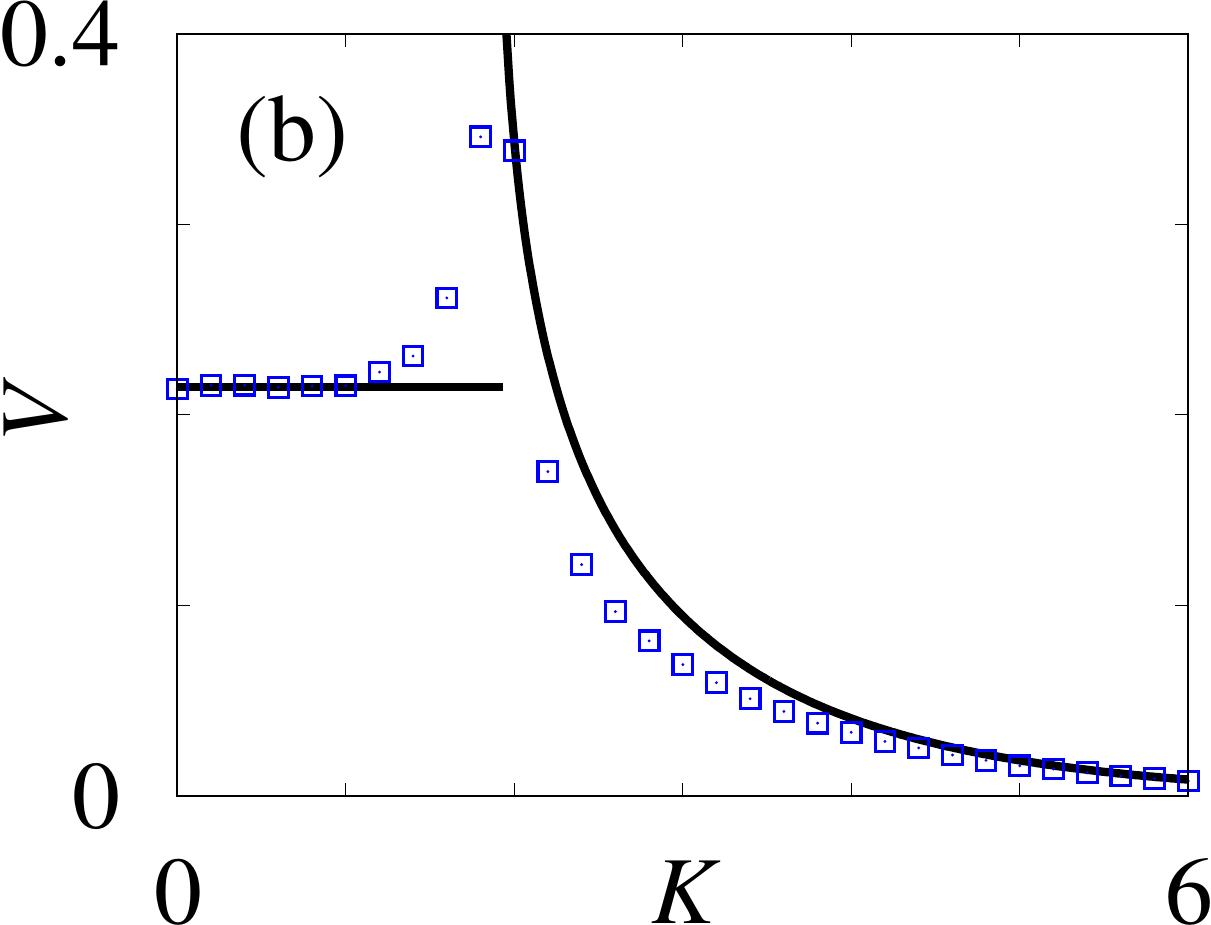}
\end{center}
\caption{(a) Averaged order parameter $\langle r(t) \rangle$, with theoretical predictions from classical mean-field theory \eqref{Eq:SC} for $K>K_{\rm{crit}}\approx 1.92$ and from our approach \eqref{Formula:incoh:r} for $K<K_{\rm{crit}}$, and (b) its variance $V$,
with theoretical predictions~\eqref{Formula:V:ps}, \eqref{Formula:incoh:r},
versus coupling strength $K$ in the KS model~\eqref{Eq:KS:original} with $\lambda = \pi/4$.}
\label{Fig:OP:OPVariance}
\end{figure}

Our main results can be formulated as follows. For partially synchronized states we predict
\begin{equation}
R_\mathrm{ps}(\tau) = K r_\infty \int\limits_{-\infty}^\infty \fr{(\overline{h^2(s)} - 1) ( 1 - |h^2(s)| ) g(\Omega_\infty + K r_\infty s) }{\overline{h^2(s)} - \exp(i K r_\infty s \sqrt{1 - s^{-2}}\:\tau)} ds
\label{Formula:R:ps}
\end{equation}
and
\begin{equation}
V_\mathrm{ps} = \fr{1}{2} R_\mathrm{ps}(0) + \mathcal{O}\left(\fr{1}{\sqrt{N}}\right)
\quad\mbox{for}\quad N\gg 1.
\label{Formula:V:ps}
\end{equation}
For completely incoherent states we find
\begin{equation}
R_\mathrm{incoh}(\tau) = \int_{-\infty}^\infty g(\omega) e^{-i\omega\tau} d\omega
\label{Formula:R:incoh}
\end{equation}
and
\begin{equation}
\sqrt{N} \langle r(t) \rangle = \fr{\sqrt{\pi}}{2}\quad\mbox{and}\quad
V_\mathrm{incoh} = 1 - \fr{\pi}{4}\quad\mbox{for}\quad N\gg 1.
\label{Formula:incoh:r}
\end{equation}
Our analytical results and their efficacy in capturing finite-size fluctuations
are illustrated in Figs.~\ref{Fig:Covariances} and~\ref{Fig:OP:OPVariance}, 
where we compare our predictions with simulations of the full KS model \eqref{Eq:KS:original} for a network of $1,000$ oscillators
with a Gaussian natural frequency distribution $g(\omega)$ with unit-variance. Fig.~\ref{Fig:Covariances} shows the covariance function $R(\tau)$ for a partially synchronized state described by \eqref{Formula:R:ps}
and for a completely incoherent state described by \eqref{Formula:R:incoh}. Fig.~\ref{Fig:OP:OPVariance}(a) shows the averaged order parameter $\langle r(t) \rangle$ as a function of the coupling strength $K$. The order parameter of the full KS model \eqref{Eq:KS:original} is very well reproduced by classical mean-field theory \eqref{Eq:SC} for values of the coupling strength $K\ge K_{\rm{crit}}$. Moreover, our extension of the mean-field results to the completely incoherent state given by \eqref{Formula:incoh:r} reproduces the order parameter for the subcritical range $K<K_{\rm{crit}}$. Fig.~\ref{Fig:OP:OPVariance}(b) shows that the variance of the order parameter is well captured by our analytical expressions~\eqref{Formula:V:ps} and \eqref{Formula:incoh:r} for coupling strength away from the bifurcation at $K_{\rm{crit}}$. 
Additional numerical results and the description
of our numerical setup are provided
in Appendix~\ref{app:addnum}
and Appendix~\ref{app:num}, respectively.

{\it Covariance function for a partially synchronized state.}
Our approach is based on the following prerequisites.
(i) Moving into the corotating frame
$\theta_i(t) = \psi_i(t) - \arg Z_\mathrm{c}(t)$ and assuming that the rotation frequency and the order parameter are given to leading order by their respective mean-field limits, Eq.~(\ref{Eq:KS:original}) becomes the so called Adler equation
\begin{equation}
\dot{\theta}_i = \omega_i - \Omega_\infty + K\: \mathrm{Im}\left( Z_\infty e^{-i \theta_i} e^{-i \lambda} \right).
\label{eq:Adler0}
\end{equation}

(ii)~We use the well-known fact (see Appendix~\ref{app:MFT} for details)
that the probability density of the solution of the Adler equation \eqref{eq:Adler0}
is given by the Ott-Antonson ansatz~\cite{OttA2008}
\begin{equation}
f_\phi(\theta,\omega) = \fr{g(\omega)}{2\pi} \left\{ 1 + \sum\limits_{n=1}^\infty
\left[ \overline{z}_\phi^n(\omega) e^{i n \theta} + z_\phi^n(\omega) e^{-i n \theta} \right] \right\},
\label{Def:f_phi}
\end{equation}
where
\begin{equation}
z_\phi(\omega) = e^{i \phi} h\left( \fr{\omega - \Omega_\infty}{K r_\infty} \right).
\label{Formula:zphi}
\end{equation}
It follows directly from (\ref{Def:f_phi}) that for every non-negative integer~$n$ we have
\begin{equation}
\int_0^{2\pi} \fr{f_\phi(\theta,\omega)}{g(\omega)} e^{i n \theta} d\theta = z_\phi^n(\omega),
\label{Formula:OA}
\end{equation}
which, upon employing (\ref{Def:Z}) and~(\ref{Eq:SC}), implies
\begin{equation}
Z_\infty = \int_{-\infty}^\infty d\omega \int_0^{2\pi} f_\phi(\theta,\omega) e^{i \theta} d\theta
= - r_\infty i e^{i \lambda} e^{i \phi}.
\label{Z_mod:infty}
\end{equation}
Moreover, if $\theta_i(t)$ is a solution of the scalar Adler equation~(\ref{eq:Adler0}),
then for any $2\pi$-periodic function $F(\theta)$ we have the ergodicity property
\begin{equation}
\langle F(\theta_j(t)) \rangle = \int_0^{2\pi} \fr{f_\phi(\theta,\omega_j)}{g(\omega_j)} F(\theta) d\theta.
\label{Ergodicity}
\end{equation}
(iii)~Finally, we establish an explicit solution of the Adler equation \eqref{eq:Adler0}. 
Introducing $\Delta_j = \omega_j - \Omega_\infty$ and $H_\infty = K Z_\infty e^{-i \lambda}$,
solutions of the Adler equation $\theta_j(t)$ with initial conditions $\theta_j(0)$
are given for $|\Delta_j| \ne |H_\infty|$ by 
\begin{equation}
e^{i \theta_j(t)} = W( t, H_\infty,\Delta_j, e^{i \theta_j(0)} ),
\label{Formula:SolAdler}
\end{equation}
with
\begin{equation}
W( t, H, \Delta, \sigma ) = H a_1(t) + \fr{a_2( t) \sigma}{1 + \overline{H} a_1( t) \sigma},
\label{MainSln}
\end{equation}
where
$$
a_1(t) = i (e^{i \xi t} - 1) / \chi_1(t)
\quad\mbox{and}\quad
a_2(t) = 4 \xi^2 e^{i \xi t} / \chi_1^2(t)
$$
with $\chi_1( t ) = (\Delta - \xi) e^{i \xi t} - (\Delta + \xi)$ and
\begin{equation}
\xi = \left\{
\begin{array}{lcl}
- i \sqrt{ |H|^2 - \Delta^2 } & \mbox{ for } & |\Delta| \le |H|,\\[2mm]
\hphantom{-i} \sqrt{ \Delta^2 - |H|^2 } & \mbox{ for } & |\Delta| > |H|.
\end{array}
\right.
\label{Def:xi}
\end{equation}
Note that for $|\Delta_j| < |H_\infty|$, the $j$th oscillator is coherent
and the dynamics of $\theta_j(t)$ converges to a fixed point as $t\to\infty$,
while for $|\Delta_j| > |H_\infty|$, the oscillator is incoherent with underlying periodic dynamics.
Details on the derivation of~\eqref{Formula:SolAdler}, \eqref{MainSln} are provided in Appendix~\ref{app:Adler}.

The covariance function (\ref{Def:R}) can be expressed in terms of the fluctuations (\ref{Def:zeta}) as 
\begin{eqnarray}
R_\mathrm{ps}(\tau) &=&  N \langle
Z_\mathrm{mod}(t) \overline{Z_\mathrm{mod}(t+\tau)}
\rangle - N |Z_\infty|^2
\nonumber\\
&=&
\fr{1}{N} \sum\limits_{j,k=1}^N \langle e^{i \theta_j(t)} e^{-i \theta_k(t+\tau)} \rangle - N |Z_\infty|^2.
\label{Formula:U:1}
\end{eqnarray}
If either $j\in\mathcal{C}$ or $k\in\mathcal{C}$ the phase asymptotically approaches a fixed point and the summands can be written as
\begin{equation}
\langle e^{i \theta_j(t)}e^{-i \theta_k(t+\tau)} \rangle = \langle e^{i \theta_j(t)} \rangle \langle e^{-i \theta_k(t)} \rangle.
\label{IndividualCorrealtions:theta}
\end{equation}
Moreover, the independence identity~\eqref{IndividualCorrealtions:theta} also holds
if $j,k\in\mathcal{R}$ and $\sqrt{\Delta_j^2 - |H_\infty|^2}$
and $\sqrt{\Delta_k^2 - |H_\infty|^2}$ are incommensurable (see Appendix~\ref{app:indep}),
and hence generically the sum in~\eqref{Formula:U:1} can be written as
\begin{eqnarray*}
&&
\sum\limits_{j,k=1}^N \langle e^{i \theta_j(t)} e^{-i \theta_k(t+\tau)} \rangle
= \sum\limits_{j,k=1}^N \langle e^{i \theta_j(t)} \rangle \langle e^{-i \theta_k(t)} \rangle \\
&&
+ \sum\limits_{j\in\mathcal{R}} \langle e^{i \theta_j(t)} e^{-i \theta_j(t+\tau)} \rangle
- \sum\limits_{j\in\mathcal{R}} \left| \langle e^{i \theta_j(t)} \rangle \right|^2.
\end{eqnarray*}
Substitution into \eqref{Formula:U:1} yields,
upon using $\langle Z_\mathrm{mod}(t)\rangle = Z_\infty$, 
\begin{equation}
    R_\mathrm{ps}(\tau) = 
\frac{1}{N}\sum\limits_{j\in\mathcal{R}} \langle e^{i \theta_j(t)} e^{-i \theta_j(t+\tau)} \rangle
- \frac{1}{N} \sum\limits_{j\in\mathcal{R}} \left| \langle e^{i \theta_j(t)} \rangle \right|^2.
\label{Formula:R:ps2}
\end{equation}
Now, using~\eqref{Ergodicity} to replace temporal averages by averages over the phases and the explicit formula~\eqref{Formula:SolAdler} for $e^{i \theta_j(t+\tau)}$, we obtain
\begin{eqnarray*}
\langle e^{i \theta_j(t)} \rangle
&=& z_\phi(\omega_j),\\[1mm]
\left\langle e^{i \theta_j(t)} e^{-i \theta_j(t+\tau)} \right\rangle
&=&
\int_0^{2\pi} e^{i \theta} \overline{W}( \tau, H_\infty,\Delta_j, e^{i \theta} ) \times \\
\times\fr{f_\phi(\theta,\omega_j)}{g(\omega_j)} d\theta
&=& \fr{\overline{H a_1(\tau)}}{\overline{z_\phi(\omega_j)}} ( |z_\phi(\omega_j)|^2 - 1 )
+ 1.
\end{eqnarray*}
(Note that for the second correlation, we used identity~\eqref{Formula:OA}
after expanding the solution \eqref{Formula:SolAdler}--\eqref{MainSln} in a geometric series.
Details of the calculations can be found in Appendix~\ref{app:Adler}.)
Altogether, this yields
$$
R_\mathrm{ps}(\tau) =
\fr{1}{N} \sum\limits_{j=1}^N 
\left[ \fr{\overline{H a_1(\tau)}}{\overline{z_\phi(\omega_j)}}
- 1 \right] \left( | z_\phi(\omega_j) |^2 - 1 \right).
$$
Finally, to obtain a compact expression, we make the technical assumption that $N\gg 1$,
and replace  averaging over indices $j$ 
by averaging over the distribution $g(\omega)$, and write
\begin{equation}
R_\mathrm{ps}(\tau) = \int\limits_{-\infty}^\infty
\left[ \fr{\overline{H a_1(\tau)}}{\overline{z_\phi(\omega)}}
- 1 \right] \left( | z_\phi(\omega) |^2 - 1 \right) g(\omega) d\omega.
\label{Formula:U1}
\end{equation}
After some algebraic manipulations (see Appendix~\ref{app:Derivation}), the covariance function~\eqref{Formula:U1} simplifies to our expression~\eqref{Formula:R:ps}. Note that if $|\omega - \Omega_\infty|\le |H_\infty| = K r_\infty$, $|z_\phi| = 1$. Therefore such a set of $\omega$ does not contribute to the integral $R_\mathrm{ps}(\tau)$. This has the desirable consequence that in order to evaluate the covariance function, no prior knowledge is required about which oscillators are and which are not synchronized. 



{\it Covariance function for a completely incoherent state.}
Our analysis of the completely incoherent state is based on the following prerequisites.
(i) In this state, the phases $\psi_j$ tend to be uniformly distributed on the interval $[0,2\pi]$,
and their probability density in the thermodynamic limit is given by
\begin{equation}
f_\mathrm{incoh}(\psi,\omega,t) = \fr{g(\omega)}{2\pi}.
\label{Def:f_incoh}
\end{equation}
(ii) The global order parameter shows no collective oscillations, and $Z(t) = 0$ as $N\to\infty$. This implies that the dynamics of each oscillator is approximately described by $\dot{\psi}_i = \omega_i$,
and therefore $\psi_i(t) = \omega_i t + \psi_i(0)$.

Consider the fluctuations
$$
\zeta_\mathrm{incoh}(t) = \sqrt{N}( Z(t) - \langle Z(t) \rangle)
$$
of the complex order parameter in this incoherent state.
The corresponding covariance function is then given by
\begin{eqnarray*}
R_\mathrm{incoh}(\tau) &=& \langle \zeta_\mathrm{incoh}(t) \overline{\zeta_\mathrm{incoh}(t+\tau)} \rangle \\
&=& \fr{1}{N} \sum\limits_{j,k = 1}^N \left\langle e^{i \psi_j(t)}e^{-i \psi_k(t+\tau)} \right\rangle - N | \langle Z(t) \rangle |^2.
\end{eqnarray*}
For different incommensurable frequencies $\omega_j$ and $\omega_k$,
the variables $e^{i \psi_j(t)}$ and $e^{i \psi_k(t)}$ are independent,
which implies again an identity of the form \eqref{IndividualCorrealtions:theta}, and we obtain
$$
R_\mathrm{incoh}(\tau) = \fr{1}{N} \sum\limits_{j = 1}^N \left\langle e^{i \psi_j(t)} e^{-i \psi_j(t+\tau)} \right\rangle
- \fr{1}{N} \sum\limits_{j = 1}^N \left| \langle e^{i \psi_j(t)} \rangle \right|^2.
$$
Moreover, using the ergodicity property and the stationary density for incoherent oscillators \eqref{Def:f_incoh}, we find
\begin{equation}
\langle e^{i \psi_j(t)} \rangle = \int_0^{2\pi} e^{i \psi} \fr{f_\mathrm{incoh}(\psi,\omega_j,t)}{g(\omega_j)} d\psi = 0
\label{Formula:Incoh}
\end{equation}
and
$$
\langle e^{i \psi_j(t)} e^{-i \psi_j(t+\tau)} \rangle = e^{- i \omega_j \tau}.
$$
Finally, in the large-$N$ limit we can replace averaging over indices $j$
by averaging over the distribution $g(\omega)$
and arrive at our expression~\eqref{Formula:R:incoh}
for the covariance function $R_{\rm{incoh}}(\tau)$.
Note that due to the normalization condition for $g(\omega)$, we have
$\langle \left| \zeta_\mathrm{incoh}(t) \right|^2 \rangle = R_\mathrm{incoh}(0) = 1$.


{\it Variance of the order parameter for a partially synchronized state.}
Rewriting \eqref{Def:zeta} as
$$
Z_\mathrm{mod}(t)
= \langle Z_\mathrm{mod}(t) \rangle + \fr{1}{\sqrt{N}} \zeta(t) = Z_\infty + \fr{1}{\sqrt{N}} \zeta(t)
$$
implies
$$
|Z_\mathrm{mod}(t)|^2 = | Z_\infty |^2 + \fr{2}{\sqrt{N}} \Real\left( \overline{Z_\infty} \zeta(t) \right) + \fr{1}{N} |\zeta(t)|^2,
$$
which upon expansion for $N\gg 1$ becomes
$$
|Z_\mathrm{mod}(t)| = | Z_\infty | + \fr{1}{\sqrt{N}| Z_\infty |} \Real\left( \overline{Z_\infty} \zeta(t) \right) + \mathcal{O}\left(\fr{1}{N}\right).
$$
This allows to express the fluctuations of the order parameter \eqref{Def:delta} as
$$
\delta(t) = \fr{1}{| Z_\infty |} \Real\left( \overline{Z_\infty} \zeta(t) \right) + \mathcal{O}\left( \fr{1}{\sqrt{N}} \right),
$$
and the variance of the order parameter becomes 
$$
V_\mathrm{ps}
= \fr{1}{2}
\Real\left( \fr{\overline{Z_\infty^2}}{| Z_\infty |^2} \tilde{R}_\mathrm{ps}(0) \right) + \fr{1}{2} R_\mathrm{ps}(0) + \mathcal{O}\left( \fr{1}{\sqrt{N}} \right),
$$
where $\tilde R_{\rm{ps}}(t)$ denotes
the pseudo-covariance
$\tilde R_{\rm{ps}}(t) = \langle \zeta(t) \zeta(t+\tau) \rangle$.
Employing calculations similar to those
for $R_\mathrm{ps}(t)$ (see Appendix~\ref{app:Pseudocov} for details),
we can show that $\tilde{R}_\mathrm{ps}(\tau) = 0$ for all $\tau\ge 0$.
The vanishing of the pseudo-covariance can be intuitively understood by evoking a symmetry between the real and imaginary parts of the fluctuations $\zeta$ and their independence. Using  $\tilde{R}_\mathrm{ps}(\tau) = 0$ for all $\tau\ge 0$, leads to our result \eqref{Formula:V:ps}.


{\it Variance of the order parameter for a completely incoherent state.}
In the large-$N$ limit, for a completely incoherent state
we have $\langle Z(t) \rangle = 0$, and therefore
$
|Z(t)| = |\zeta_\mathrm{incoh}(t)| / \sqrt{N}.
$
Using the definition of $\zeta_\mathrm{incoh}(t)$, we write
$$
\zeta_\mathrm{incoh}(t) = \fr{1}{\sqrt{N}} \sum\limits_{j=1}^N \cos \psi_j(t) + \fr{i}{\sqrt{N}} \sum\limits_{j=1}^N \sin\psi_j(t).
$$
In the completely incoherent state all phases $\psi_j(t)$ are uniformly distributed in $[0,2\pi]$. This implies that in the large-$N$ limit fluctuations satisfy a central limit theorem, and $\zeta_\mathrm{incoh}(t)$ is a two-dimensional random vector
distributed according to a bivariate normal distribution. The mean and variances are readily determined. For each $t\ge 0$ the function $\cos \psi_j(t)$ is a random variable with zero expected value $\langle \cos\psi_j \rangle_\mathrm{ens} = 0$ and variance
$$
\left\langle \cos^2\psi_j \right\rangle_\mathrm{ens} =
\fr{1}{2\pi} \int_0^{2\pi} \cos^2 \psi\: d\psi = \fr{1}{2},
$$
where $\langle \cdot \rangle_\mathrm{ens}$ denotes the ensemble average. Similarly, $\sin\psi_j(t)\sim\mathcal{N}(0,1/2)$. Moreover, the random variables $\cos\psi_j(t)$ and $\sin\psi_j(t)$ are uncorrelated since
$$
\langle \cos\psi_j \sin\psi_j \rangle_\mathrm{ens}
= \fr{1}{2\pi} \int_0^{2\pi} \cos\psi \sin\psi\:d\psi = 0.
$$
Hence $\Real(\zeta_\mathrm{incoh}(t))\sim\mathcal{N}(0,1/2)$ and $\Imag(\zeta_\mathrm{incoh}(t))\sim\mathcal{N}(0,1/2)$, and $\left| \zeta_\mathrm{incoh}(t) \right|$ has a Rayleigh distribution with expected value
$$
\langle \left| \zeta_\mathrm{incoh}(t) \right| \rangle = \sqrt{\pi}/2
$$
and variance
$$
V_\mathrm{incoh} = \langle \left| \zeta_\mathrm{incoh}(t) \right|^2 \rangle - \langle \left| \zeta_\mathrm{incoh}(t) \right| \rangle^2 = 1 - \pi/4.
$$
Note that the value $\langle \left| \zeta_\mathrm{incoh}(t) \right|^2 \rangle = 1$, which follows from the last two identities, agrees with the value $\langle \left| \zeta_\mathrm{incoh}(t) \right|^2 \rangle = R_\mathrm{incoh}(0)$ obtained above.


{\it Discussion and outlook.}
In this Letter, we have proposed a fully analytical approach
to capture the statistical behavior of finite-size fluctuations
in the deterministic KS model~\eqref{Eq:KS:original}.
In particular, our approach provides

(i) expressions for the covariance function of fluctuations of the complex order parameter
and the variance of the order parameter entirely in terms of the KS model parameters,

(ii) expressions for both sub- and super-critical coupling strengths,

(iii) expressions that do not require prior knowledge
about which oscillators are synchronized and which are not.

Our theory relies on two main assumptions:
(a) finite-size fluctuations of the deterministic KS model~\eqref{Eq:KS:original}
are governed by a Gaussian process and (b) the phase dynamics can be approximated
by the Adler equation~\eqref{eq:Adler0}, where we neglect fluctuations
and approximate the mean values of the rotation frequency
and the order parameter by their mean-field limits.
Remarkably, the statistical independence of fluctuations of distinct oscillators,
which were essential for finding explicit expressions,
follows directly from our solution of the Adler equation.
The assumption of Gaussian statistics of finite-size fluctuations would naturally follow,
if one could show that the KS model supports an underlying functional central limit theorem.
This would in turn require large system sizes $N$ -- an assumption
we had already employed in~\eqref{Formula:U1} when replacing sums by integrals.

Simulations for the full KS model showed
that our theory performs well away from criticality.
On the other hand, our approximations were less accurate
when approaching the critical coupling strength $K_{\rm{crit}}$,  
demarcating the transition from a desynchronized to a partially synchronized state.
The explicit expressions for the covariance function rely
on the explicit solution of the Adler equation.
Hence, the quality of our theory crucially depends on the Adler equation
being a good approximation of the full system.
This is only guaranteed if fluctuations can be neglected in Eq.~(\ref{Eq:KS:original}),
and breaks down when approaching criticality
(see Appendix~\ref{app:lim} for details).
We envisage extensions of our framework to the near critical case
using a perturbative approach around the expressions derived in this work.

Our approach is sufficiently general to allow for further applications
to other interacting particle systems such as $\theta$-neuron models
and more complex network topologies
going beyond the all-to-all coupling considered in this Letter. 
The mathematical tools developed here can potentially be adapted
to study finite-size fluctuations in noisy coupled oscillator systems~\cite{MasKK2010,Luc2014,Gottwald17,TyuGKP2018,PikDG2019,DelgadinoEtAl21,ZagliEtAl23,Bue2025}
finite-size sampling effects \cite{PetP2018,PetGP2019,FialkowskiEtAl23}
and dynamics of complexified KS models~\cite{ThuSST2023,LeeBBSTT2024}.
Further, it is known that the incoherent state and partially synchronized states
in the KS model are weakly chaotic attractors,
where the chaoticity can be attributed to the presence of finite-size fluctuations.
Therefore, it is natural to expect that knowing the characteristics of finite-size fluctuations,
may aid in computing  finite-size corrections of Lyapunov exponents~\cite{CarGP2018}.

In summary, our results fully specify the statistical behavior
of finite-size fluctuations in the KS model
which have been numerically shown to be Gaussian processes.
This opens up the way to perform systematic stochastic model reductions 
for such interacting particle systems from first principles
and to construct reduced stochastic equations
for designated collective variables \cite{Givonetal04,SmithGottwald20}.


\section*{Acknowledgements}
The work of O.E.O. was partly supported by the Deutsche Forschungsgemeinschaft under Grant No. OM 99/2-3.
O.E.O acknowledges the financial support and hospitality of the Sydney Mathematical Research Institute (SMRI), where parts of this work was completed on an International Visitor Program.
The authors are grateful for insightful discussions with A. Pikovsky.


\appendix

\section{Thermodynamic limit for the KS model~(\ref{Eq:KS:original})}
\label{app:MFT}

Most of the dynamical states found in the KS model~(\ref{Eq:KS:original}) with finite $N$, have complex chaotic behavior. However, in the thermodynamic limit $N\to\infty$, they admit a simple analytic representation. If $N\gg 1$, the state of the phase oscillators $\{\psi_i(t)\}$ can be described by a probability density function $f(\psi,\omega,t)$, which obeys the continuity equation
\begin{equation}
\pf{f}{t} + \pf{}{\psi}(f v) = 0,
\label{Eq:Continuity}
\end{equation}
where
$$
v(\psi,\omega,t) = \omega + \fr{K}{2i} \left[ e^{-i\lambda} \mathcal{Z} (t) e^{-i\psi} - e^{i \lambda} \overline{\mathcal{Z}}(t) e^{i \psi} \right]
$$
is the continuum version of the velocity field in Eq.~(\ref{Eq:KS:original}), and
\begin{equation}
\mathcal{Z}(t) = \int_{-\infty}^\infty d\omega \int_0^{2\pi} f(\psi,\omega,t) e^{i \psi} d\psi
\label{Def:W}
\end{equation}
is the continuum version of the complex order parameter~\eqref{Def:Z}.

It is well-known~\cite{OttA2008,OttA2009} that the long-term dynamics of the continuity equation (\ref{Eq:Continuity}) asymptotically approaches the Ott-Antonsen manifold for $t\to\infty$, consisting of distributions of the form
$$
f(\psi,\omega,t) = \fr{g(\omega)}{2\pi} \left\{ 1 + \sum\limits_{n=1}^\infty
\left[ \overline{z}^n(\omega,t) e^{i n \psi} + z^n(\omega,t) e^{-i n \psi} \right] \right\},
$$
where $z(\omega,t)$ satisfies the inequality $|z|\le 1$
and solves the integro-differential equation
\begin{equation}
\pf{z}{t} = i \omega z(\omega,t) + \fr{K}{2} e^{-i\lambda} \mathcal{G}z - \fr{K}{2} e^{i \lambda} z^2(\omega,t) \mathcal{G}\overline{z}
\label{Eq:OA}
\end{equation}
with the integral operator
$$
(\mathcal{G} z)(t) = \int_{-\infty}^\infty g(\omega) z(\omega,t) d\omega.
$$
Remarkably, all statistically stationary states of the KS model~(\ref{Eq:KS:original}) lie on the Ott-Antonsen manifold~\cite{OmeW2012,OmeW2013}. In particular, the zero solution $z(\omega,t) = 0$ of Eq.~(\ref{Eq:OA}) corresponds to the completely incoherent state in~(\ref{Eq:KS:original}), while all stationary partially synchronized states in~(\ref{Eq:KS:original}) are represented by rotating waves
\begin{equation}
z(\omega,t) = h\left( \fr{\omega - \Omega_\infty}{K r_\infty} \right) e^{i \Omega_\infty t},
\label{Ansatz:Wave}
\end{equation}
where $h(s)$ is defined by~\eqref{Def:h} and $(r_\infty,\Omega_\infty)\in(0,1)\times\mathbb{R}$ is a pair of numbers satisfying the self-consistency equation~\eqref{Eq:SC}.

The physical meaning of the parameters $r_\infty$ and $\Omega_\infty$ can be understood by inserting the probability density function $f(\psi,\omega,t)$ corresponding to~(\ref{Ansatz:Wave}) into Eq.~(\ref{Def:W}). This yields
\begin{equation}
\mathcal{Z}(t) = - r_\infty i e^{i \lambda} e^{i \Omega_\infty t},
\label{Z_infty}
\end{equation}
and therefore
\begin{equation}
|\mathcal{Z}(t)| = r_\infty
\label{Z_infty_}
\end{equation}
is the magnitude of the order parameter, while $\Omega_\infty$ is the angular speed of its phase.

If in the KS model~(\ref{Eq:KS:original})
we move into the corotating frame
\begin{equation}
\theta_i(t) = \psi_i(t) - \Omega_\infty t + \phi
\quad\mbox{with some}\quad\phi\in\mathbb{R},
\label{Def:Frame:phi}
\end{equation}
the probability density function in the new frame will take on the time-independent form~\eqref{Def:f_phi}, which is used for the calculations in the main text.
Importantly, for an appropriately chosen value of $\phi$,
we can guarantee that (\ref{Z_mod:infty})
yields the thermodynamic limit of the modified order parameter $Z_\mathrm{mod}(t)$.


\section{Explicit solutions of the Adler equation}
\label{app:Adler}

Let us consider the Adler equation written in complex form,  
\begin{equation}
\dot{\theta} = \Delta + \Imag\left( H e^{-i \theta} \right)
= \Delta + \fr{H e^{-i \theta} - \overline{H} e^{i \theta}}{2i}
\label{Eq:Adler}
\end{equation}
with arbitrary $\Delta \in \mathbb{R}$ and $H\in \mathbb{C}$.  
\begin{proposition}
If $|\Delta|\ne|H|$, the solution of the Adler equation (\ref{Eq:Adler}) is of the form
$$
e^{i \theta(t)} = W( t, H, \Delta, e^{i \theta(0)} )
:= \fr{ i H ( e^{i \xi t} - 1 ) - \chi_2(t) e^{i \theta(0)} }{ i \overline{H} ( e^{i \xi t} - 1 ) e^{i \theta(0)} + \chi_1(t) },
$$
where $\xi$ is defined in Eq. \eqref{Def:xi}, and
$$
\chi_1( t ) = (\Delta - \xi) e^{i \xi t} - (\Delta + \xi),
\quad
\chi_2( t ) = (\Delta + \xi) e^{i \xi t} - (\Delta - \xi).
$$
In addition, in the degenerate case $|\Delta|=|H|$, the solution of the Adler equation (\ref{Eq:Adler}) is given by
$$
e^{i \theta(t)} = W_0( t, H, \Delta, e^{i \theta(0)} )
:= z_* + \fr{e^{i \theta(0)} - z_*}{1 + ( e^{i \theta(0)} - z_* ) \overline{H} t / 2},
$$
where $z_* = i \Delta / H$. (For brevity of notation, we do not specify that $\xi$, $\chi_1(t)$, $\chi_2(t)$ and $z_*$ depend on $\Delta$ and $H$.)
\label{Prop:Adler}
\end{proposition}

{\bf Proof:}
Multiplying Eq.~(\ref{Eq:Adler}) by $i e^{i \theta}$,
we obtain a complex differential equation for $z = e^{i \theta}$:
\begin{equation}
\dot{z} = \fr{1}{2} H + i \Delta z - \fr{1}{2} \overline{H} z^2.
\label{Eq:Adler:Complex}
\end{equation}
For $|\Delta|\ne |H|$, the right-hand side of this equation can be written as
$$
\fr{1}{2} H + i \Delta z - \fr{1}{2} \overline{H} z^2
= - \fr{1}{2} \overline{H} ( z - z_+ ) ( z - z_- ),
$$
where $z_\pm = i ( \Delta \pm \xi ) / \overline{H}$.
Using the method of separation of variables and the initial condition $z(0) = \sigma$, we obtain
$$
z(t) = \fr{z_+ ( \sigma - z_- ) e^{i \xi t} - z_- ( \sigma - z_+ )}{ ( \sigma - z_- ) e^{i \xi t} - ( \sigma - z_+ )},
$$
or equivalently $e^{i \theta(t)} = W( t, H, \Delta, e^{i \theta(0)} )$.

In the degenerate case $|\Delta|=|H|$, we have $z_+ = z_- = z_*$. Integrating Eq.~(\ref{Eq:Adler:Complex}), we obtain $e^{i \theta(t)} = W_0( t, H, \Delta, e^{i \theta(0)} )$.
\qed
\\

\begin{remark}
In the case $|\Delta| \ne |H|$, we can write
$$
W( t, H, \Delta, \sigma ) = H a_1(t) + \fr{a_2( t) \sigma}{1 + \overline{H} a_1( t) \sigma},
$$
with
$$
a_1(t) = \fr{i (e^{i \xi t} - 1)}{\chi_1(t)}
\quad\mbox{and}\quad
a_2(t) = \fr{4 \xi^2 e^{i \xi t}}{\chi_1^2(t)}.
$$
Moreover, in this case, we have $|H a_1(t)| < 1$, and hence
$$
W(t, H, \Delta, \sigma) = H a_1(t) + a_2(t) \sum\limits_{n=0}^\infty (-1)^n \overline{H}^n a_1^n(t) \sigma^{n+1}
$$
is an absolutely convergent series for $|\sigma|\le 1$.
\label{Remark:W}
\end{remark}

\begin{remark}
If $(r_\infty,\Omega_\infty)\in(0,1)\times\mathbb{R}$ satisfy
the self-consistency equation~\eqref{Eq:SC}, it can be easily verified
that $z_\varphi(\omega)$ defined by~\eqref{Formula:zphi} is a fixed point
of the Adler equation~(\ref{Eq:Adler}) with $\Delta = \omega - \Omega_\infty$
and $H = - i K r_\infty e^{i \phi}$. Therefore,
$$
W\left(t, - i K r_\infty e^{i \phi}, \omega - \Omega_\infty, z_\phi(\omega) \right) = z_\phi(\omega)
$$
for all $t,\omega,\phi\in\mathbb{R}$.
\label{Remark:W_z0}
\end{remark}

\begin{remark}
Under the assumptions of Remark~\ref{Remark:W_z0},
using Remark~\ref{Remark:W} and formula~\eqref{Formula:OA}, we can show
\begin{eqnarray*}
&&
\int_0^{2\pi} e^{i \theta} \overline{W}( t, H, \Delta, e^{i \theta} ) \fr{f_\phi(\theta,\omega)}{g(\omega)} d\theta
= \int_0^{2\pi} \left[ \overline{H a_1(t)} e^{i \theta} \vphantom{\sum\limits_{n=0}^\infty} \right. \\
&&
\left. + \overline{a_2(t)} \sum\limits_{n=0}^\infty (-1)^n H^n \overline{a_1^n(t)} e^{-i n \theta} \right]
\fr{f_\phi(\theta,\omega)}{g(\omega)} d\theta \\
&&
= \overline{H a_1(t)} z_\phi(\omega) + \overline{a_2(t)} \sum\limits_{n=0}^\infty (-1)^n H^n \overline{a_1^n(t)} \overline{z}_\phi^n(\omega) \\
&&
= \overline{H a_1(t)} \left( z_\phi(\omega) - \fr{1}{\overline{z_\phi(\omega)}} \right) + \fr{1}{\overline{z_\phi(\omega)}} \overline{W( t, H, \Delta, z_\phi(\omega) )} \\
&&
= \fr{\overline{H a_1(t)}}{\overline{z_\phi(\omega)}} ( |z_\phi(\omega)|^2 - 1 ) + 1.
\end{eqnarray*}
\label{Remark:Integral}
\end{remark}


\section{Derivation of Eq.~\eqref{Formula:R:ps} from Eq.~\eqref{Formula:U1}}
\label{app:Derivation}

Using~\eqref{Def:h} it is easy to verify that
\begin{equation}
|H|\,  h\left( \fr{\Delta}{|H|} \right) = \left\{
\begin{array}{lcl}
\Delta - \xi\:\mathrm{sgn}(\Delta) & \mbox{ for } & |\Delta| > |H|,\\[2mm]
\Delta + \xi & \mbox{ for } & |\Delta|\le |H|,
\end{array}
\right.
\label{Formula:H:h}
\end{equation}
where $\xi$ is given by~\eqref{Def:xi}.
Note that formula~\eqref{Formula:H:h} holds for all $\Delta\in\mathbb{R}$ and $H\in\mathbb{C}$.

\begin{figure*}
\begin{center}
\begin{overpic}
[width=0.32\textwidth,angle=0]{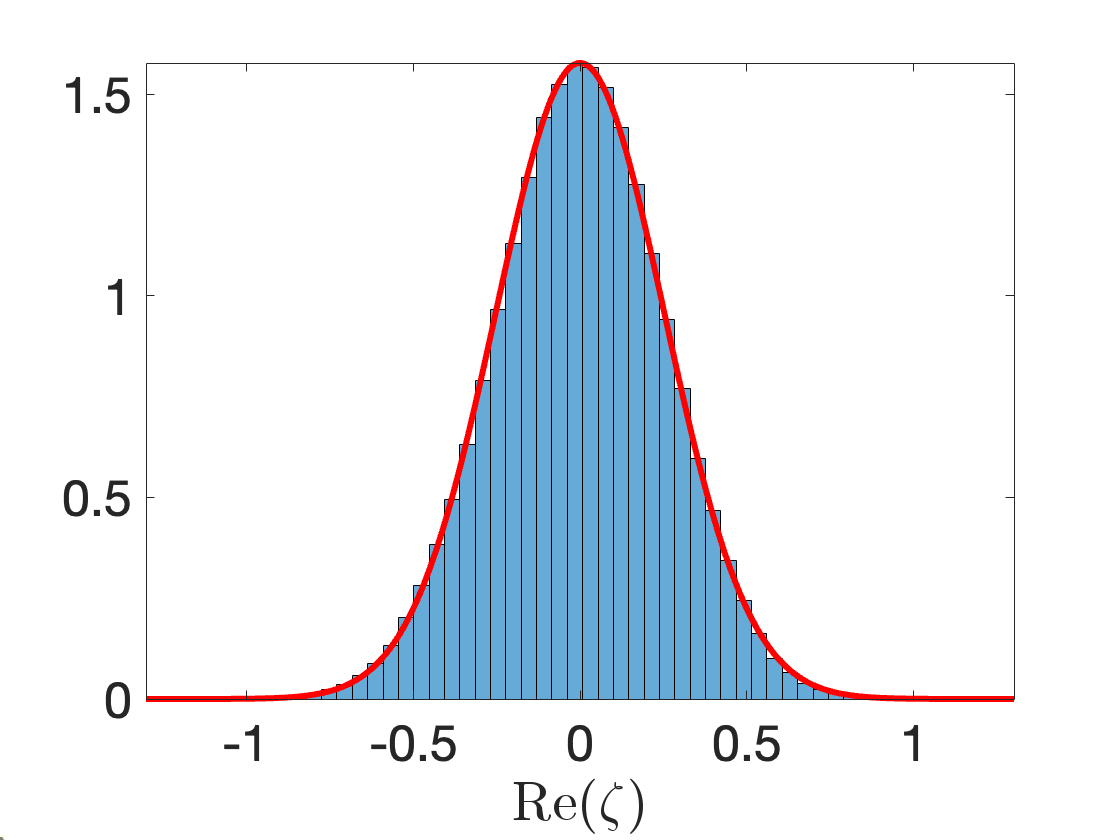}
     \put(47,-7){\small (a)}
\end{overpic}
\begin{overpic}
[width=0.32\textwidth,angle=0]{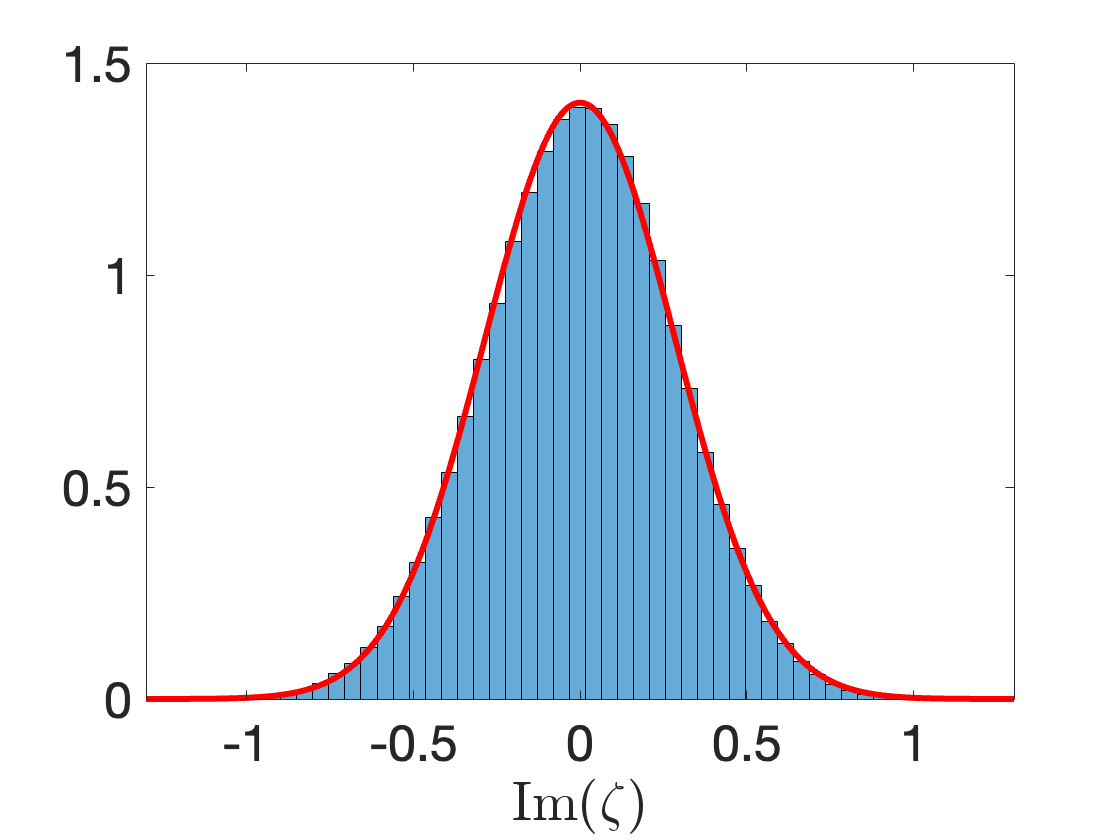}
     \put(47,-7){\small (b)}
\end{overpic}\begin{overpic}
[width=0.32\textwidth,angle=0]{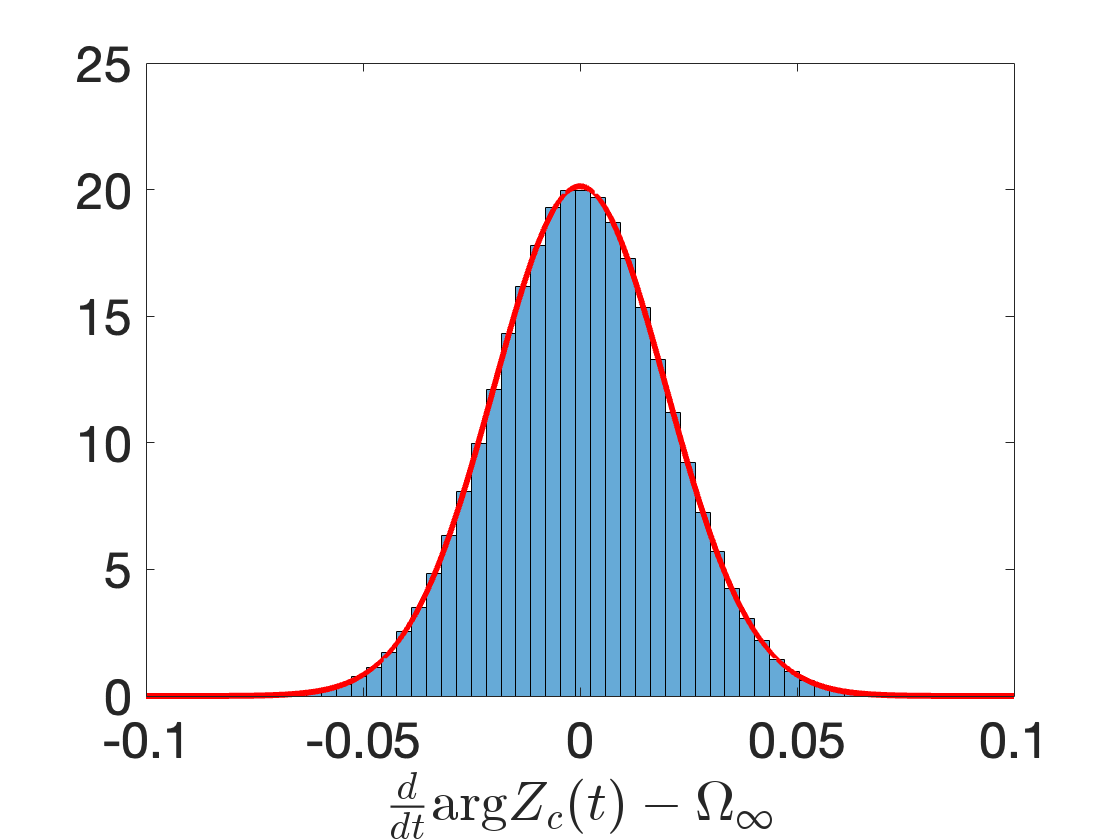}
     \put(47,-7){\small (c)}
\end{overpic}
\end{center}
\caption{
Empirical histograms of the real (a) and imaginary (b) part of the fluctuations $\zeta$
and of $d[{\rm{arg}}Z_\mathrm{c}(t)]/dt$ centered around $\Omega_\infty\approx -1.71$
(c) for the KS model (\ref{Eq:KS:original}) with $N=1,000$ oscillators,
phase lag $\lambda=\pi/4$, and $K=3$.
The continuous lines (online red) show a best-fit Gaussian with the same variance. 
}
\label{Fig:Gaussianity}
\end{figure*}
Now, we consider the integrand of Eq.~\eqref{Formula:U1}, 
recalling that $H = K Z_\infty e^{-i \lambda} = - i K r_\infty e^{i \phi}$.
Recognizing that in this case
$$
\fr{H}{z_\phi(\omega)} = - \fr{i K r_\infty e^{i \phi}}{e^{i \phi} h\left(\fr{\omega-\Omega_\infty}{K r_\infty}\right)}
= - \fr{i |H|}{h\left(\fr{\Delta}{|H|}\right)}
$$
does not depend on $\phi$, we find
\begin{eqnarray}
&&
\fr{H a_1(\tau)}{z_\phi(\omega)} - 1 =
- \fr{i |H|}{h\left(\fr{\Delta}{|H|}\right)} \fr{i (e^{i \xi t} - 1)}{\chi_1(t)} - 1 \nonumber\\[2mm]
&&
=
\fr{|H|^2}{|H| h\left(\fr{\Delta}{|H|}\right)} \fr{e^{i \xi t} - 1}{(\Delta - \xi) e^{i \xi t} - (\Delta + \xi)} - 1.
\label{Formula:H:a:z_phi}
\end{eqnarray}
For $\Delta > |H|$, upon using~\eqref{Formula:H:h} and the identity $\Delta^2 - \xi^2 = |H|^2$,
\eqref{Formula:H:a:z_phi} is written as
\begin{eqnarray*}
&&
\fr{H a_1(\tau)}{z_\phi(\omega)} - 1 =
\fr{|H|^2 ( e^{i \xi t} - 1 )}{(\Delta - \xi)^2 e^{i \xi t} - (\Delta + \xi) (\Delta - \xi)} - 1 \\[2mm]
&&
= \fr{e^{i \xi t} - 1}{h^2(\Delta/|H|) e^{i \xi t} - 1} - 1
= \fr{1 - h^2(\Delta/|H|)}{h^2(\Delta/|H|) - e^{-i \xi t}}.
\end{eqnarray*}
Similarly, for $\Delta \le |H|$ we obtain
$$
\fr{H a_1(\tau)}{z_\phi(\omega)} - 1
= \fr{1 - h^2(\Delta/|H|)}{h^2(\Delta/|H|) - e^{i \xi t}},
$$
and thus in both cases we have
$$
\fr{H a_1(\tau)}{z_\phi(\omega)} - 1
= \fr{1 - h^2(\Delta/|H|)}{h^2(\Delta/|H|) - e^{-i \Delta \sqrt{1 - |H|^2/\Delta^2} t}}.
$$
Substituting into Eq.~\eqref{Formula:U1}, we obtain Eq.~\eqref{Formula:R:ps}.


\section{Pseudo-covariance function for a partially synchronized state}
\label{app:Pseudocov}

Using the definition of the pseudo-covariance function $\tilde R_{\rm{ps}}(t) = \langle \zeta(t) \zeta(t+\tau) \rangle$
and employing the established independence of the functions $e^{i \theta_j(t)}$ and $e^{i \theta_k(t)}$ for $j\ne k$,
by analogy with \eqref{Formula:R:ps2}, we obtain
$$
\tilde{R}_\mathrm{ps}(\tau) = \frac{1}{N}
\sum\limits_{j\in\mathcal{R}} \langle e^{i \theta_j(t)} e^{i \theta_j(t+\tau)} \rangle
- \frac{1}{N}
\sum\limits_{j\in\mathcal{R}} \langle e^{i \theta_j(t)} \rangle^2.
$$
Using the ergodicity property~\eqref{Ergodicity}
and performing calculations as in Remark~\ref{Remark:Integral}, we find
\begin{eqnarray*}
&&
\langle e^{i \theta_j(t)} e^{i \theta_j(t+\tau)} \rangle
=
\int\limits_0^{2\pi} e^{i \theta} W( \tau, H_\infty,\Delta_j, e^{i \theta} ) \fr{f_\phi(\theta,\omega_j)}{g(\omega_j)} d\theta \\
&&
= z_\phi(\omega_j) W( \tau, H_\infty, \Delta_j, z_\phi(\omega_j) ) = z_\phi^2(\omega_j).
\end{eqnarray*}
Recalling that $\langle e^{i \theta_j(t)} \rangle = z_\phi(\omega_j)$,
we obtain $\tilde{R}_\mathrm{ps}(\tau) = 0$.
Note that this relation is valid regardless of the choice of $\phi$
in the definition of the corotating frame~\eqref{Def:Frame:phi}.


\section{Details of the numerical setup}
\label{app:num}

Given a distribution $g(\omega)$, we generate a set of $N$ natural frequencies $\omega_j$ according to
$$
\int_{-\infty}^{\omega_j} g(\omega) d\omega = \fr{j}{N + 1},
\qquad
j=1,\dots,N.
$$
This equiprobable sampling avoids finite-size effects such as randomly clustered natural frequencies leading to small synchronized clusters with non-zero mean frequencies and their respective interactions~\cite{PetP2018,FialkowskiEtAl23}. For each pair of parameters $K$ and $\lambda$, we perform simulations using a standard Runge-Kutta solver with constant time step $dt = 0.02$. We start from initial conditions chosen randomly from the interval $[0,2\pi]$, discard a transient period of $T_\mathrm{transient} = 10^5$ time units and use the subsequent interval of length $T_\mathrm{max} = 10^5$ time units to calculate statistical quantities such as means, variances, and covariances.


\section{Numerical evidence for the assumption on the Gaussianity of fluctuations}
\label{app:CLT}

Previous work \cite{YueG2024,IrvG2025} has shown numerically
that finite-size fluctuations in the Kuramoto-Sakaguchi model (\ref{Eq:KS:original})
constitute a Gaussian process, consistent with an underlying central limit theorem.
For completeness, we provide here further numerical evidence.
Fig.~\ref{Fig:Gaussianity}(a) and (b) show the empirical histogram
of the real and the imaginary part of the fluctuations $\zeta$
obtained from a long time simulation over $50,000$ time units
with a sampling time of $dt=0.05$ of the KS model together
with a best fit Gaussian clearly establishing the Gaussian character of the fluctuations.
For $K=3$ we find a value of the skewness for the real and imaginary part
of the fluctuations $\zeta$ of $-0.058$ and $-0.051$, respectively,
and a kurtosis of $2.98$ for both. We further show in Fig.~\ref{Fig:Gaussianity}(c)
the empirical histogram for the temporal derivative of the mean phase
of the entrained oscillators $\tfrac{d}{dt}{\rm{arg}} Z_c(t)$,
which also shows a clear Gaussian statistics,
consistent with the phase diffusion discussed below.
\begin{figure*}
\begin{center}
\begin{overpic}
    [width=0.32\textwidth,angle=0]{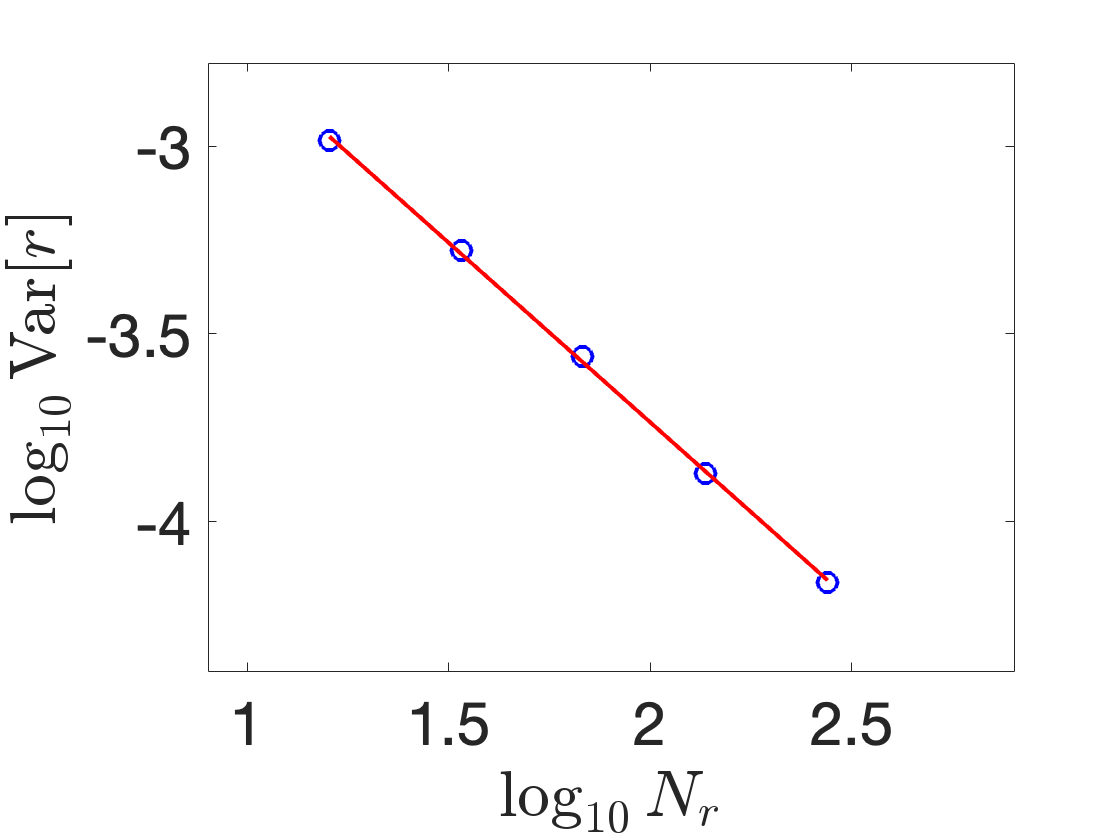}
     \put(47,-7){\small (a)}
\end{overpic}
\begin{overpic}
[width=0.32\textwidth,angle=0]{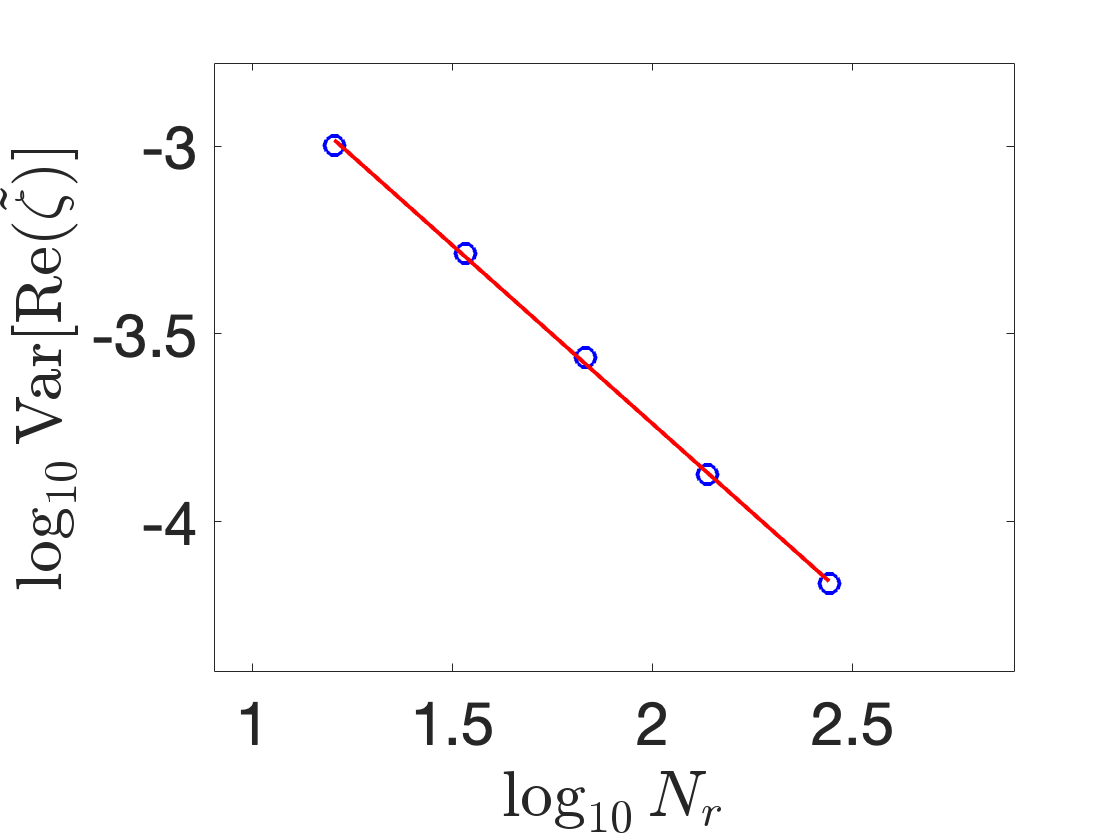}
     \put(47,-7){\small (b)}
\end{overpic}\begin{overpic}
[width=0.32\textwidth,angle=0]{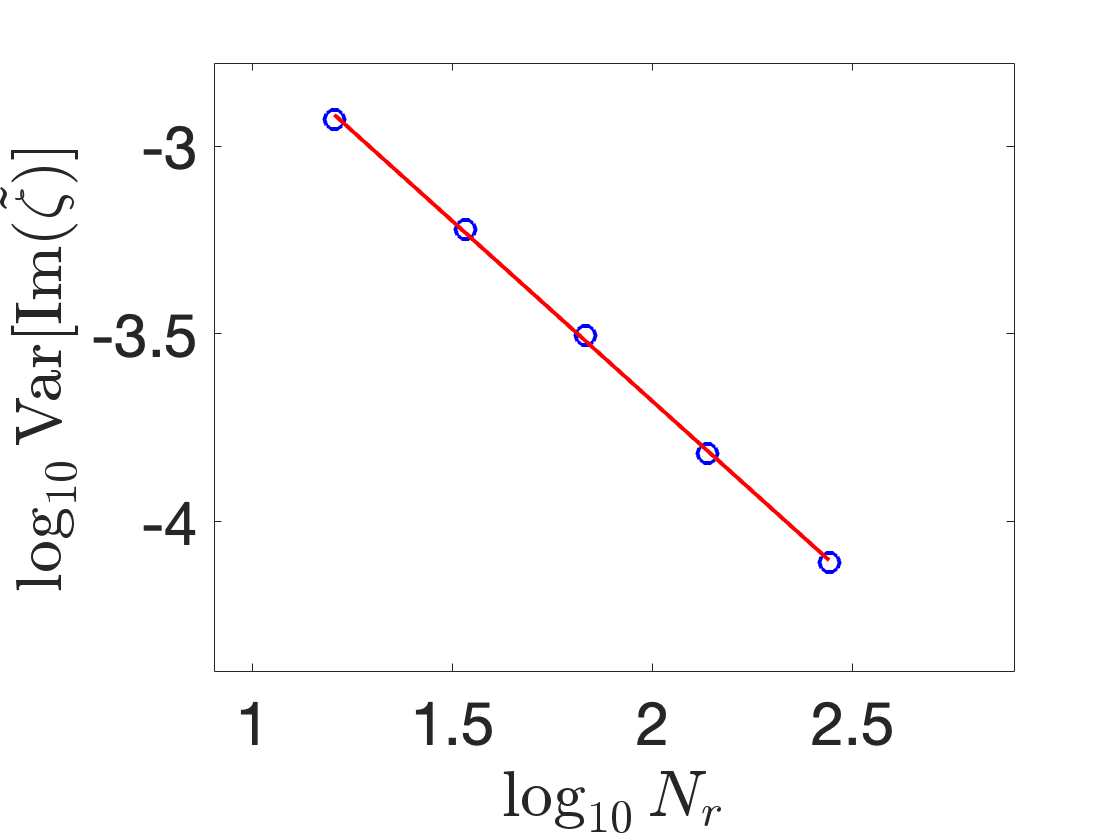}
     \put(47,-7){\small (c)}
\end{overpic}
\end{center}
\caption{
Variance of the order parameter $r(t)$ (a) and of the real (b) and imaginary (c) part of the fluctuations $\tilde \zeta(t) =  Z_{\rm{mod}}(t)-\langle Z_{\rm{mod}} \rangle$ as a function of the non-entrained oscillators $N_r$ for the KS model (\ref{Eq:KS:original}) with phase lag $\lambda=\pi/4$ and $K=3$ for $N=60, 125, 250, 500, 1000$. The continuous lines (online red) are lines of best fit indicating a scaling law of $N_r^{-0.96}$, $N_r^{-0.96}$ and $N_r^{-0.95}$, respectively.
}
\label{Fig:V_N}
\end{figure*}

To further establish numerical evidence for an underlying central limit theorem
we show in Fig.~\ref{Fig:V_N} that the variance of the fluctuations $\zeta/\sqrt{N}$
scales with the number of non-entrained oscillators $N_r$ as $1/N_r$,
as expected for Gaussian fluctuations described by an underlying central limit theorem.
We show the scaling for the variance of the order parameter in Fig.~\ref{Fig:V_N}(a)
and of the real and imaginary part of the fluctuations $\zeta$ in Fig.~\ref{Fig:V_N} (b)
and (c), respectively.
\begin{figure}[b]
\begin{center}
\includegraphics[width=0.9\columnwidth,angle=0]{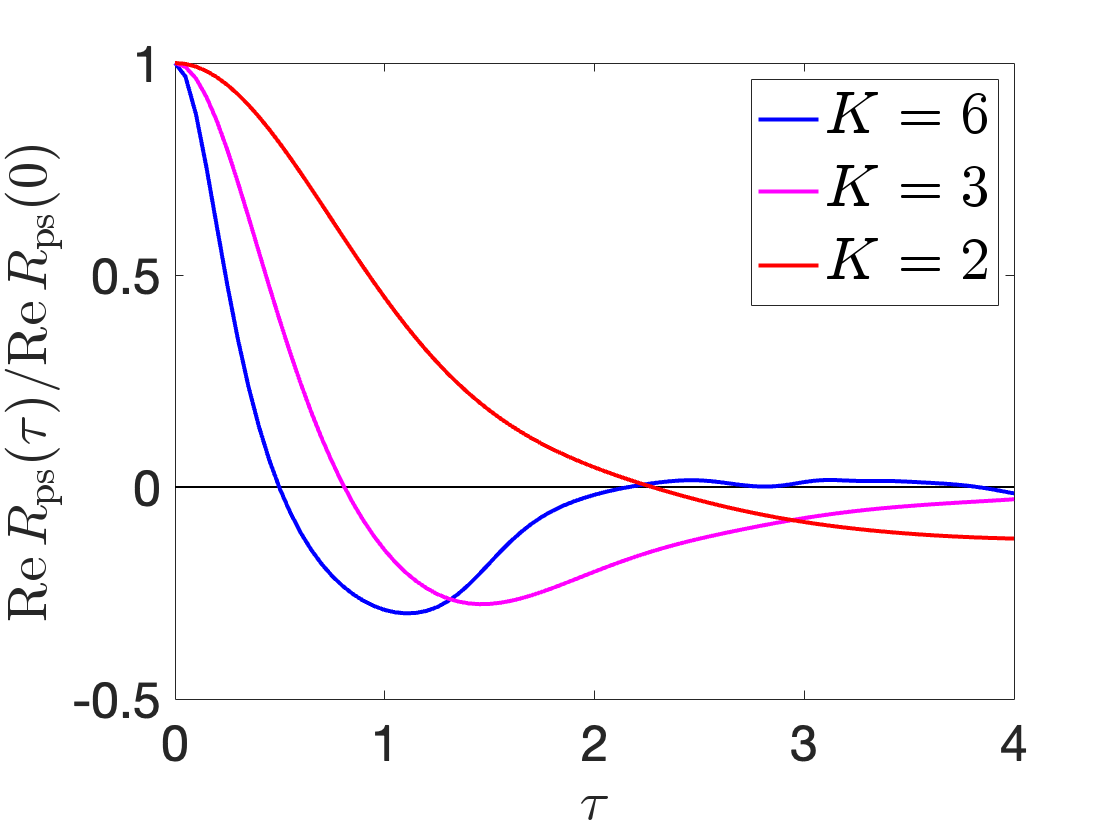}
\end{center}
\caption{
Normalized covariance $R_{\rm{ps}}(\tau)/R_{\rm{ps}}(0)$ (real part only)
of the KS model~(\ref{Eq:KS:original}) for partially synchronized states
with $N=1,000$ oscillators at $\lambda = \pi/4$
for various values of the coupling strength $K$.
}
\label{Fig:Cov_K}
\end{figure}

We remark that the central limit theorem holds
for all values of the coupling strength $K>K_{\rm{crit}}$.
For $K=1.98$ we find a value of the skewness for the real and imaginary part
of the fluctuations $\zeta$ of $-0.024$ and $-0.007$,
and a kurtosis of $2.98$ and $3.04$, respectively, suggestive of a Gaussian distribution.
The Gaussian character of the empirical probability density function of the fluctuations alone,
however, is not sufficient for the existence of a functional central limit theorem
which is concerned with convergence to a stochastic process
rather than to just random variables.
We show in Figure~\ref{Fig:Cov_K} plots of the covariance function of the real part
of the fluctuations, $\mathrm{Re}\, R_\mathrm{ps}(\tau)$
which decays to zero for all $K> K_{\rm{crit}}$
with the decay of correlation time increasing when approaching criticality.


\section{Validity and limitations of the theory}
\label{app:lim}

\begin{figure}[h]
\begin{center}
\includegraphics[height=0.9\columnwidth,angle=270]{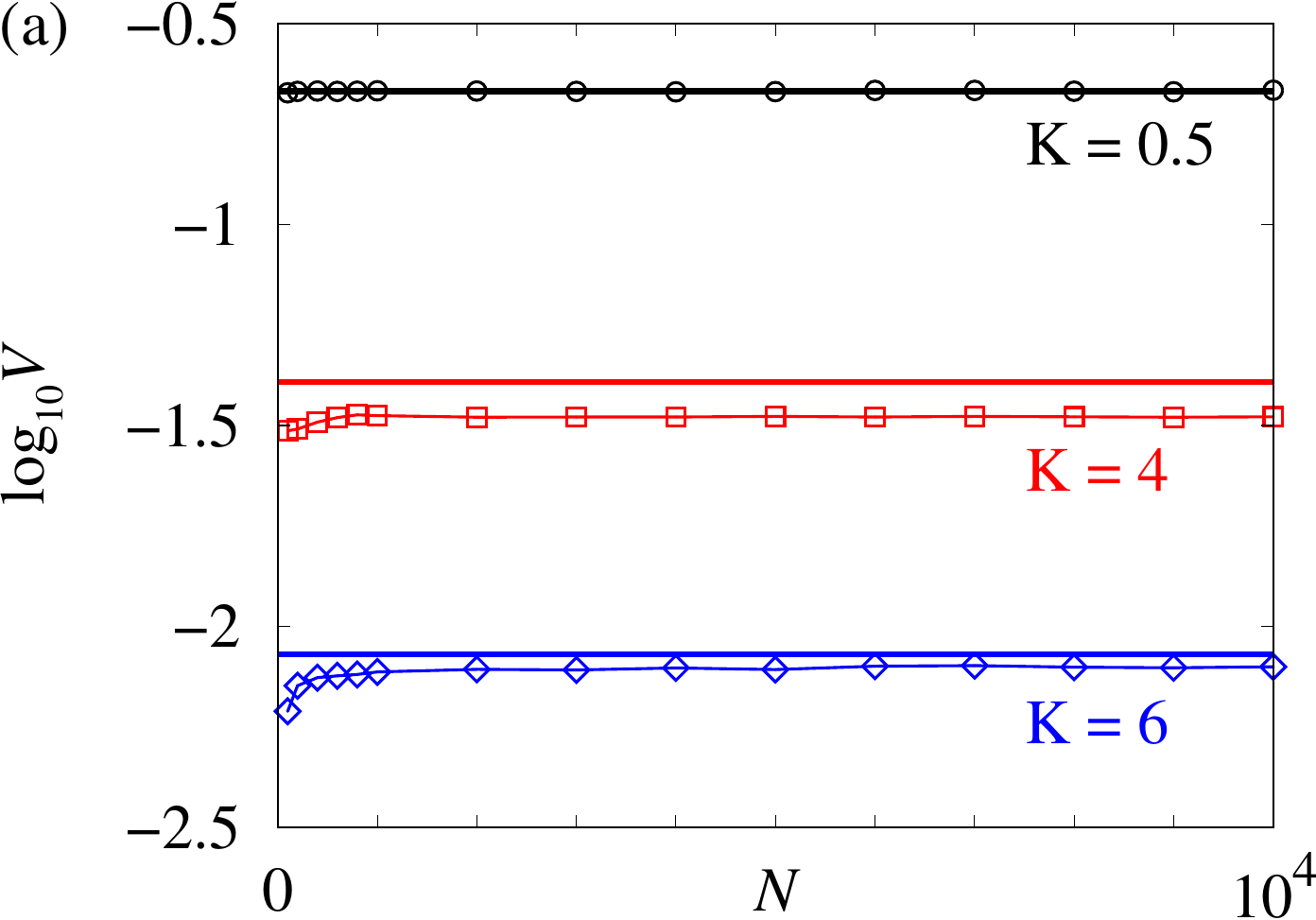}\\[4mm]
\includegraphics[height=0.9\columnwidth,angle=270]{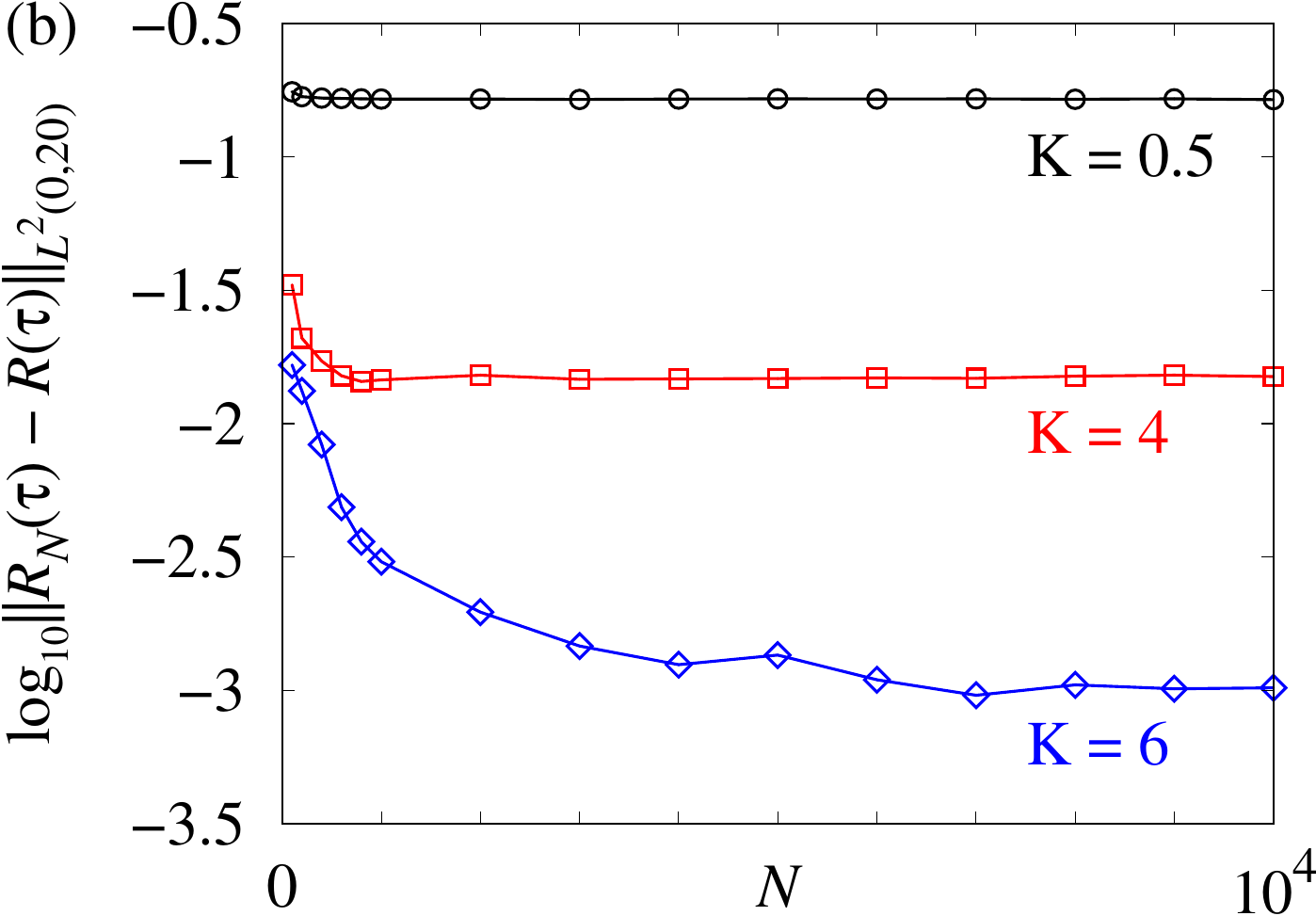}
\end{center}
\caption{
(a) Scaling of the variance~$V$
with system size $N$ 
for three different values of the coupling strength $K$, with $\lambda=\pi/4$.
Symbols show empirical values
obtained from numerical simulations
of the KS model~(\ref{Eq:KS:original}). 
Thick lines show the theoretical predictions given by formulas~(\ref{Formula:V:ps}) and~(\ref{Formula:incoh:r}).  
(b) $L^2$-distance between the empirical
covariance $R_N(\tau)$ and its theoretical prediction $R(\tau)$ given by formulas~(\ref{Formula:R:ps}) and~(\ref{Formula:R:incoh}). 
Parameters as in panel~(a).
}
\label{Fig:VarL2ScalingN}
\end{figure}
\begin{figure}[ht]
\begin{center}
\includegraphics[height=0.9\columnwidth,angle=270]{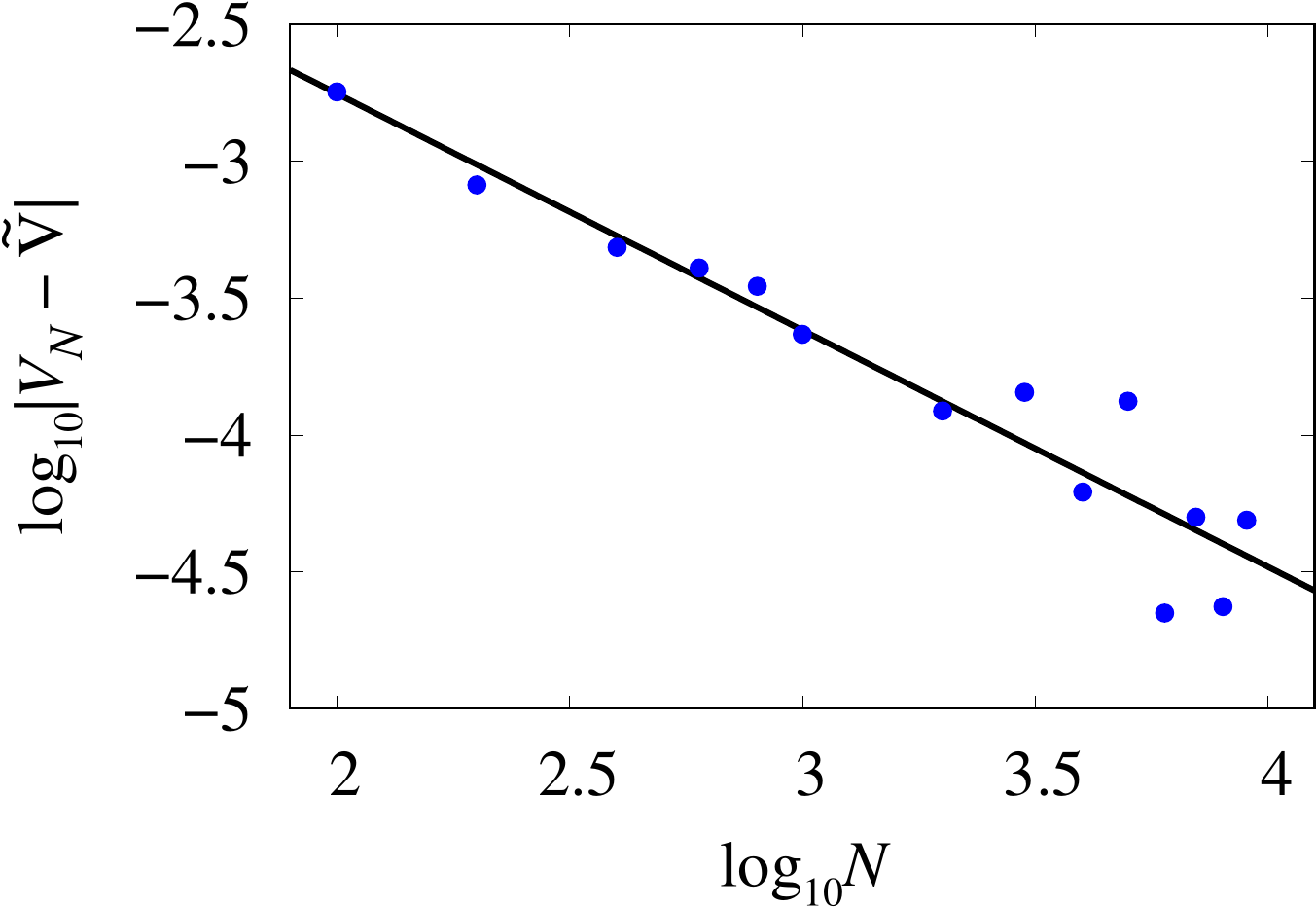}
\end{center}
\caption{
Scaling of the discrepancy between the empirical value of the variance $V_N$ and the average~$\tilde{V}$
of the last four points
for $K = 6$ and $\lambda=\pi/4$.
The line shows a best linear fit suggesting $|V_N-\tilde V| \sim 1/N^{0.86}$.
}
\label{Fig:ScalingK6_0}
\end{figure}
\begin{figure}[ht]
\begin{center}
\includegraphics[height=0.9\columnwidth,angle=270]{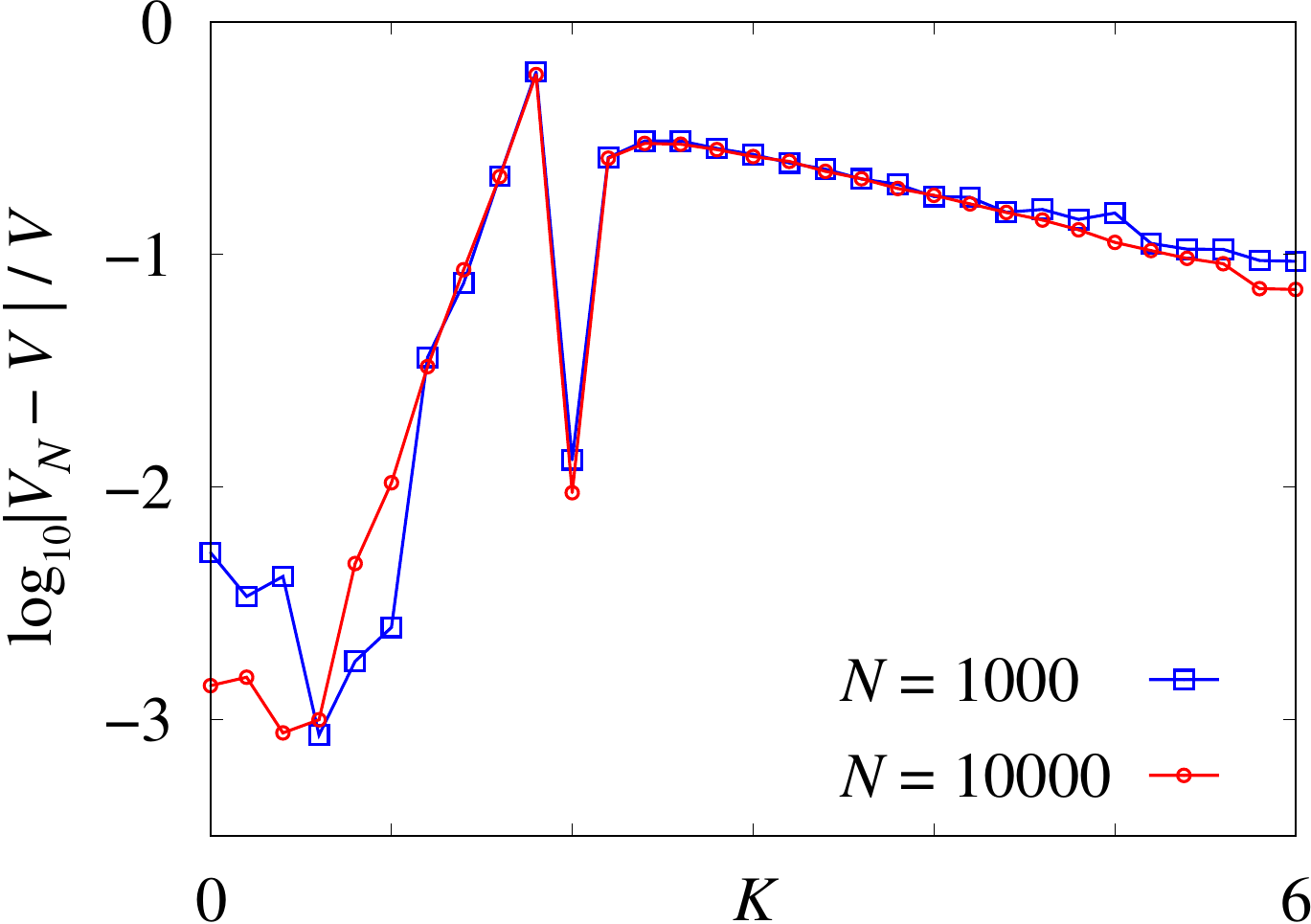}
\end{center}
\caption{
Relative discrepancy between the empirical value of the variance $V_N$ and its theoretical prediction~$V$
given by formulas~(\ref{Formula:V:ps}) and~(\ref{Formula:incoh:r}) as function of the coupling strength $K$. The critical coupling strength is approximately given by $K_\mathrm{crit}\approx 1.92$. 
Parameters: $\lambda=\pi/4$.
}
\label{Fig:VarScalingK_rel}
\end{figure}

The analytical results obtained in the main text
are approximate in two respects.
On the one hand, if we fix the parameters $K$
and $\lambda$ in the KS model~(\ref{Eq:KS:original})
and vary the system size~$N$,
we observe that the empirical value of the variance~$V_N$
changes slightly with increasing $N$
and then remains almost constant for $N\ge 1,000$,
see Fig.~\ref{Fig:VarL2ScalingN}(a).
Moreover, for large values of~$K$
these changes exhibit a power law scaling
shown in Fig.~\ref{Fig:ScalingK6_0}. Recall that the classical scaling $1/N$ of the variance has already been explicitly used in the definition of the fluctuations (cf. (\ref{Def:zeta}) and (\ref{Def:delta})). 
On the other hand, even for large~$N$
there is a small discrepancy
between the theoretical predictions~(\ref{Formula:V:ps}) and~(\ref{Formula:incoh:r})
and the empirical characteristics of the fluctuations
in the KS model~(\ref{Eq:KS:original}). This discrepancy, however, 
decreases the further the parameter $K$
is from its critical value at $K\approx 1.92$,
see Fig.~\ref{Fig:VarScalingK_rel}.
A similar scaling behavior also holds for the $L^2$-distance between
the empirical covariance $R_N(\tau)$
and its theoretical prediction $R(\tau)$
given by formulas~(\ref{Formula:R:ps}) and~(\ref{Formula:R:incoh}),
see Fig.~\ref{Fig:VarL2ScalingN}(b).
Thus, we conjecture that our analytical results
are accurate only in the asymptotic regime
of large system sizes $N$ and equation parameters $K$ and $\lambda$, far from the synchronization transition point.
Outside this regime, we expect our results
to provide only good approximations of the actual values.

\begin{figure*}[t]
\begin{center}
\begin{overpic}
    [width=0.32\textwidth,angle=0]{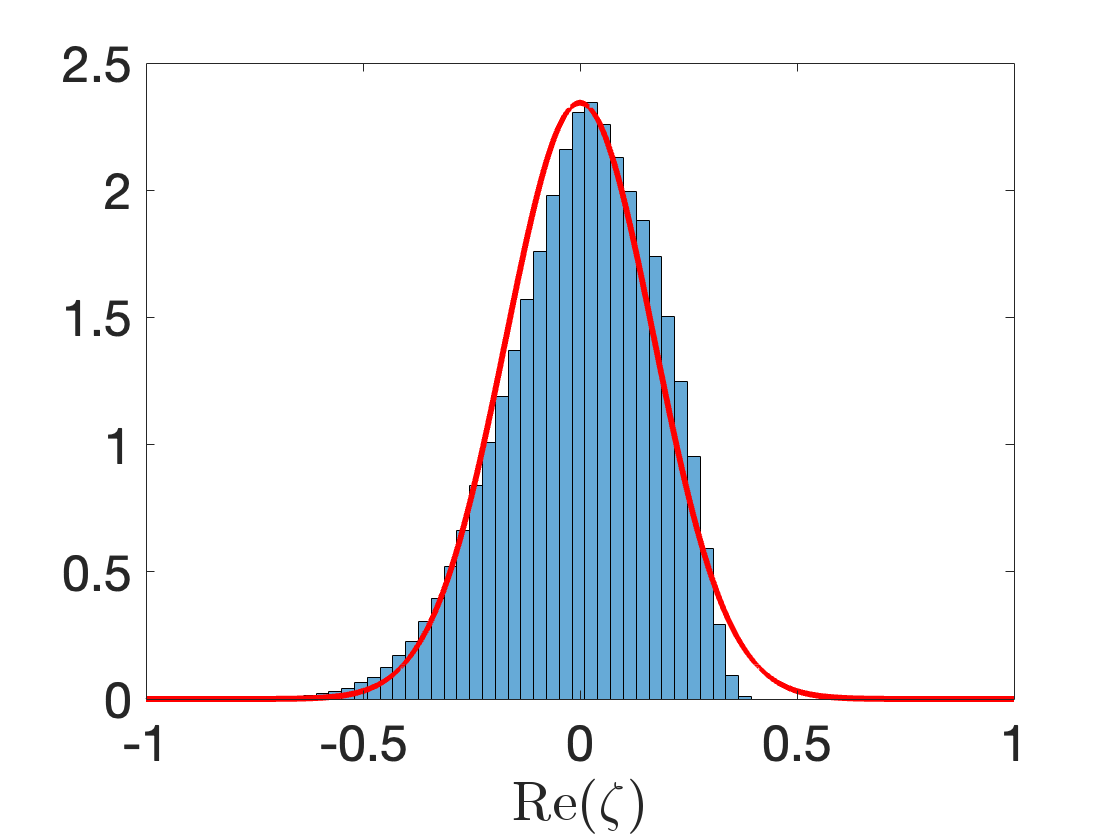}
     \put(47,-7){\small (a)}
\end{overpic}
\begin{overpic}
[width=0.32\textwidth,angle=0]{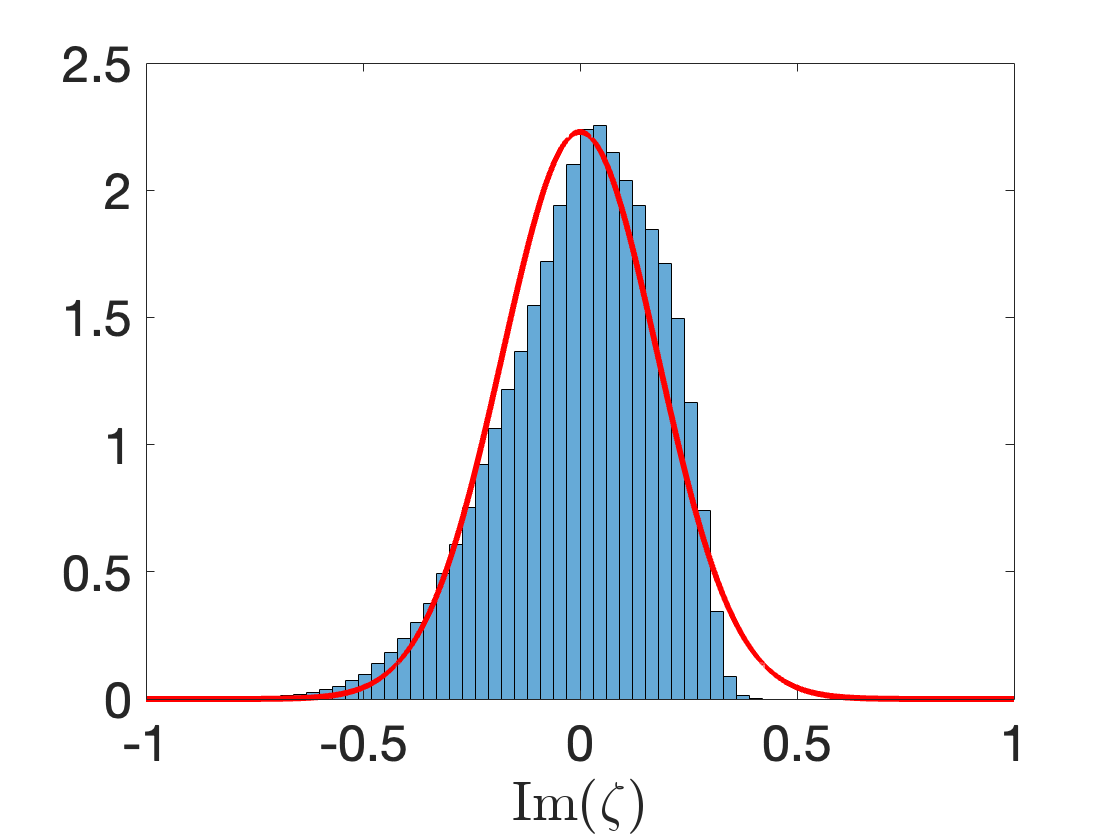}
     \put(47,-7){\small (b)}
\end{overpic}\begin{overpic}
[width=0.32\textwidth,angle=0]{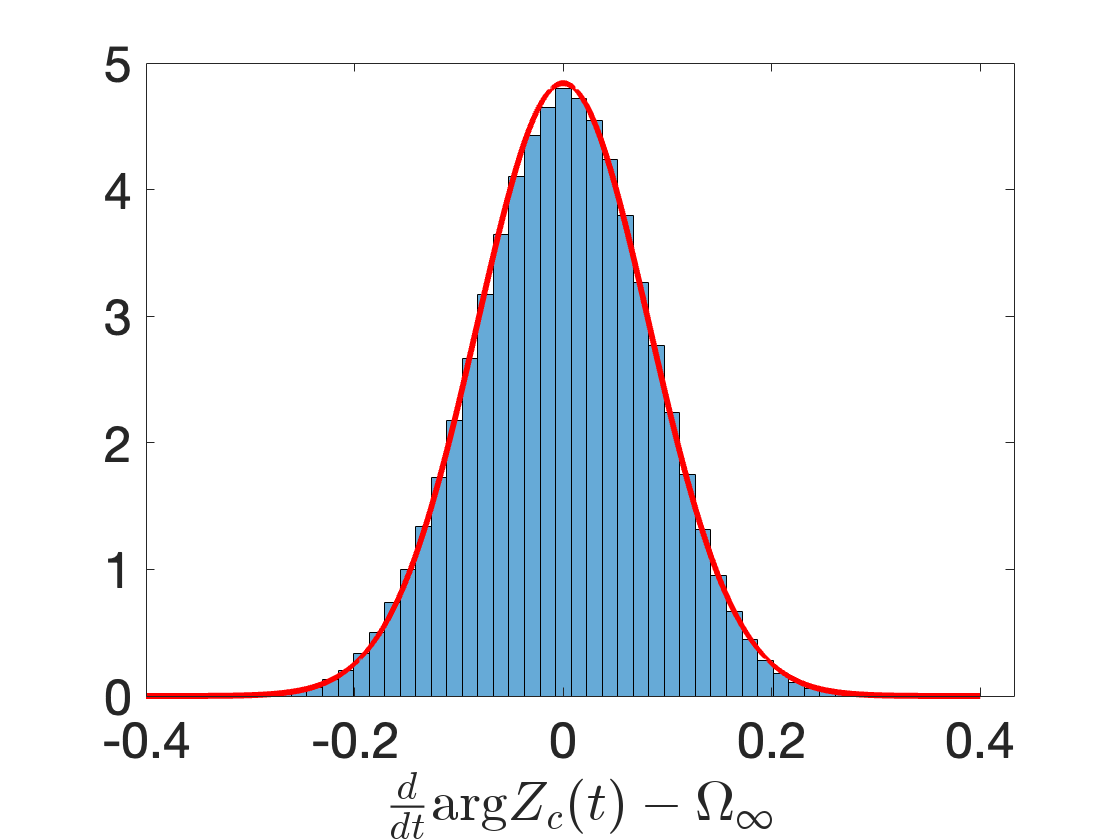}
     \put(47,-7){\small (c)}
\end{overpic}\end{center}
\caption{
Empirical histograms of the real (a) and imaginary (b) part of the fluctuations $\zeta$ and of $d[{\rm{arg}}Z_\mathrm{c}(t)]/dt$ centered around $\Omega_\infty\approx  -2.39$ (c) for the KS model (\ref{Eq:KS:original}) with $N=50$ oscillators,
phase lag $\lambda=\pi/4$, and $K=4$.
The continuous lines (online red) show a best-fit Gaussian with the same variance.
}
\label{Fig:GaussianityN50}
\end{figure*}

In the following we will further investigate the requirement for sufficiently large system sizes $N$ and the requirement of being sufficiently far from criticality, and how it relates to our assumptions of Gaussian fluctuations and the replacement of the time-varying rotation frequency and order parameter by their stationary thermodynamic limits in the Adler equation.
From the previous Appendix~\ref{app:CLT} it is clear that in order for the central limit theorem to be valid we require sufficiently large values of $N$. Indeed, for small values of $N=50$ we only have $N_r=6$ for $K=4$. Figure~\ref{Fig:GaussianityN50}(a) and (b) clearly show that the empirical histograms of the real and imaginary parts of fluctuations $\zeta$ are far from Gaussian. The speed of the phase of the synchronized order parameter $d[{\rm{arg}}Z_\mathrm{c}(t)]/dt$ is still Gaussian, illustrating the phase diffusion discussed below.
\begin{figure}
\begin{center}
\includegraphics[width=0.45\columnwidth,angle=270]{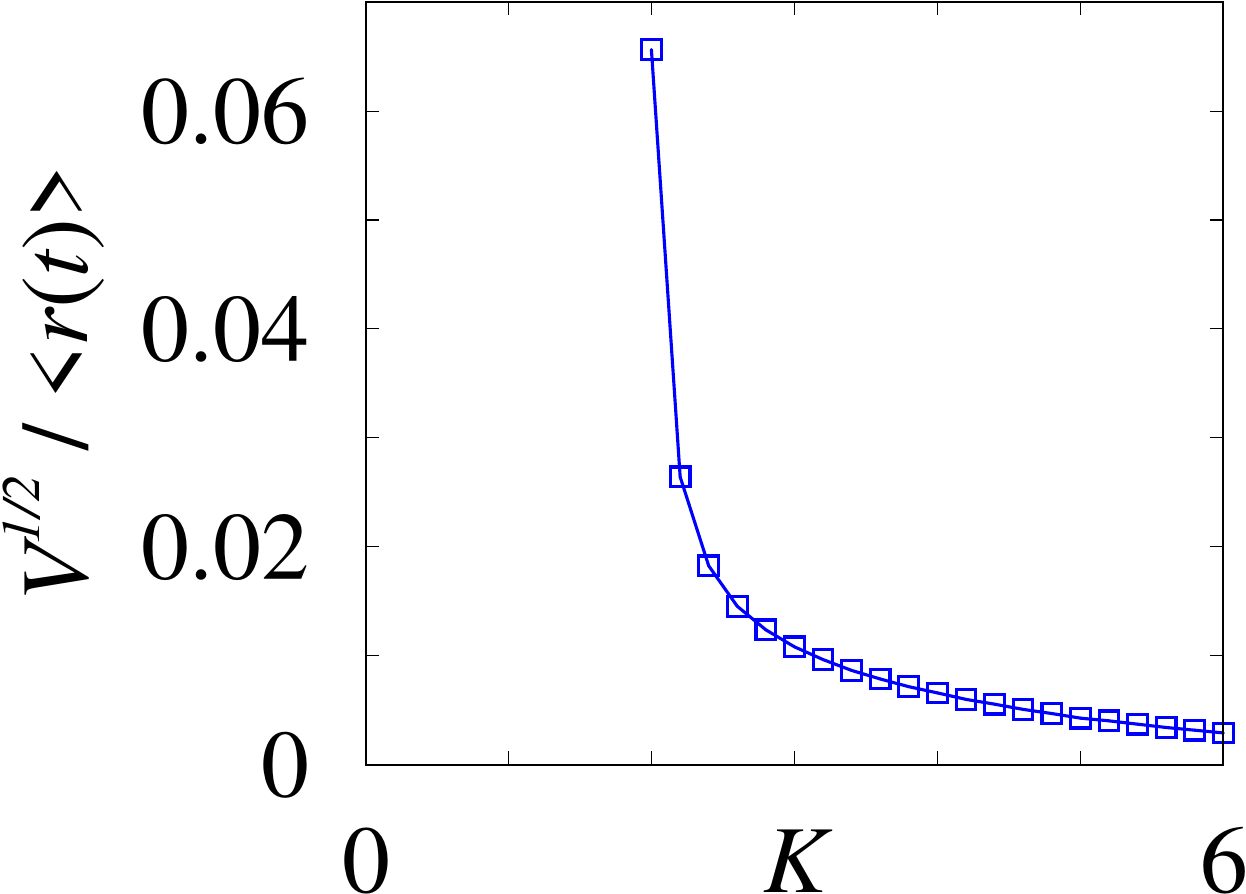}
\end{center}
\caption{
Ratio of the standard deviation of the fluctuations $\sqrt{V}$ of the order parameter $r(t)$ and its mean $\langle r(t)\rangle$ for $N=1,000$ oscillators at $\lambda=\pi/4$ as a function of the coupling strength $K$.
}
\label{Fig:ratio_K}
\end{figure}

The explicit expressions for the covariance functions crucially depends
on the explicit solution of the Adler equation (\ref{eq:Adler0}).
To obtain the Adler equation we neglected fluctuations of the modified order parameter,
replacing $Z_\mathrm{mod}(t)$ by its constant thermodynamic limit.
This is only valid provided the fluctuations are small in size
compared to the thermodynamic mean values.
Approaching criticality, however, the size of fluctuations increases.
Figure~\ref{Fig:ratio_K} shows the ratio of the standard deviation
of the fluctuations $\sqrt{V}$ of the order parameter $r(t)$
and its mean as a function of the coupling strength $K$.
It is clearly seen how fluctuations increase rapidly in size
when approaching criticality at $K_{\rm{crit}}\approx 1.92$.
This implies that the Adler equation and its solution become less valid closer to criticality.
Hence, the quality of our theory depends on the value of the coupling strength $K$.
Indeed, Figure 2(c) in the main text illustrates that our theory works well
for coupling strengths sufficiently far from criticality,
and gets worse the closer we approach criticality.


\section{Additional numerical results}
\label{app:addnum}
\begin{figure}
\begin{center}
\includegraphics[height=0.48\columnwidth,angle=270]{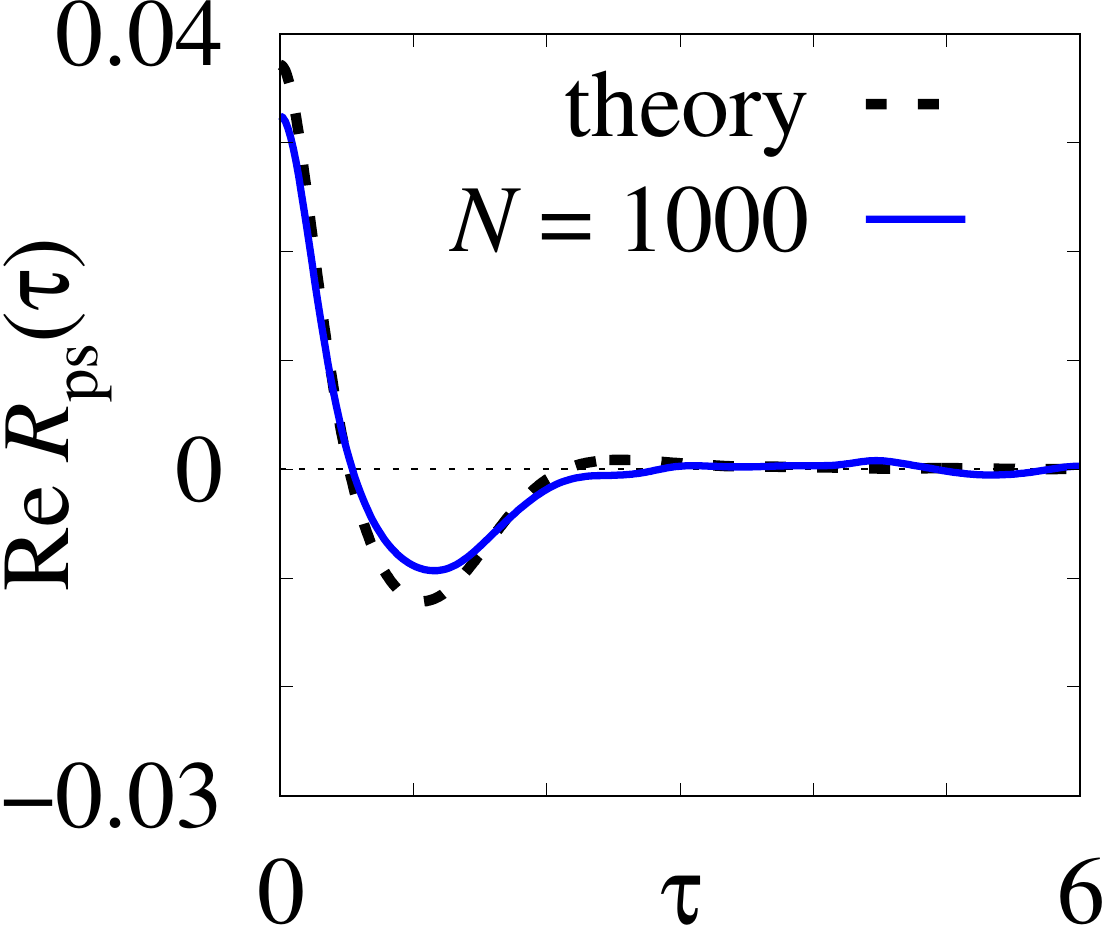}\hspace{0.02\columnwidth}
\includegraphics[height=0.48\columnwidth,angle=270]{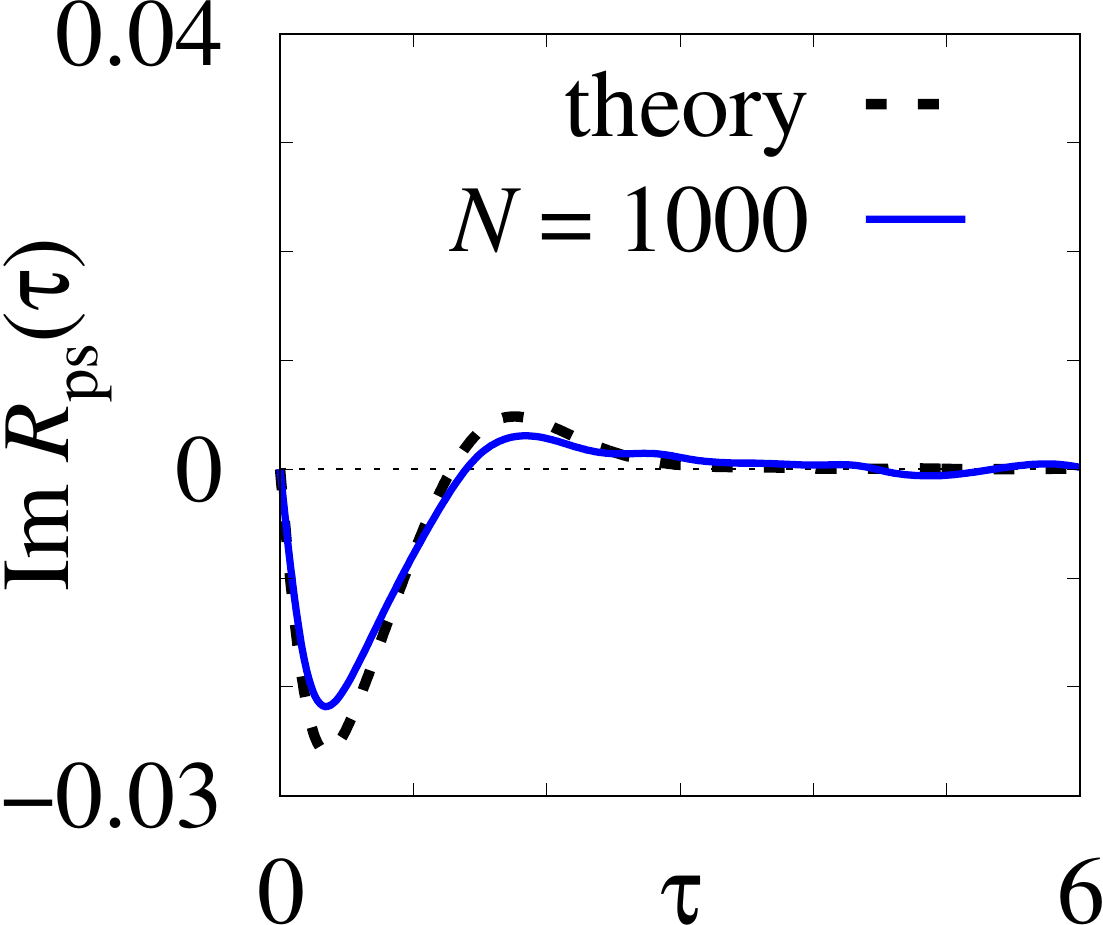}\\[1mm]
\includegraphics[height=0.48\columnwidth,angle=270]{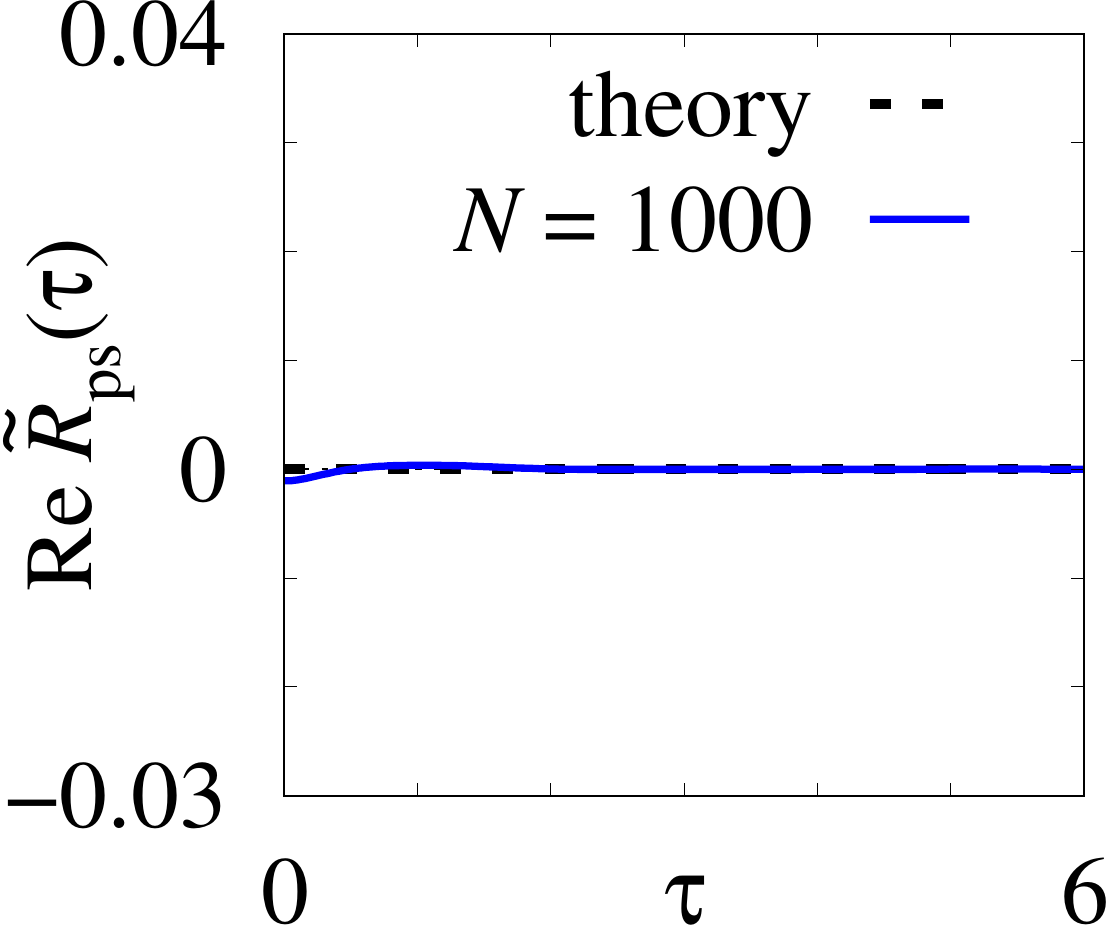}\hspace{0.02\columnwidth}
\includegraphics[height=0.48\columnwidth,angle=270]{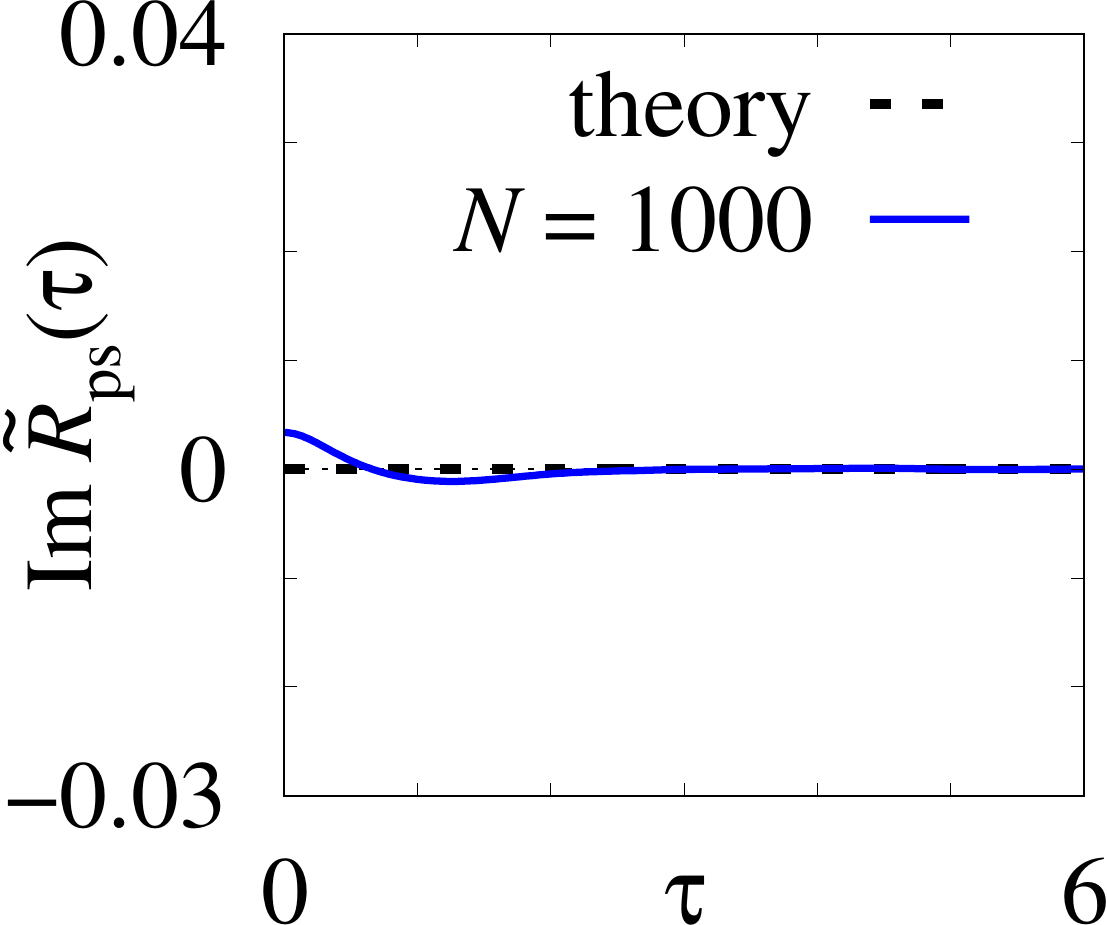}
\end{center}
\caption{Covariance $R(\tau)$ and pseudo-covariance $\tilde{R}(\tau)$
of the complex order parameter fluctuation $\zeta(t)$
for a partially synchronized state at $K = 5$ and $\lambda = \pi/4$
in the KS model \eqref{Eq:KS:original}.
Numerical simulations (solid curve) vs. theoretical prediction \eqref{Formula:R:ps} (dashed curve).}
\label{Fig:Covariances:K5}
\end{figure}

\begin{figure}
\begin{center}
\includegraphics[height=0.48\columnwidth,angle=270]{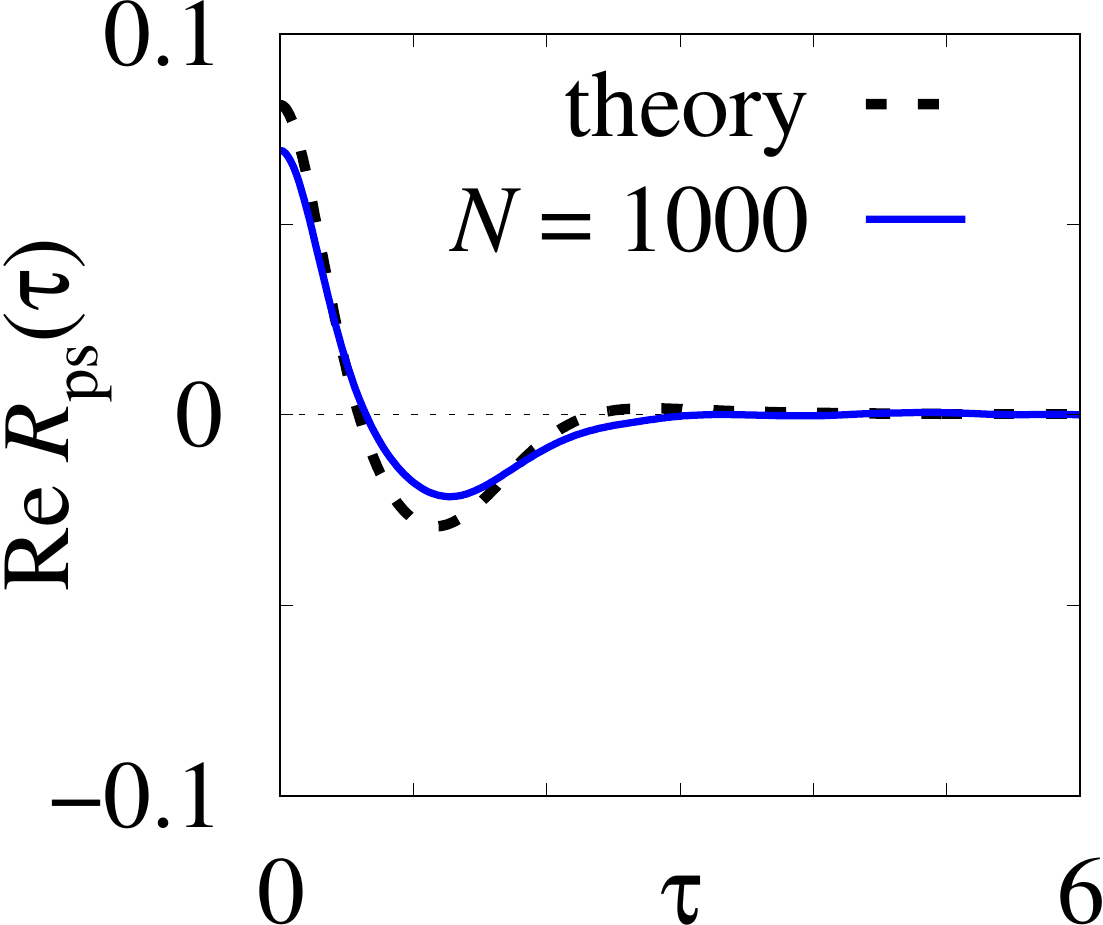}\hspace{0.02\columnwidth}
\includegraphics[height=0.48\columnwidth,angle=270]{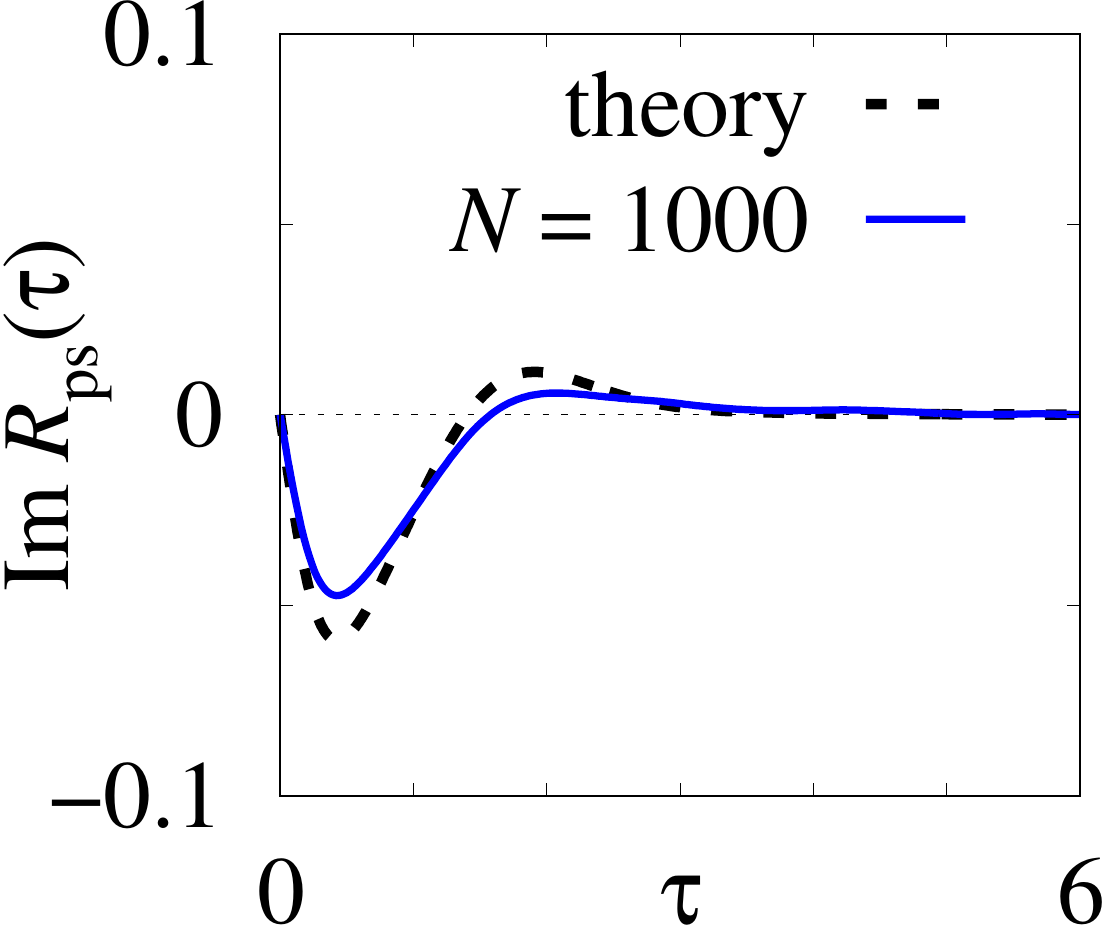}\\[1mm]
\includegraphics[height=0.48\columnwidth,angle=270]{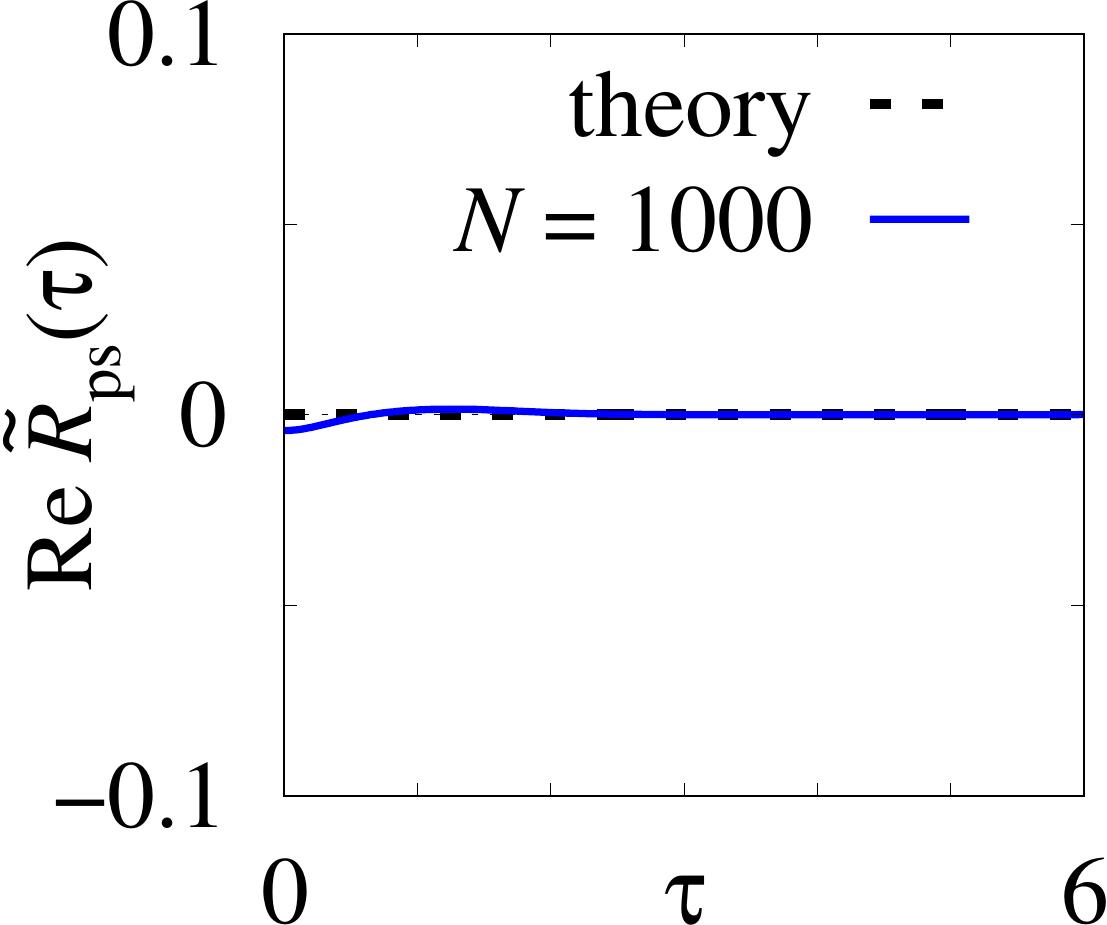}\hspace{0.02\columnwidth}
\includegraphics[height=0.48\columnwidth,angle=270]{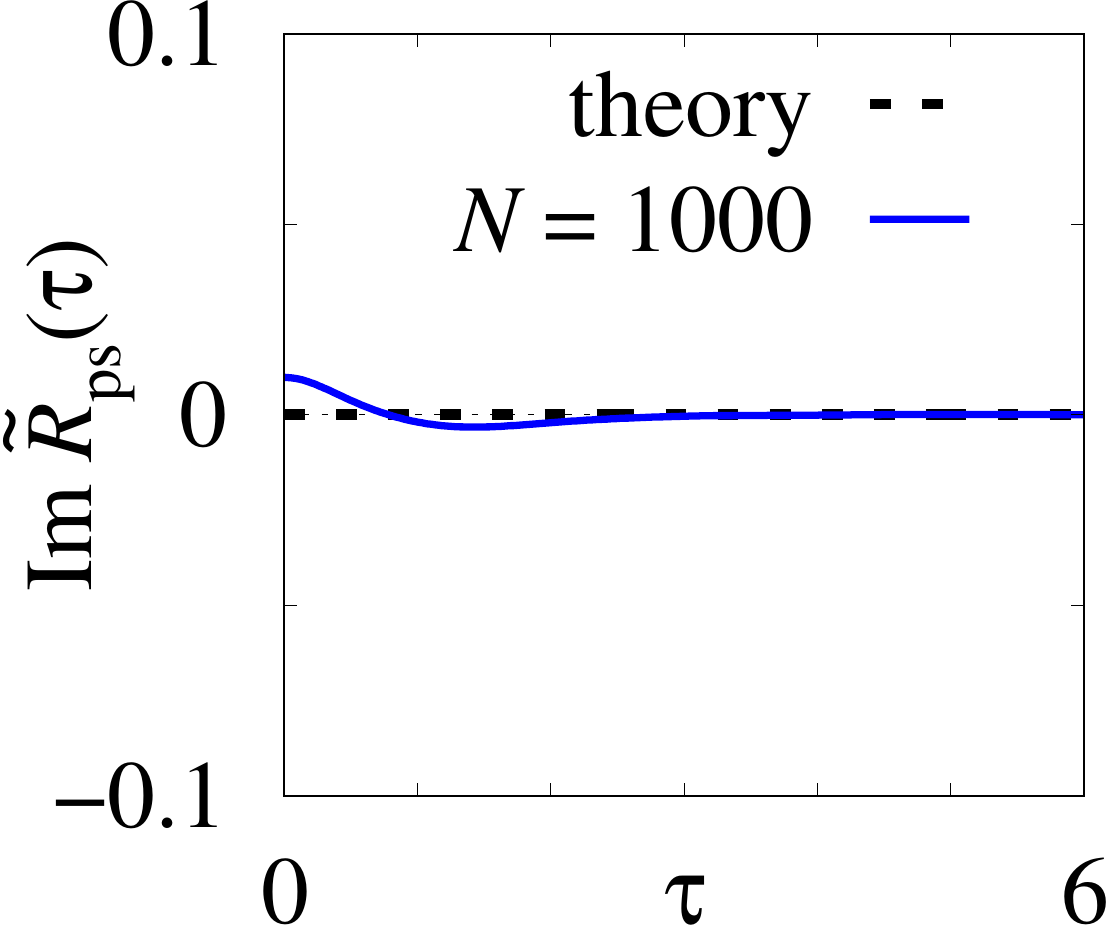}
\end{center}
\caption{The same fluctuation characteristics as in Fig.~\ref{Fig:Covariances:K5}
but for a partially synchronized state at $K = 4$ and $\lambda = \pi/4$.}
\label{Fig:Covariances:K4}
\end{figure}

\begin{figure}
\begin{center}
\includegraphics[height=0.48\columnwidth,angle=270]{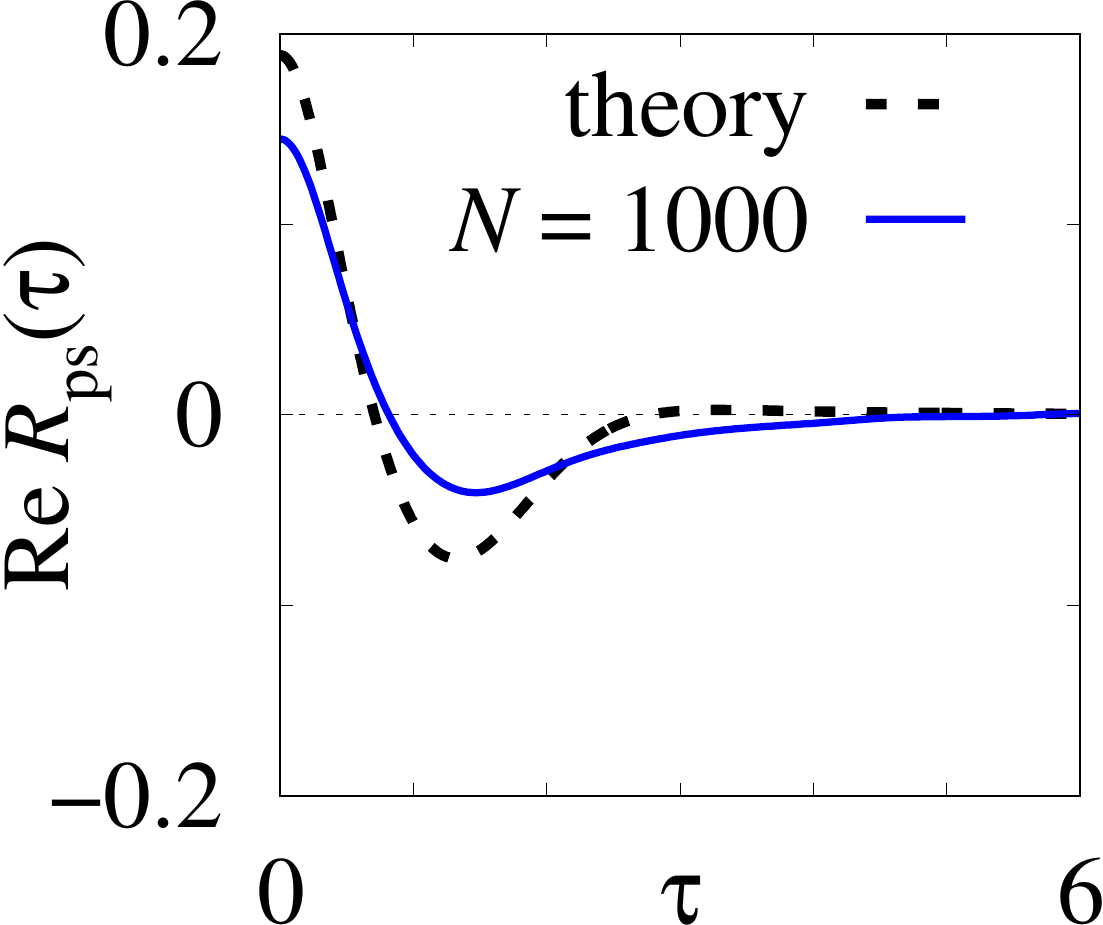}\hspace{0.02\columnwidth}
\includegraphics[height=0.48\columnwidth,angle=270]{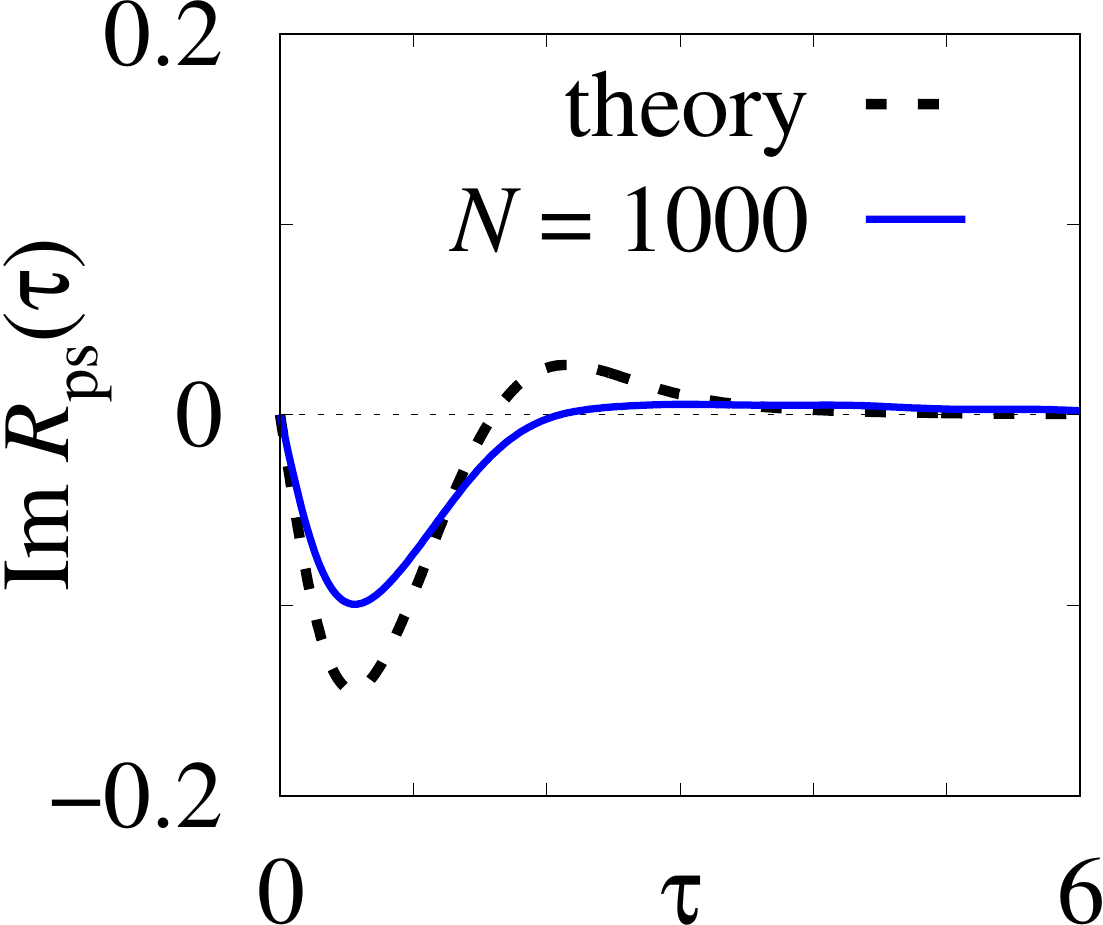}\\[1mm]
\includegraphics[height=0.48\columnwidth,angle=270]{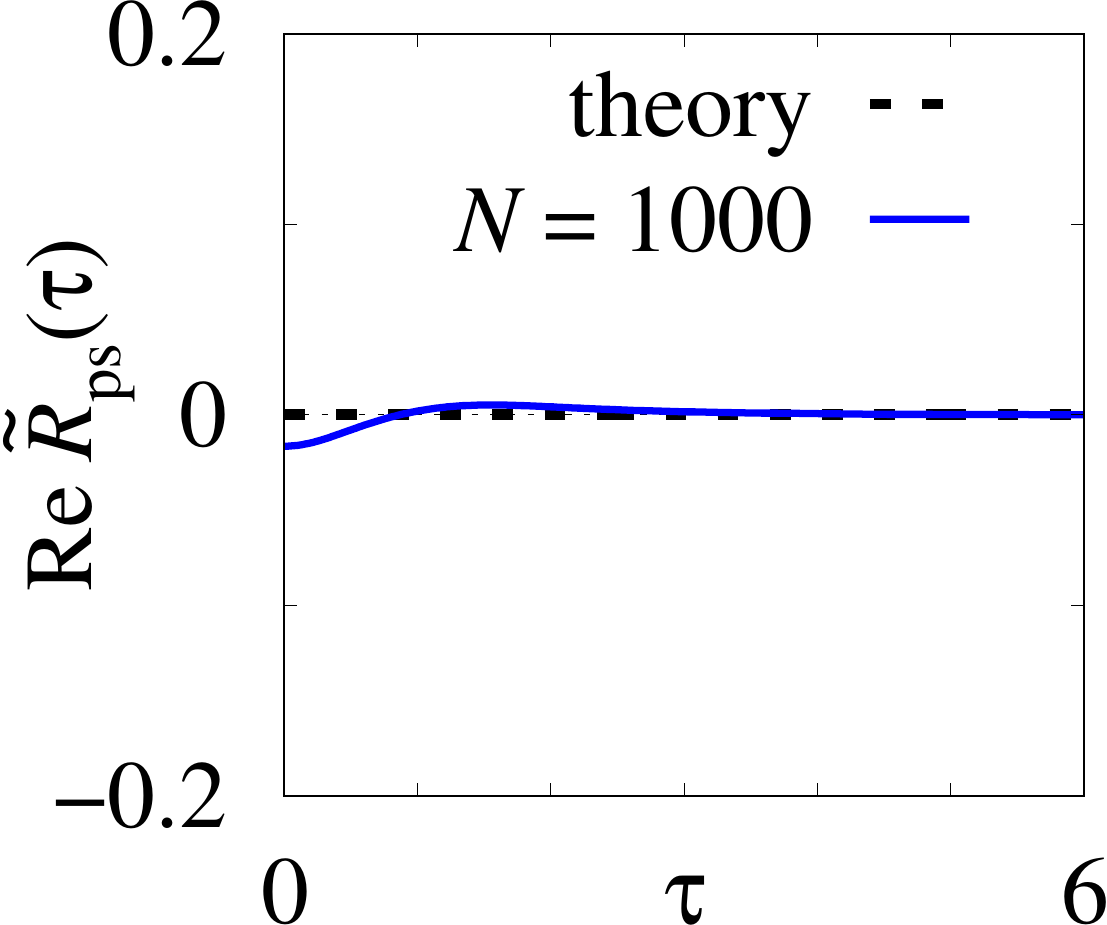}\hspace{0.02\columnwidth}
\includegraphics[height=0.48\columnwidth,angle=270]{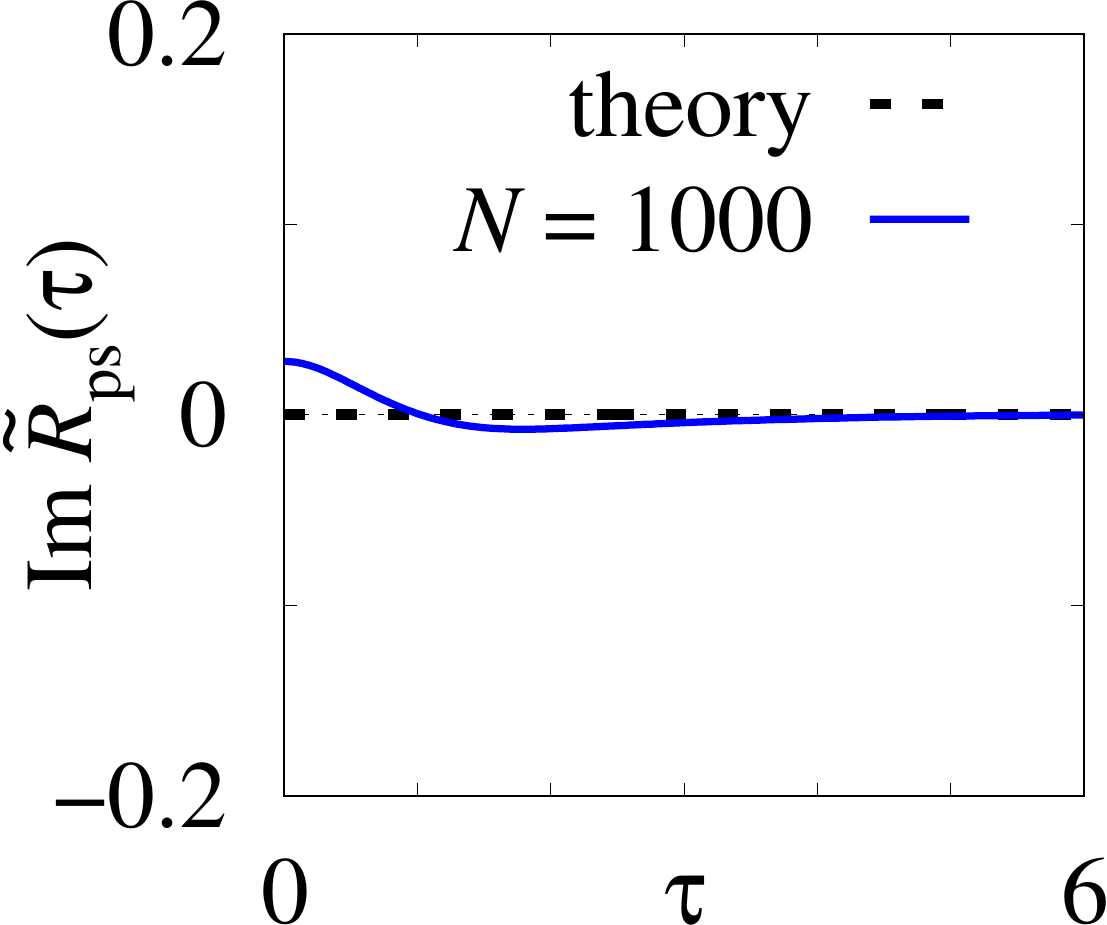}
\end{center}
\caption{The same fluctuation characteristics as in Fig.~\ref{Fig:Covariances:K5}
but for a partially synchronized state at $K = 3$ and $\lambda = \pi/4$.}
\label{Fig:Covariances:K3}
\end{figure}

\begin{figure}
\begin{center}
\includegraphics[height=0.48\columnwidth,angle=270]{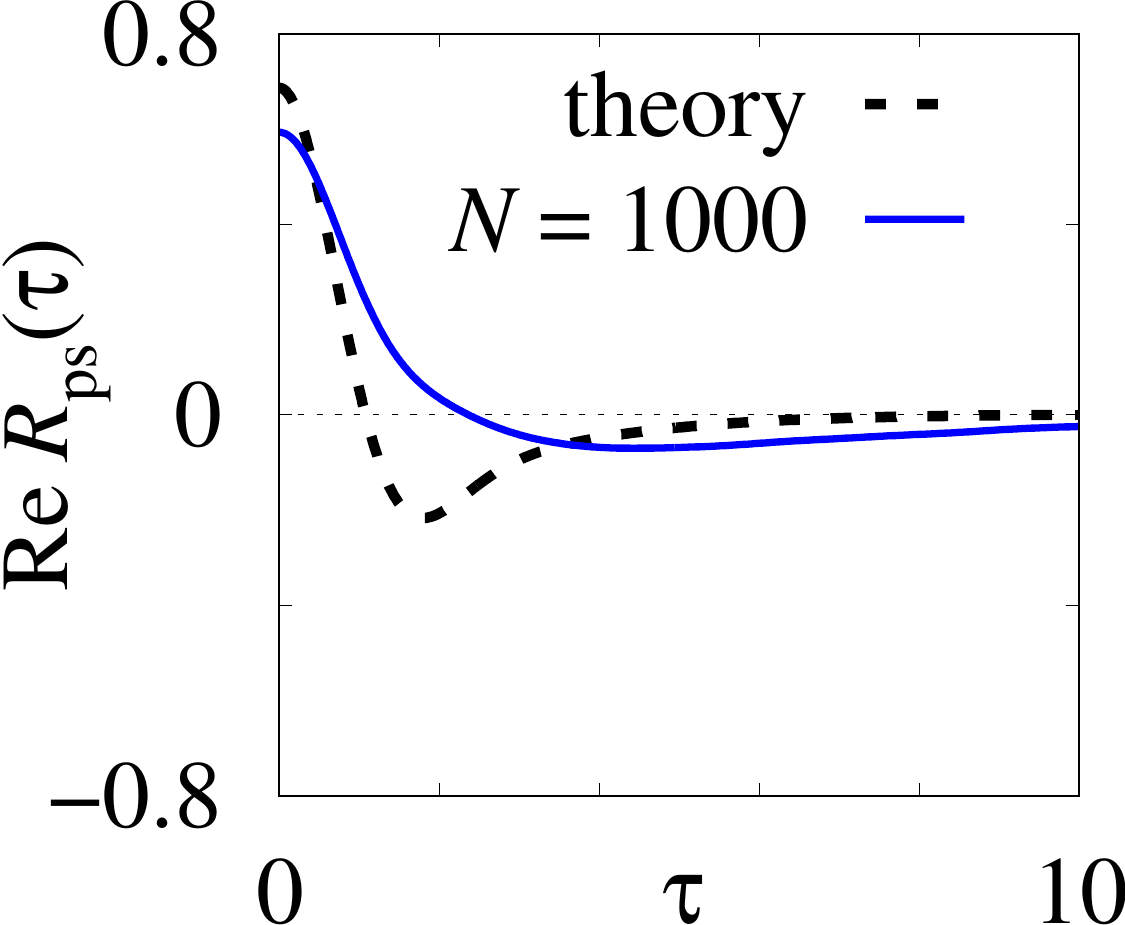}\hspace{0.02\columnwidth}
\includegraphics[height=0.48\columnwidth,angle=270]{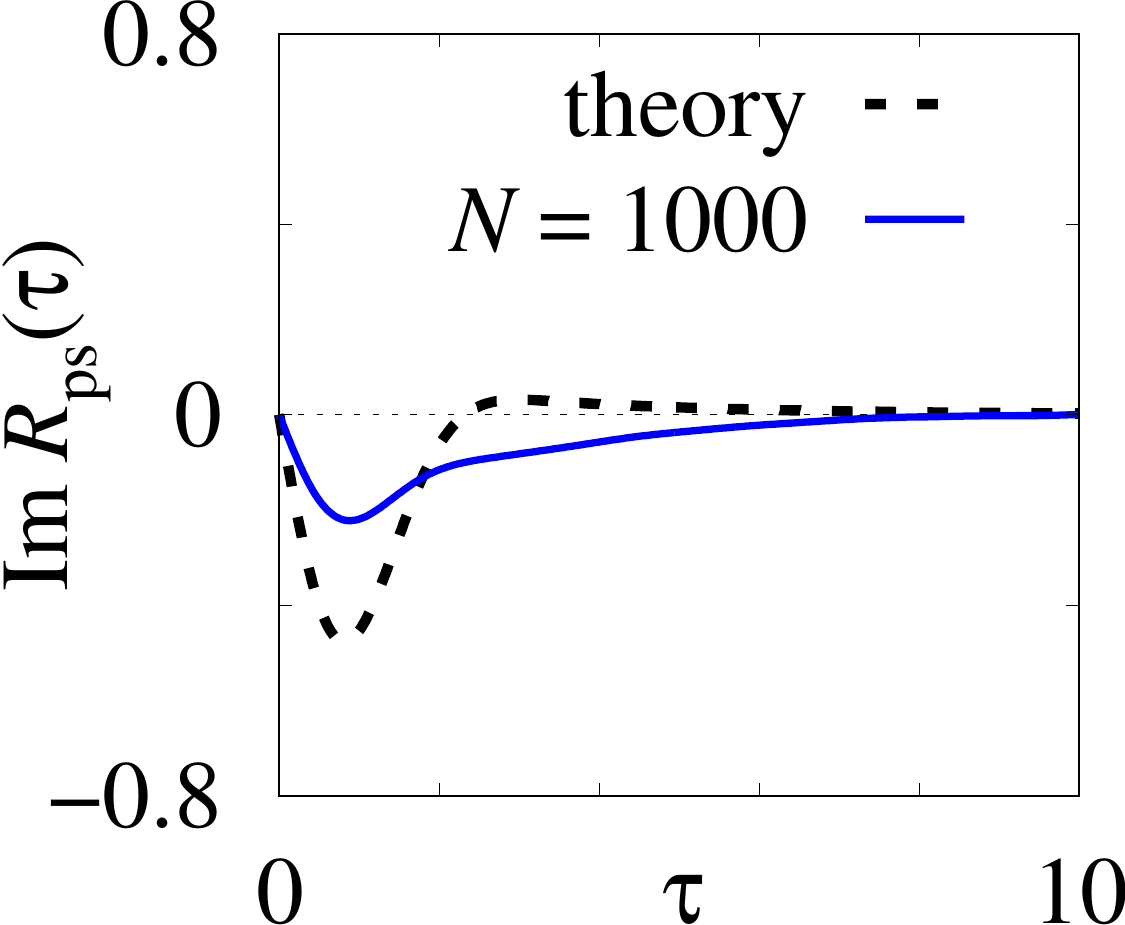}\\[1mm]
\includegraphics[height=0.48\columnwidth,angle=270]{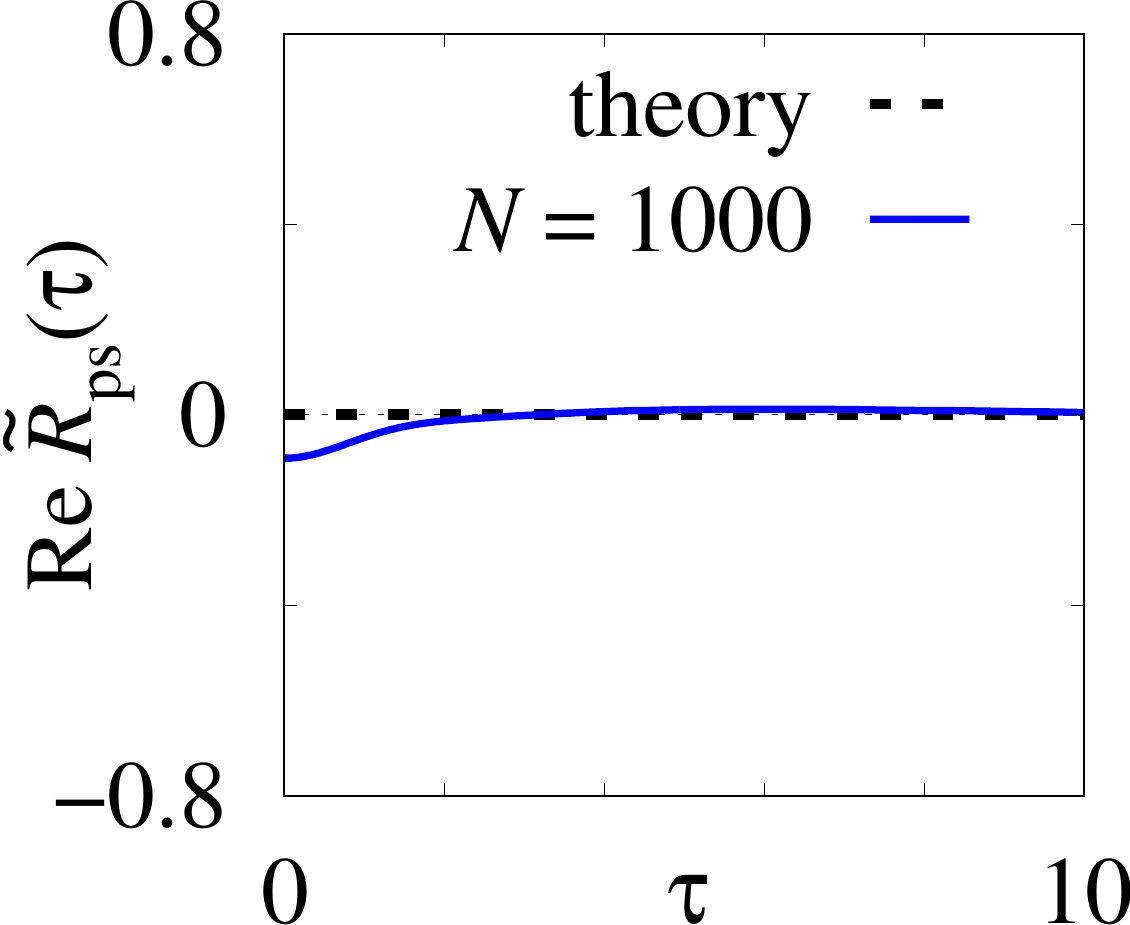}\hspace{0.02\columnwidth}
\includegraphics[height=0.48\columnwidth,angle=270]{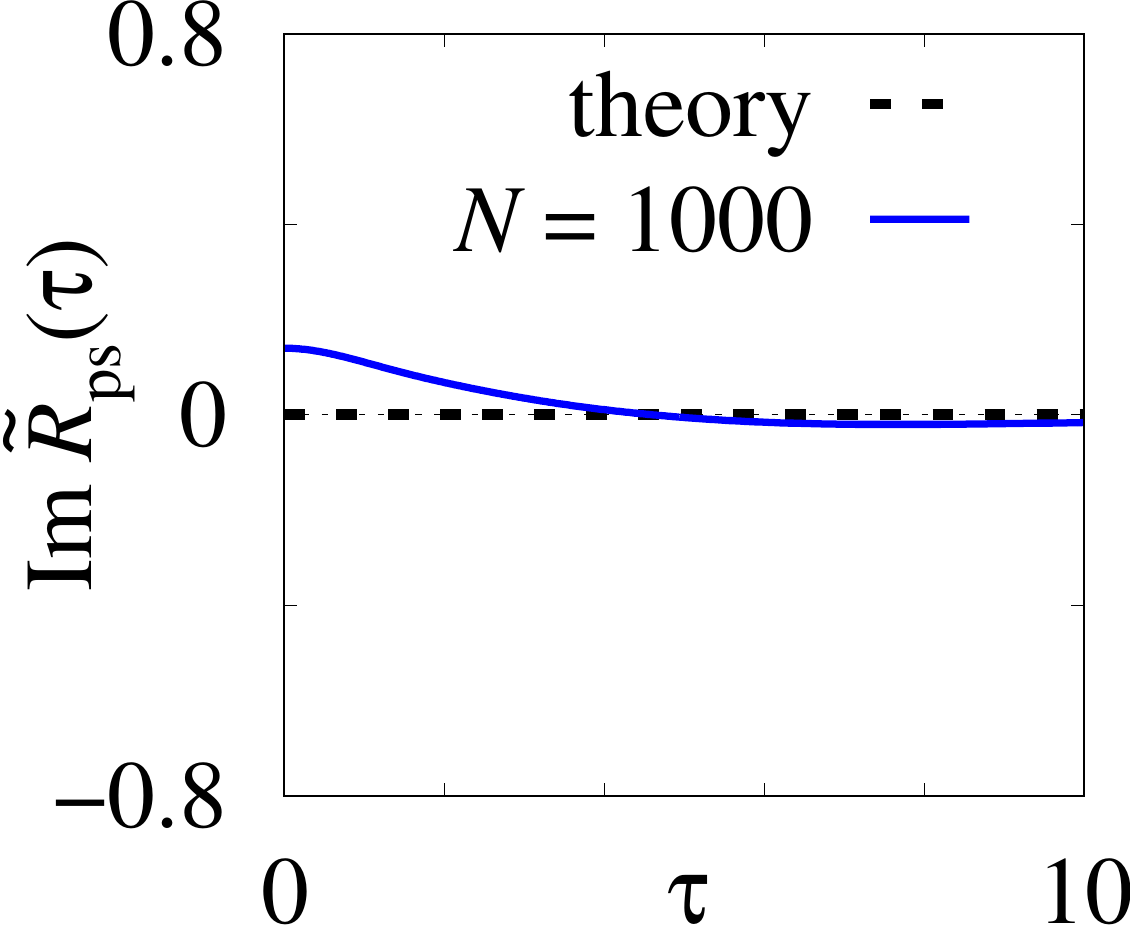}
\end{center}
\caption{The same fluctuation characteristics as in Fig.~\ref{Fig:Covariances:K5}
but for a partially synchronized state at $K = 2$ and $\lambda = \pi/4$.}
\label{Fig:Covariances:K2}
\end{figure}

\begin{figure}
\begin{center}
\includegraphics[height=0.48\columnwidth,angle=270]{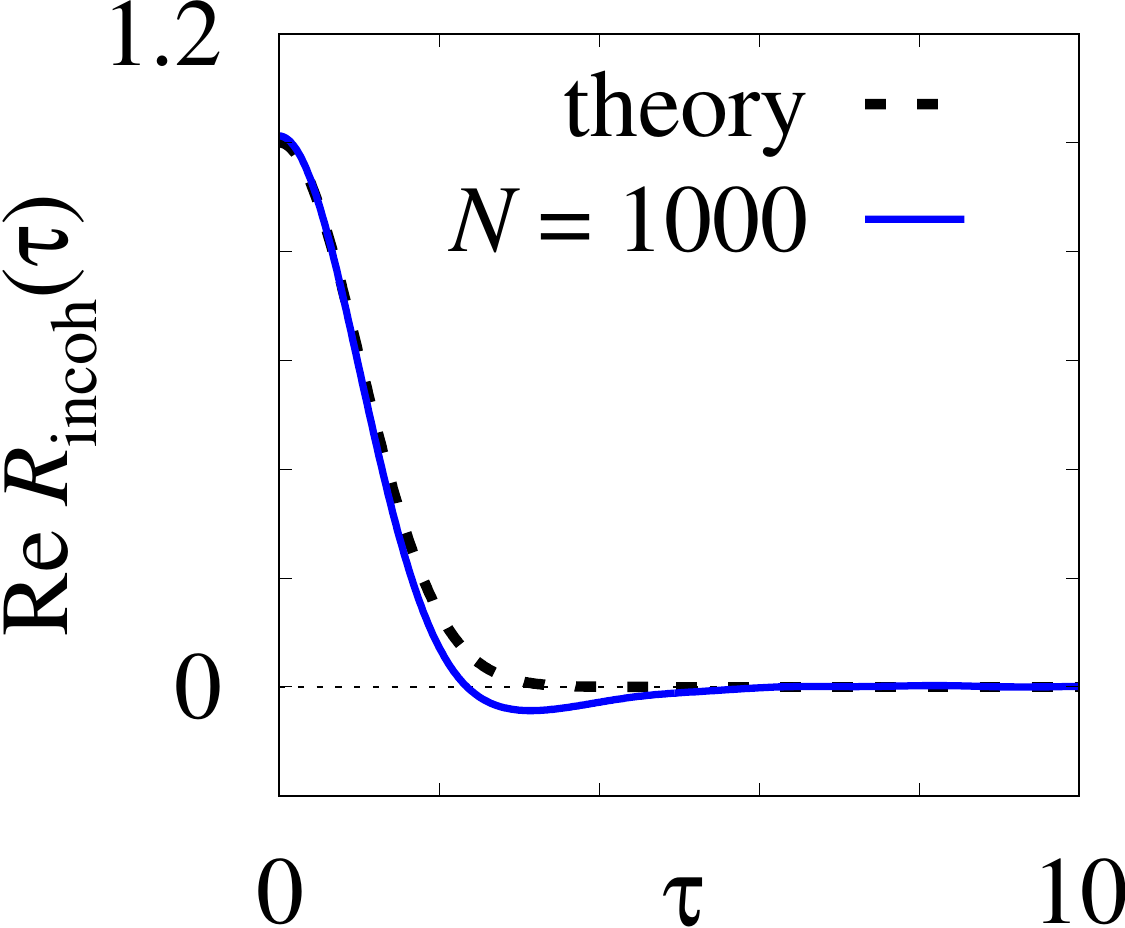}\hspace{0.02\columnwidth}
\includegraphics[height=0.48\columnwidth,angle=270]{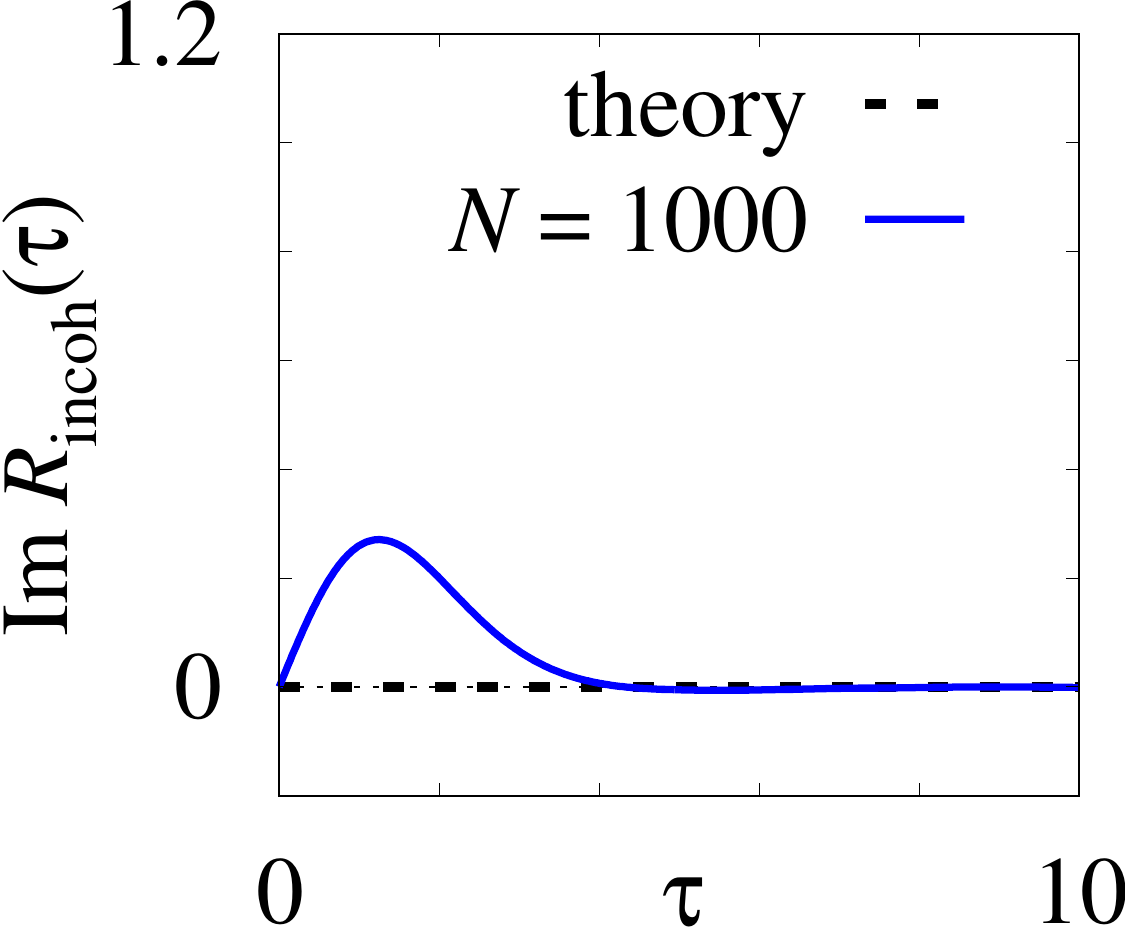}\\[1mm]
\includegraphics[height=0.48\columnwidth,angle=270]{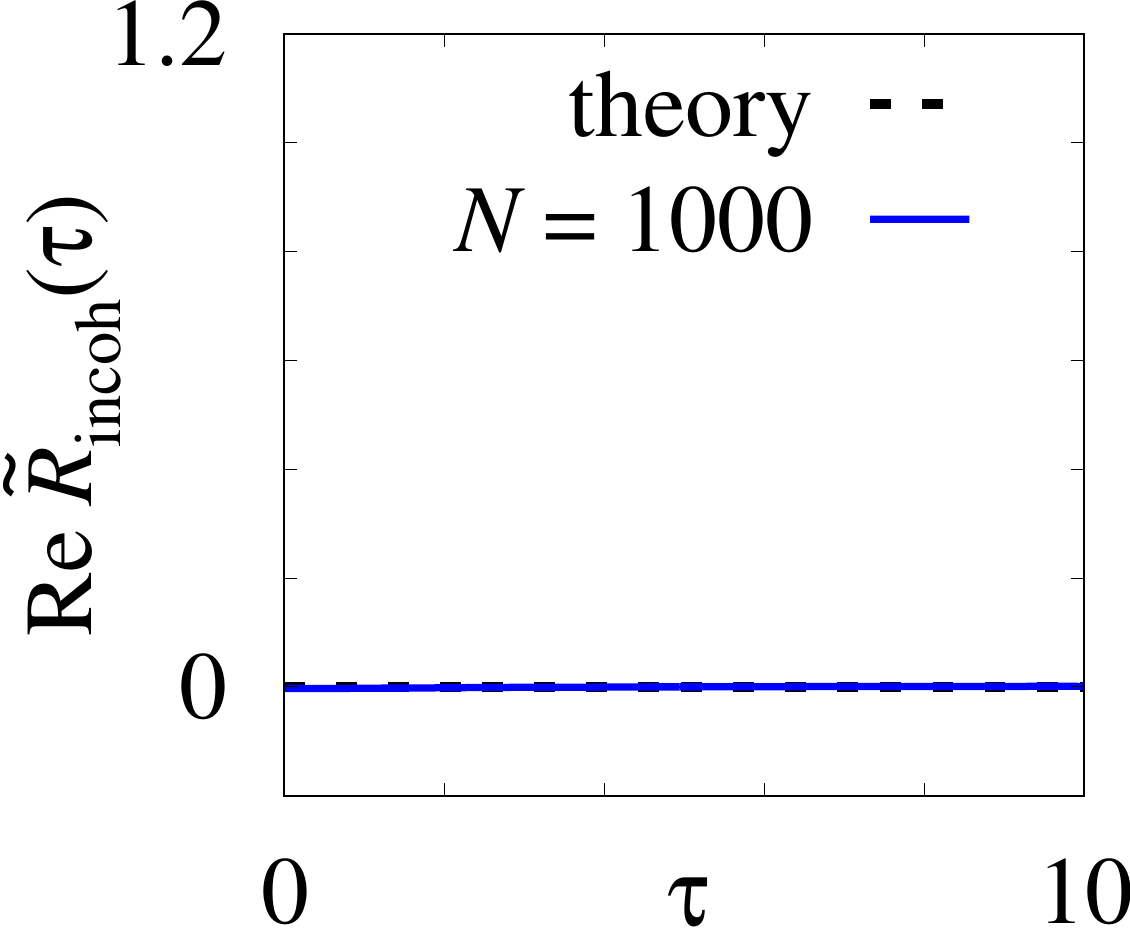}\hspace{0.02\columnwidth}
\includegraphics[height=0.48\columnwidth,angle=270]{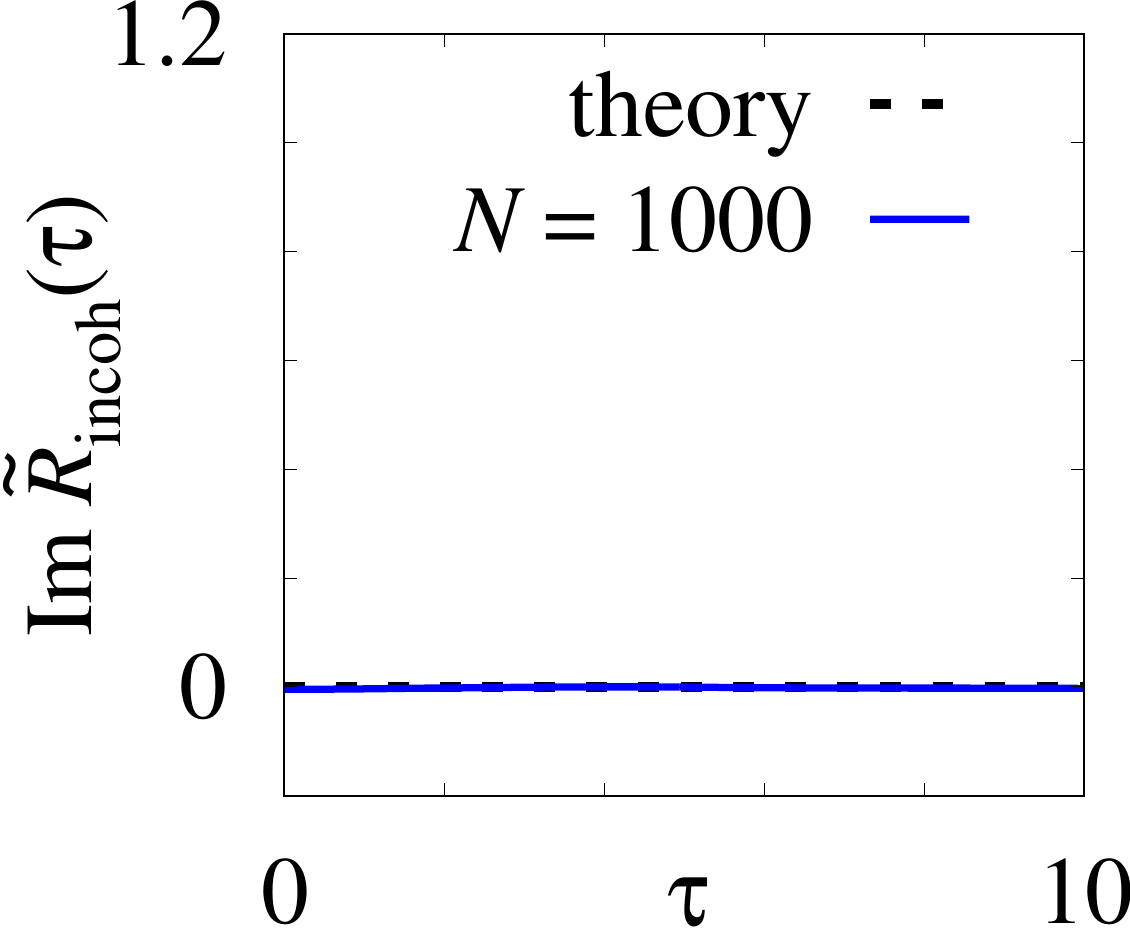}
\end{center}
\caption{Covariance $R(\tau)$ and pseudo-covariance $\tilde{R}(\tau)$
of the complex order parameter fluctuation $\zeta(t)$
for a completely incoherent state at $K = 1$ and $\lambda = \pi/4$
in the KS model \eqref{Eq:KS:original}.
Numerical simulations (solid curve) vs. theoretical prediction \eqref{Formula:R:incoh} (dashed curve).}
\label{Fig:Covariances:K1}
\end{figure}

To demonstrate how the accuracy of our method changes
as we approach the critical coupling strength $K_\mathrm{crit} \approx 1.92$,
we show in Figs.~\ref{Fig:Covariances:K5}--\ref{Fig:Covariances:K2} additional results
for partially synchronized states at $K = 5, 4, 3, 2$ for $\lambda = \pi/4$, 
and in Fig.~\ref{Fig:Covariances:K1} additional results
for a completely incoherent state at $K = 1$ and $\lambda = \pi/4$.
We plot results from numerical simulations of the the KS model~(\ref{Eq:KS:original})
and of our theory, Eq.~(\ref{Formula:R:ps}) for the partially synchronized case
and Eq.~(\ref{Formula:R:incoh}) for the incoherent state, respectively.

\begin{figure*}[t]
\begin{center}
\includegraphics[height=0.25\textwidth,angle=270]{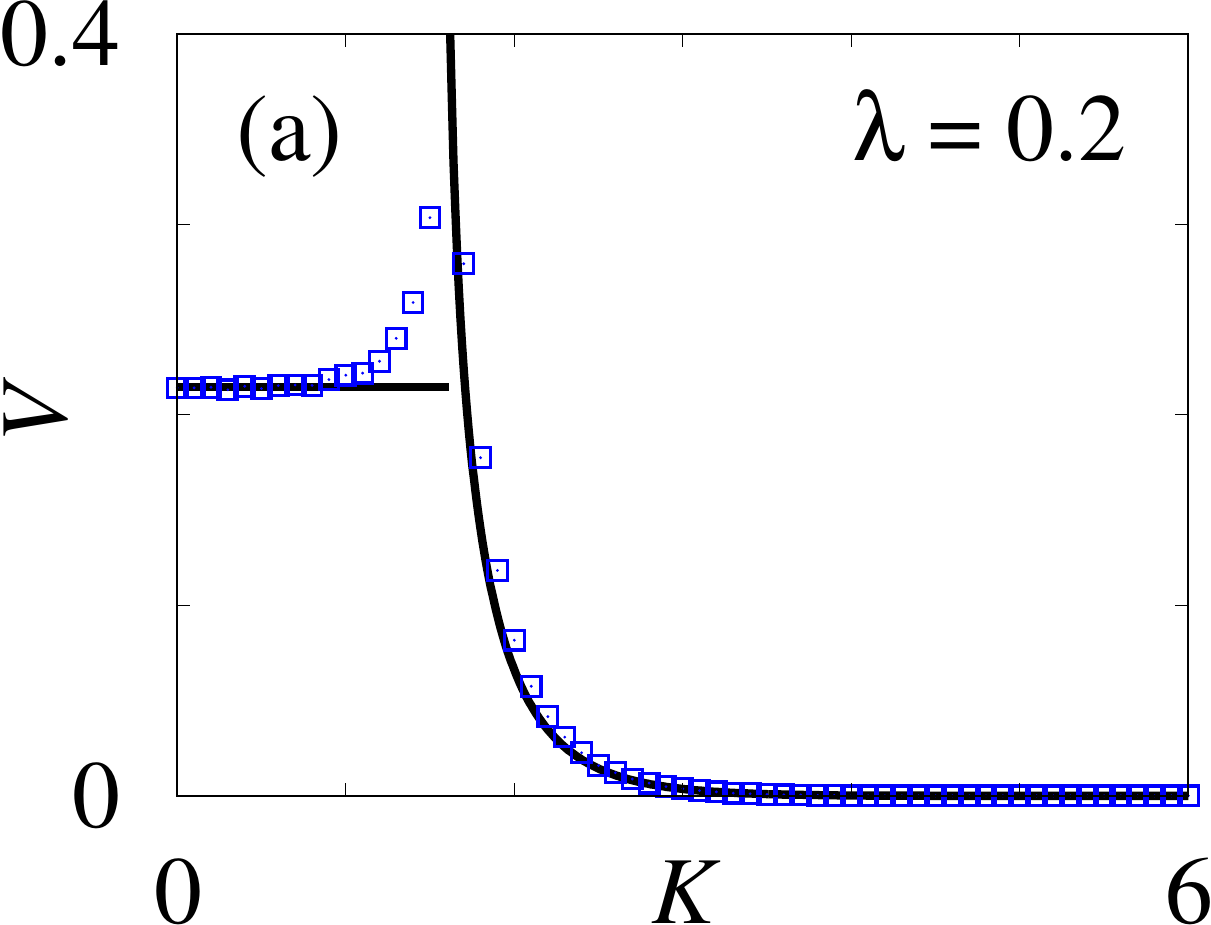}\hspace{0.05\columnwidth}
\includegraphics[height=0.25\textwidth,angle=270]{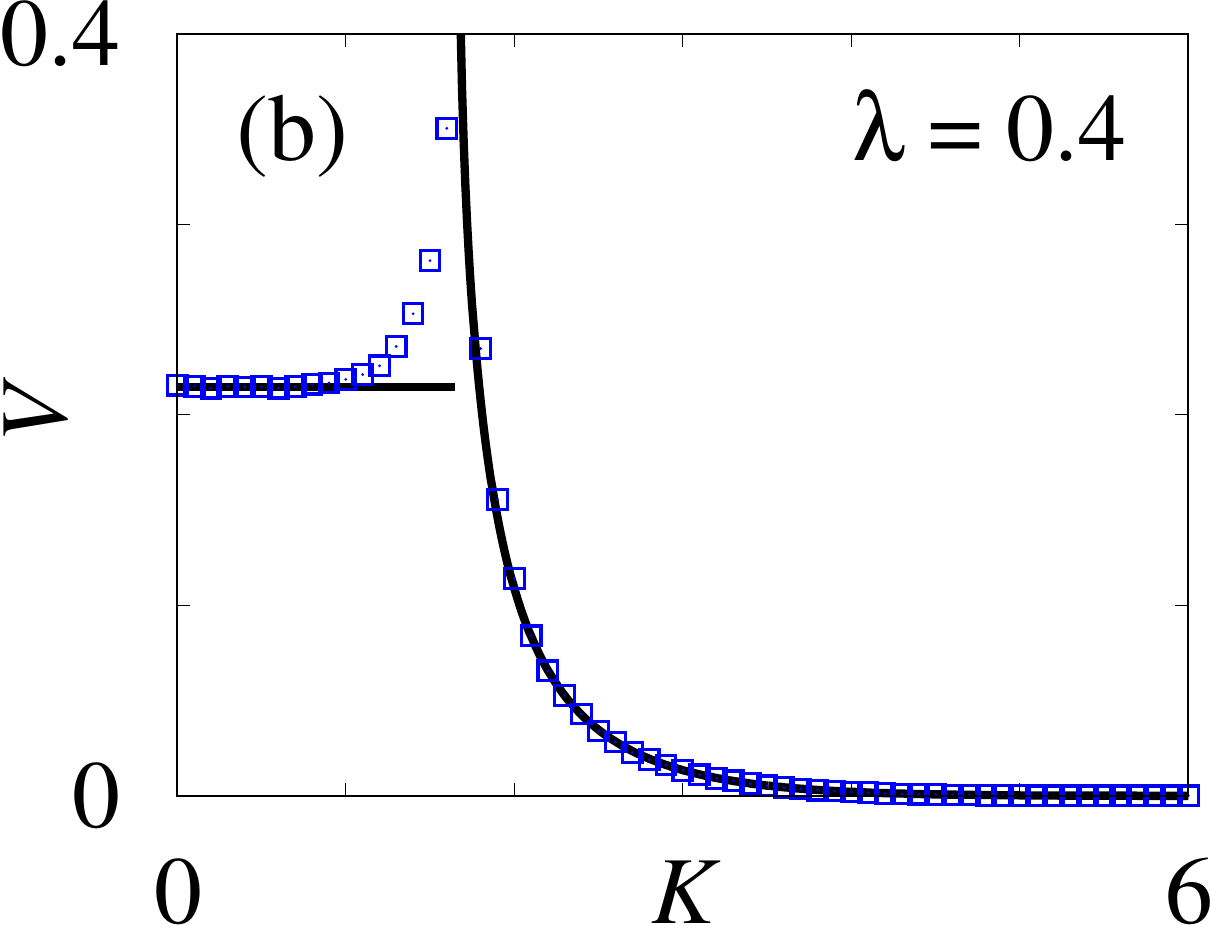}\hspace{0.05\columnwidth}
\includegraphics[height=0.25\textwidth,angle=270]{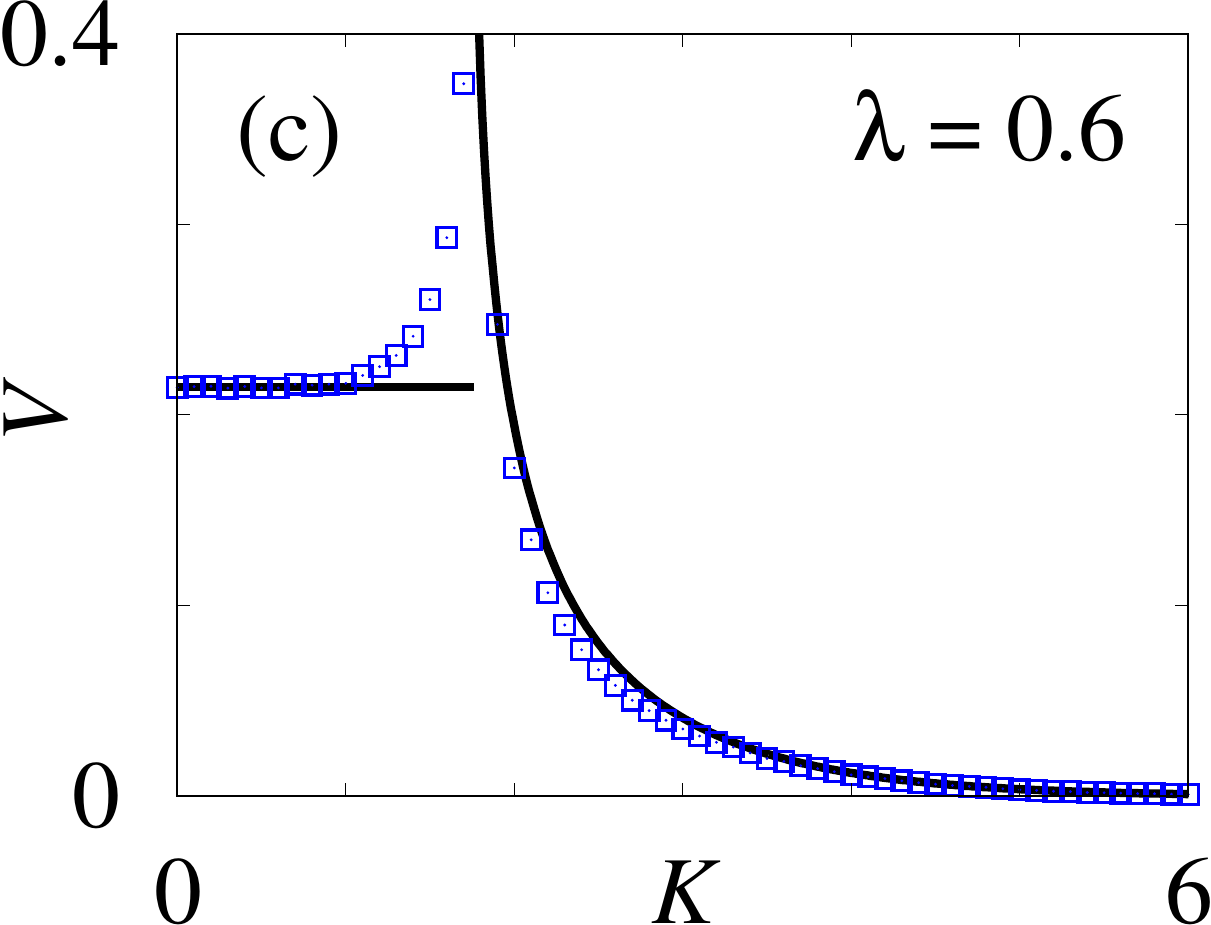}\\[3mm]
\includegraphics[height=0.25\textwidth,angle=270]{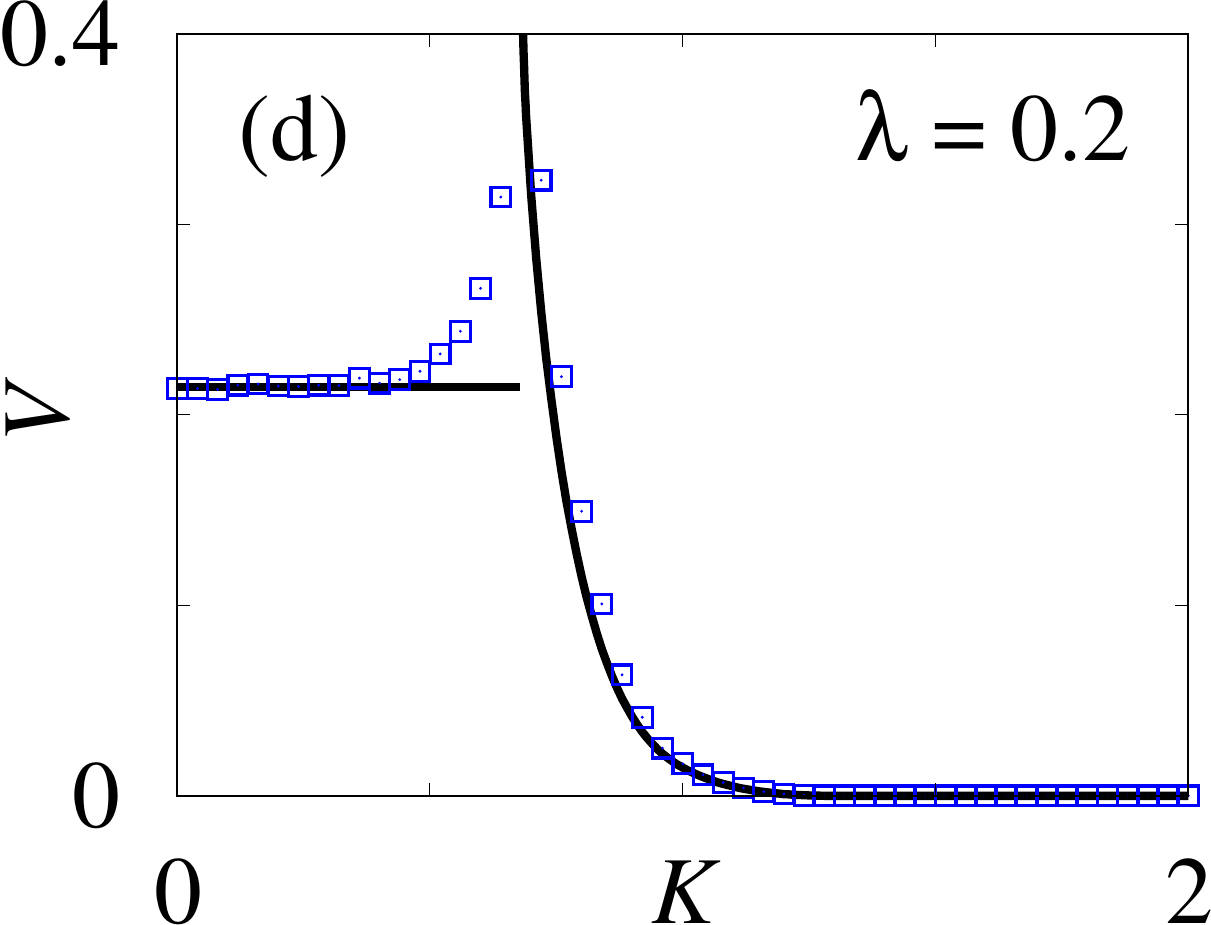}\hspace{0.05\columnwidth}
\includegraphics[height=0.25\textwidth,angle=270]{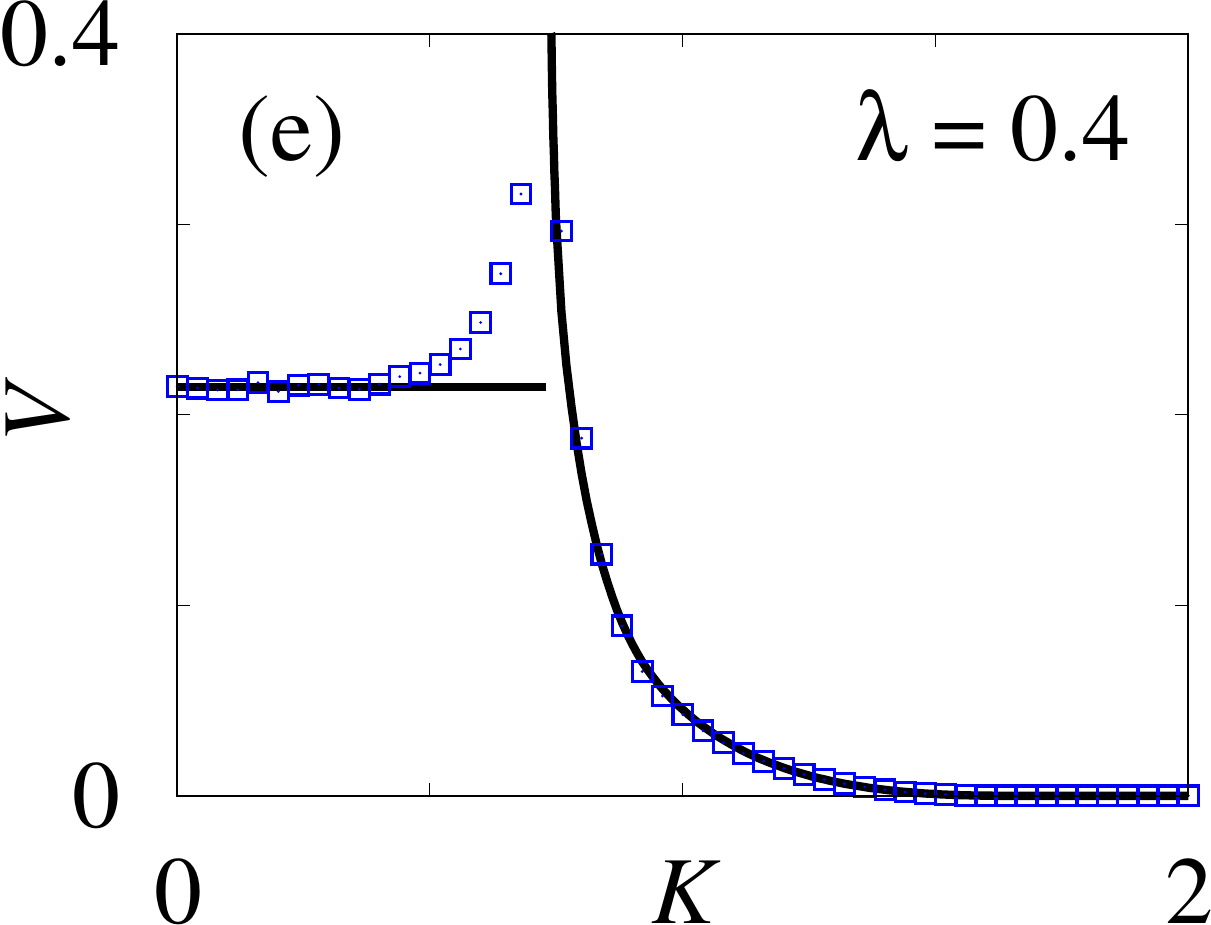}\hspace{0.05\columnwidth}
\includegraphics[height=0.25\textwidth,angle=270]{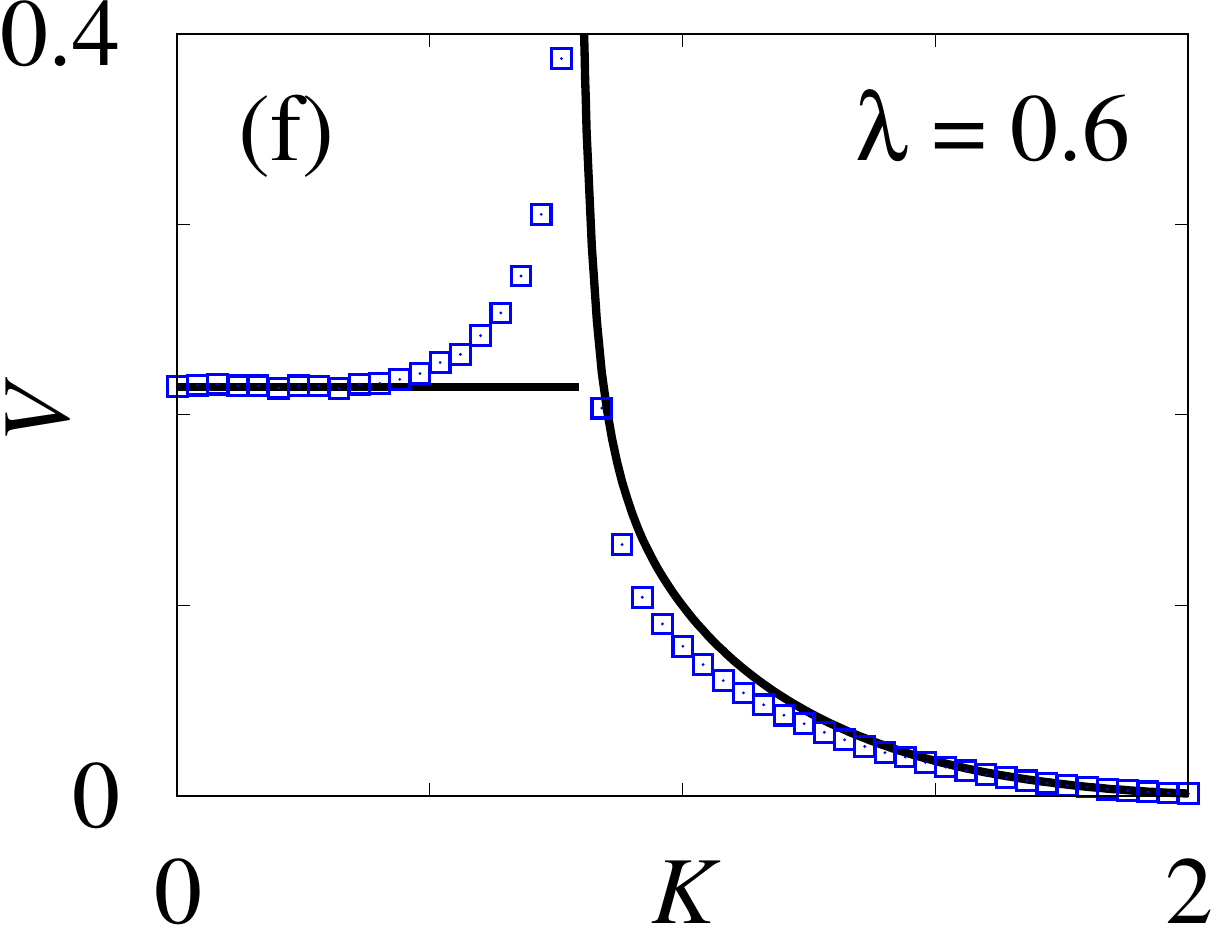}\\[3mm]
\includegraphics[height=0.25\textwidth,angle=270]{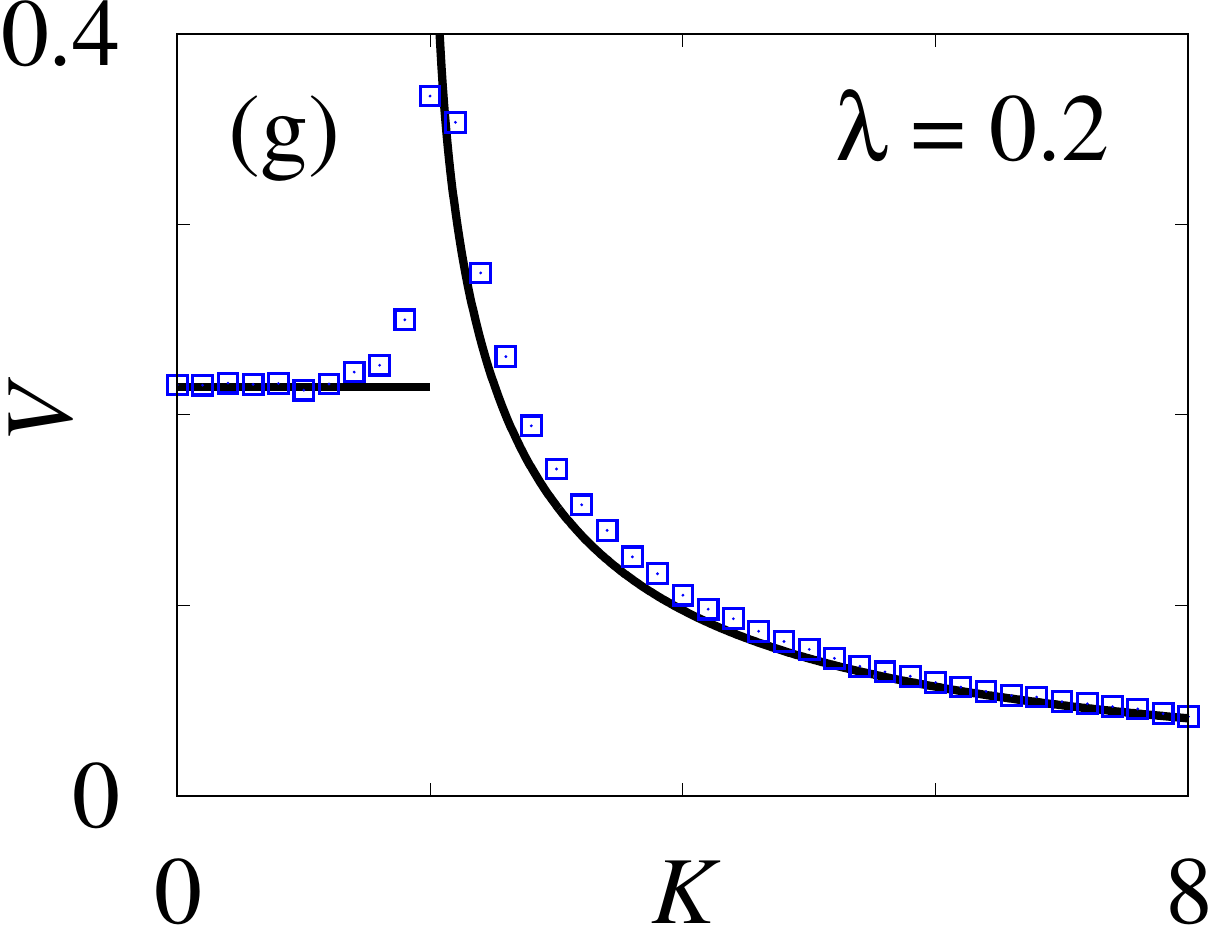}\hspace{0.05\columnwidth}
\includegraphics[height=0.25\textwidth,angle=270]{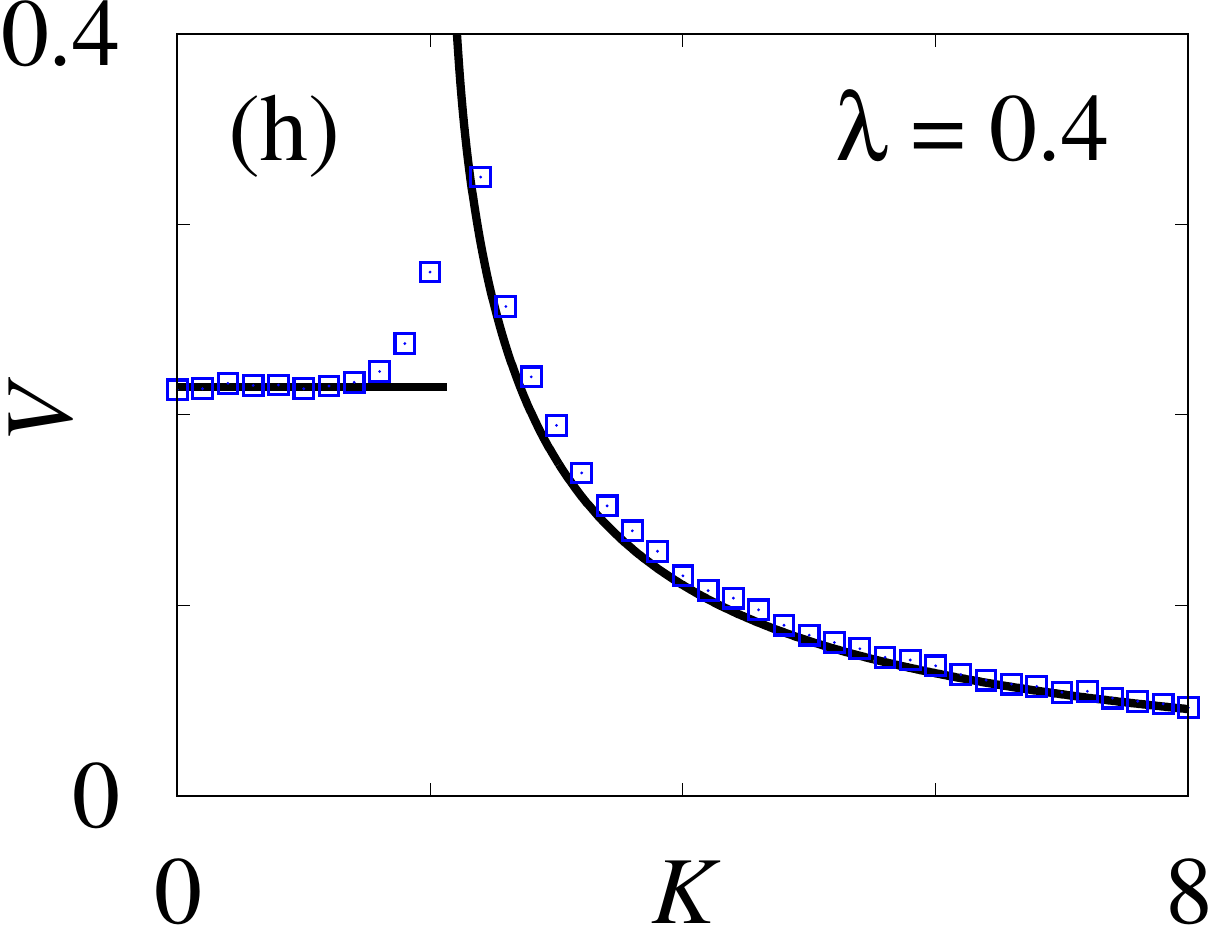}\hspace{0.05\columnwidth}
\includegraphics[height=0.25\textwidth,angle=270]{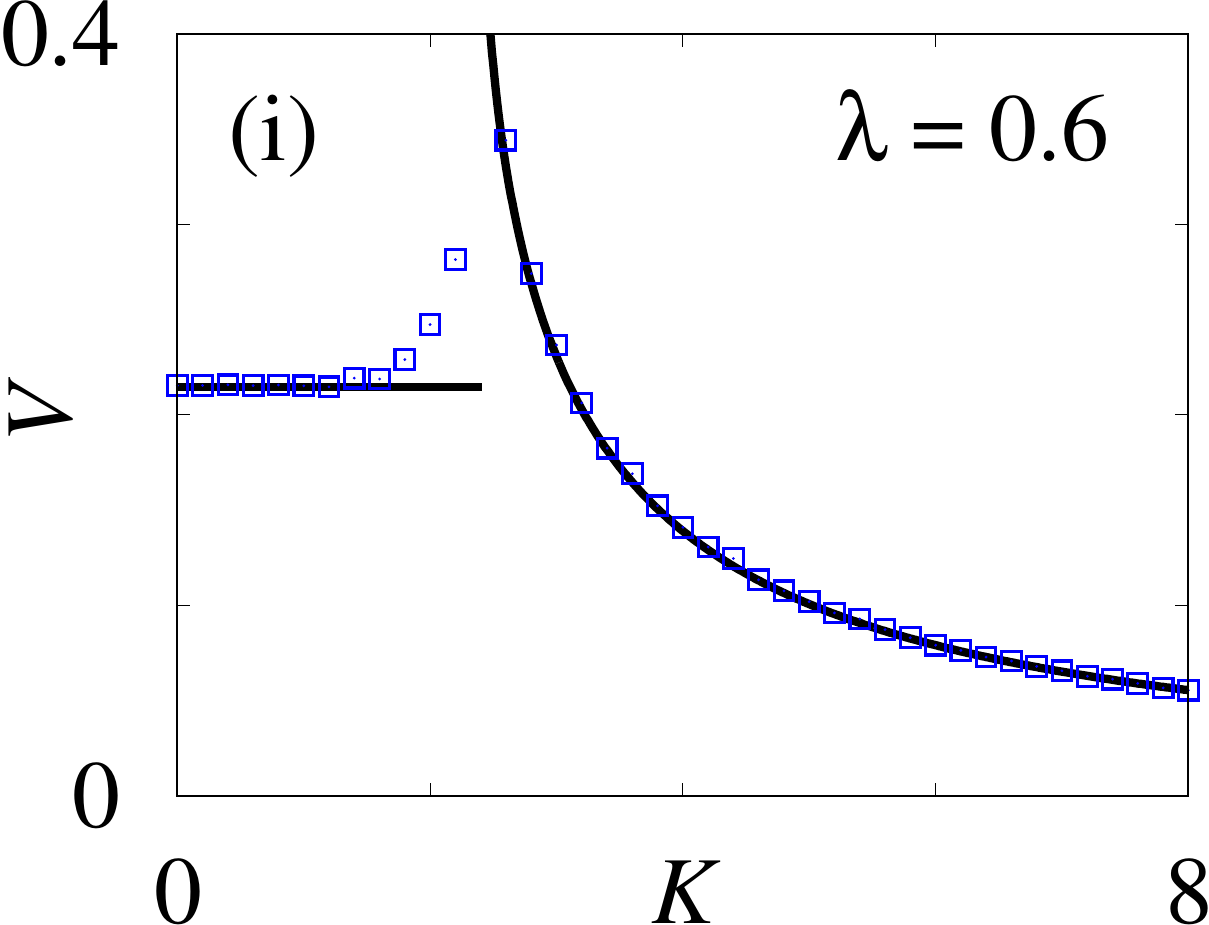}
\end{center}
\caption{
Variance~$V$ of the order parameter~$r(t)$
as a function of the coupling strength~$K$
for different values of the phase lag $\lambda$ for three frequency distributions $g(\omega)$:
Gaussian (a)--(c), Triangular (d)--(f),
and Lorentzian (g)--(i).
Numerical simulations for the KS model~(\ref{Eq:KS:original})
with $N = 1,000$ (blue squares)
vs. theoretical predictions (\ref{Formula:V:ps}) and (\ref{Formula:incoh:r}) (solid curves).
}
\label{Fig:Others}
\end{figure*}

In addition, to demonstrate the effectiveness
of our theoretical formulas for other system parameters,
we present results
for different values of the phase lag $\lambda$
and different frequency distributions $g(\omega)$.
Fig.~\ref{Fig:Others} shows how the variance $V$ 
of order parameter depends on the coupling strength $K$
for a unit variance Gaussian distribution (top row), 
for a triangular distribution (middle row) 
$$
g(\omega) = \left\{
\begin{array}{lcl}
1 - |\omega| & \mbox{ if } & |\omega|\le 1,\\[1mm]
0 & \mbox{ if } & |\omega| > 1,
\end{array}
\right.
$$
and for a Lorentzian distribution (bottom row) 
$$
g(\omega) = \frac{1}{\pi (1 + \omega^2)},
$$
for several values of the phase lag $\lambda$. 
It is seen that the quality of our approximations does not significantly depend on neither the frequency distribution nor on the value of the phase lag.

\begin{figure*}[t]
\begin{center}
\includegraphics[width=0.2\textwidth,angle=270]{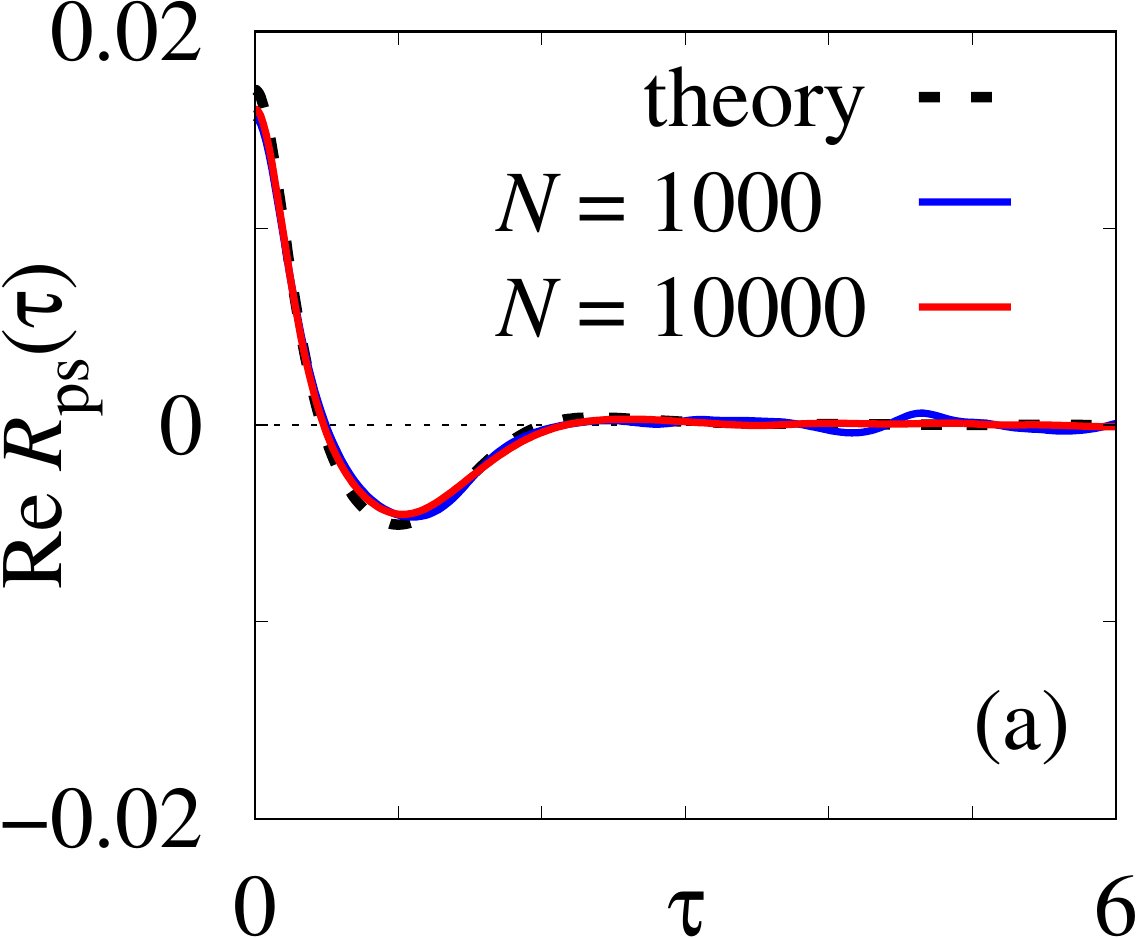}\hspace{0.05\textwidth}
\includegraphics[width=0.2\textwidth,angle=270]{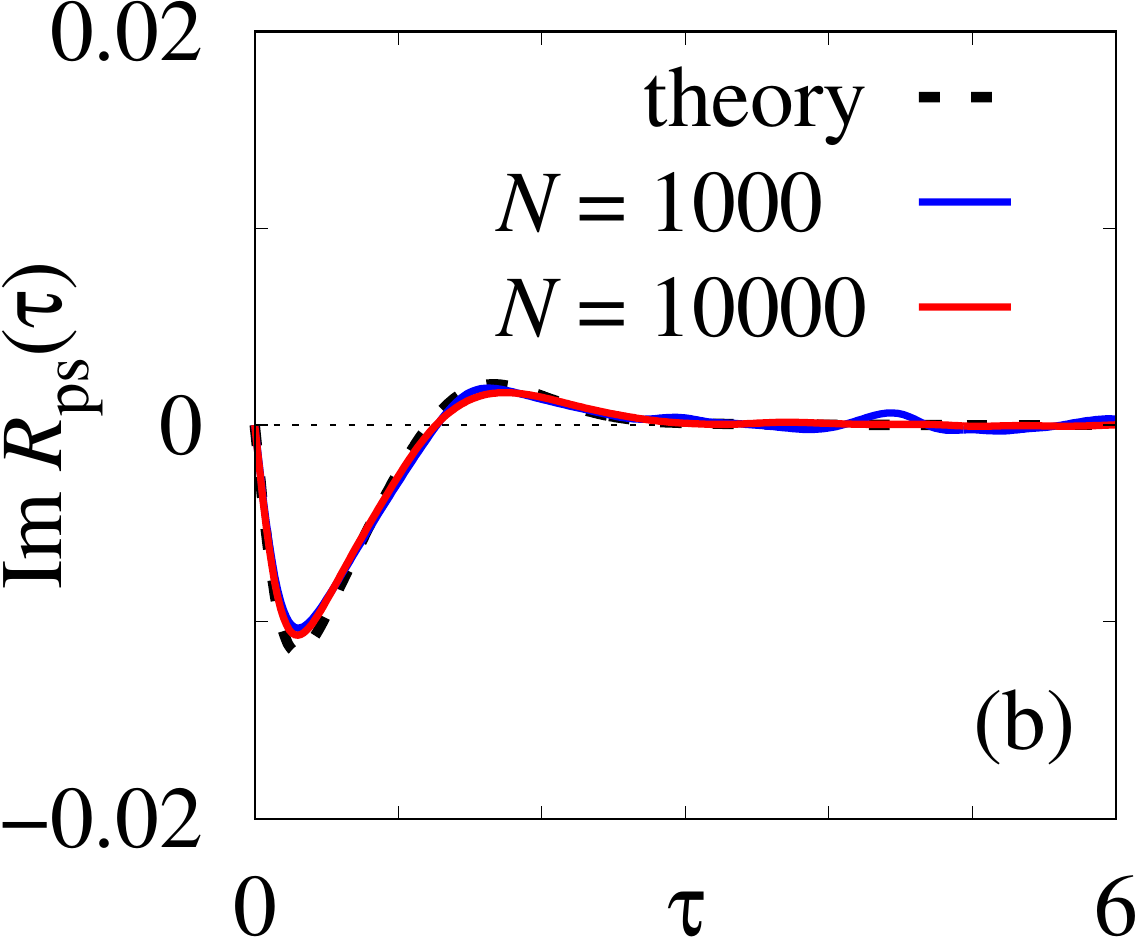}\hspace{0.05\textwidth}
\includegraphics[width=0.2\textwidth,angle=270]{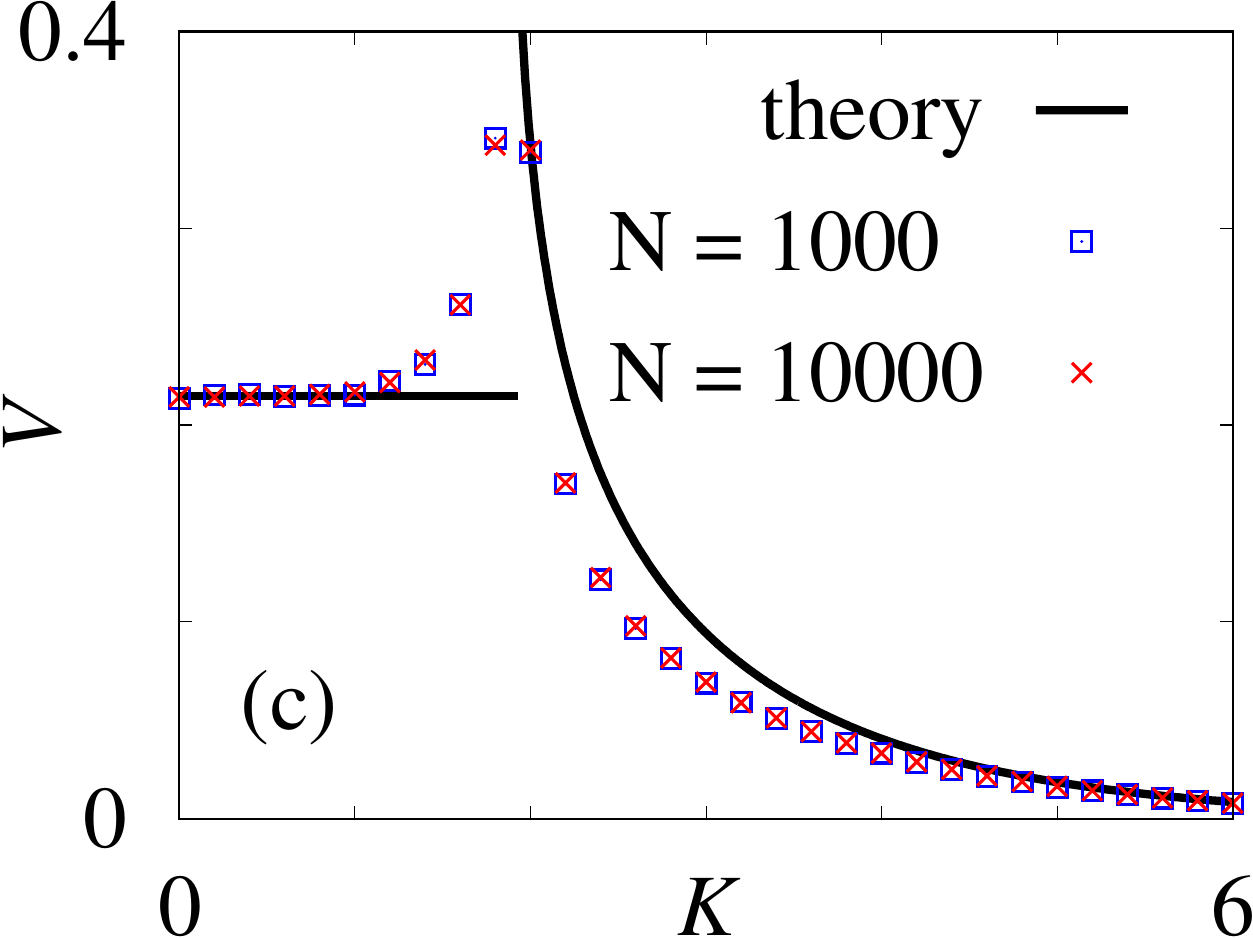}
\end{center}
\caption{
Comparison of the results from numerical simulations
of the KS model~(\ref{Eq:KS:original}) for $N=1,000$ and $N=10,000$.
(a)-(b): The solid curves show the covariance $R_\mathrm{ps}(\tau)$
of fluctuations $\zeta(t)$ of the modified order parameter $Z_\mathrm{mod}(t)$
for a partially synchronized state at $K = 6$ and $\lambda=\pi/4$.
Dashed curves show the theoretical prediction~(\ref{Formula:R:ps}).
(c): Variance of the order parameter $V$ as a function of the coupling strength $K$ (squares and crosses).
The theoretical predictions~(\ref{Formula:R:ps}) and (\ref{Formula:V:ps}) are shown as solid curves.
}
\label{Fig:Comparison:N10000}
\end{figure*}

Finally, in Figure~\ref{Fig:Comparison:N10000} we show
that the results for the full KS model~(\ref{Eq:KS:original}) are indistinguishable by eye
when a larger network with $10,000$ oscillators is simulated,
indicative that $N=1,000$ are sufficient to study the thermodynamic limit.


\section{Phase diffusion}
\label{app:phasediff}
\begin{figure}[t]
\begin{center}
\includegraphics[width=0.9\columnwidth,angle=0]{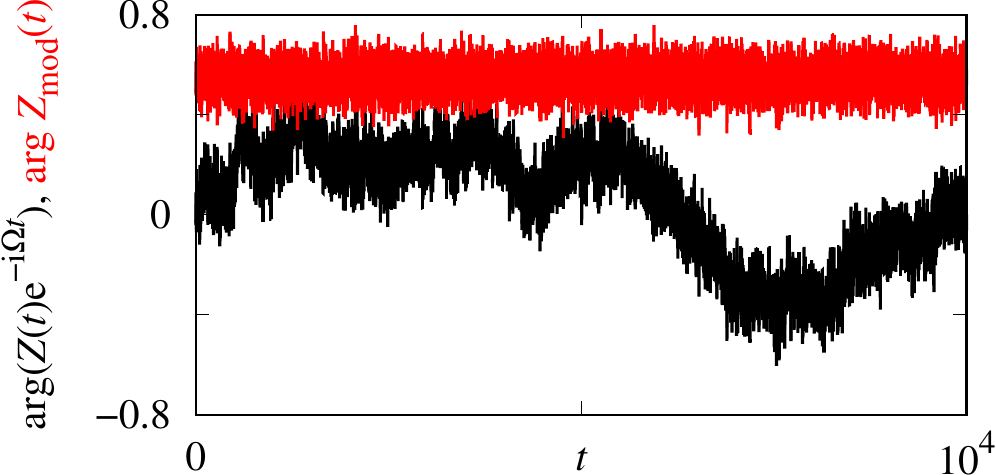}
\end{center}
\caption{
Phase diffusion in the KS model~(\ref{Eq:KS:original}). 
The argument of the complex order parameter
in the uniformly corotating frame (black curve)
exhibits non-stationary random walk behavior.
Contrary, the argument of the modified
order parameter $Z_\mathrm{mod}(t)$ (red curve)
exhibits stationary statistical behaviour. 
Shown are numerical results for the KS model~(\ref{Eq:KS:original}) 
with $N = 1,000$, $K = 2$, $\lambda = \pi/4$,
and a unit width Gaussian frequency distribution $g(\omega)$.
}
\label{Fig:Diffusion}
\end{figure}

Mean-field theory for partially synchronized states
predicts that the argument of the complex order parameter 
$$
Z(t) = \frac{1}{N} \sum\limits_{j=1}^N e^{i \psi_j(t)},
$$
uniformly increases or decreases with time.
To subtract this drift, we can move into the corotating frame of reference and consider $\arg(Z(t)e^{-i\Omega t})$,
where $\Omega$ is the average speed. However, this quantity experiences non-stationary irregular behaviour,
resembling Brownian motion, as shown in Fig.~\ref{Fig:Diffusion}. 
This phenomenon is known as {\it phase diffusion}~\cite{PikR1999,PikBI2026}.
Due to this phase diffusion, the function $Z(t)e^{-i \Omega t}$ is difficult to characterize.
Therefore, in our study we work with a modified order parameter
$$
Z_\mathrm{mod}(t) = Z(t) \frac{\overline{Z_\mathrm{c}(t)}}{|Z_\mathrm{c}(t)|},
$$
where
$$
Z_\mathrm{c}(t) = \frac{1}{N} \sum\limits_{j\in\mathcal{C}} e^{i \psi_j(t)}
$$
is the order parameter, defined as the sum over the coherent
(locked) oscillators only.
In numerical simulations,
we identify the $j$th oscillator as coherent
if its natural frequency $\omega_j$ satisfies
the locking condition
$|\omega_j - \Omega| \le K \langle |Z(t)| \rangle$.
It turns out, that the argument
of the modified order parameter $Z_\mathrm{mod}(t)$
does not exhibit phase diffusion phenomenon,
and the behavior of this complex function
is nicely approximated
by a two-dimensional Gaussian process~\cite{YueG2024}.
This numerical observation is crucial for our analytical approach.


\section{Pairwise covariances of rogue oscillators}
\label{app:indep}
In Proposition~\ref{Prop:Adler}, we derived an explicit formula
for the solutions of the Adler equation~(\ref{Eq:Adler}).
Simple calculations show that this formula has the equivalent form
$$
e^{i \theta(t)} e^{- i \theta(0)}
= \fr{ \chi_+ e^{i \xi t} - \chi_- }{- \overline{\chi}_- e^{i \xi t} + \overline{\chi}_+}
= \fr{ \chi_+ - \chi_- e^{-i \xi t} }{- \overline{\chi}_- + \overline{\chi}_+ e^{-i \xi t}},
$$
where
$$
\chi_- = i H e^{- i \theta(0)} - ( \Delta - \xi ),\quad
\chi_+ = i H e^{- i \theta(0)} - ( \Delta + \xi ).
$$
We use this form to justify the following statements.

\begin{proposition}
If $\Delta > |H|$, we have $| \chi_- / \chi_+ | < 1$
and the following expansion holds
\begin{eqnarray}
e^{i \theta(t)}
&=& - \fr{\chi_-}{\overline{\chi}_+} e^{i \theta(0)}
+ \fr{ |\chi_+|^2 - |\chi_-|^2 }{ \overline{\chi}_+ \overline{\chi}_- } e^{i \theta(0)}
\sum\limits_{n=1}^\infty \left( \fr{\overline{\chi}_-}{\overline{\chi}_+} \right)^n e^{i n \xi t} 
\nonumber\\[2mm]
&=& \left\langle e^{i \theta(t)} \right\rangle - \fr{2 i \xi}{\overline{H}}
\sum\limits_{n=1}^\infty \left( \fr{i \overline{H} e^{i \theta(0)} + \Delta - \xi}{i \overline{H} e^{i \theta(0)} + \Delta + \xi} \right)^n e^{i n \xi t},
\label{Formula:exp_theta:plus}
\end{eqnarray}
where $\left\langle e^{i \theta(t)} \right\rangle$
denotes the time average of $e^{i \theta(t)}$.
On the other hand, if $\Delta < - |H|$, then $| \chi_+ / \chi_- | < 1$ and
\begin{eqnarray*}
e^{i \theta(t)}
&=& - \fr{\chi_+}{\overline{\chi}_-} e^{i \theta(0)}
+ \fr{ |\chi_-|^2 - |\chi_+|^2 }{ \overline{\chi}_+ \overline{\chi}_- } e^{i \theta(0)}
\sum\limits_{n=1}^\infty \left( \fr{\overline{\chi}_+}{\overline{\chi}_-} \right)^n e^{-i n \xi t} \\[2mm]
&=& \left\langle e^{i \theta(t)} \right\rangle + \fr{2 i \xi}{\overline{H}}
\sum\limits_{n=1}^\infty \left( \fr{i \overline{H} e^{i \theta(0)} + \Delta + \xi}{i \overline{H} e^{i \theta(0)} + \Delta - \xi} \right)^n e^{-i n \xi t}.
\end{eqnarray*}
\label{Proposition:Expansions}
\end{proposition}

{\bf Proof:} For brevity, we denote $\zeta = i H e^{- i \theta(0)}$.
Then $e^{i \theta(0)} = i \overline{\zeta} / \overline{H}$.
Next, using the definition $\xi = \sqrt{ \Delta^2 - |H|^2 }$, we obtain
\begin{eqnarray*}
| \chi_+ |^2 &=& ( \zeta - \Delta - \xi ) ( \overline{\zeta} - \Delta - \xi ) \\[2mm]
&=& ( \Delta + \xi ) ( 2 \Delta - \zeta - \overline{\zeta} ),\\[2mm]
| \chi_- |^2 &=& ( \zeta - \Delta + \xi ) ( \overline{\zeta} - \Delta + \xi ) \\[2mm]
&=& ( \Delta - \xi ) ( 2 \Delta - \zeta - \overline{\zeta} ),
\end{eqnarray*}
and
\begin{eqnarray*}
\left( |\chi_+|^2 - |\chi_-|^2 \right) e^{i \theta(0)} &=&
2 \xi ( 2 \Delta - \zeta - \overline{\zeta} ) i \overline{\zeta} / \overline{H} \\[2mm]
&=& - 2 i \xi ( \overline{\zeta}^2 - 2 \Delta \overline{\zeta} + |H|^2 ) / \overline{H},\\[2mm]
\overline{\chi}_+ \overline{\chi}_- &=&
( \overline{\zeta} - \Delta - \xi ) ( \overline{\zeta} - \Delta + \xi ) \\[2mm]
&=& \overline{\zeta}^2 - 2 \Delta \overline{\zeta} + |H|^2.
\end{eqnarray*}
Therefore,
$$
\fr{ |\chi_+|^2 - |\chi_-|^2 }{ \overline{\chi}_+ \overline{\chi}_- } e^{i \theta(0)}
= -\fr{2 i \xi}{\overline{H}}.
$$
On the other hand, we have
$$
\fr{\overline{\chi}_-}{\overline{\chi}_+} = \fr{\overline{\zeta} - \Delta + \xi}{\overline{\zeta} - \Delta - \xi} = \fr{i \overline{H} e^{i \theta(0)} + \Delta - \xi}{i \overline{H} e^{i \theta(0)} + \Delta + \xi}.
$$
Summarizing, this gives formula~(\ref{Formula:exp_theta:plus}).
The other formula in the proposition can be checked similarly.~\qed

\begin{proposition}
Suppose that $\theta_j(t)$ and $\theta_k(t)$
are two solutions of Eq.~(\ref{Eq:Adler})
with $\Delta = \Delta_j$ and $\Delta = \Delta_k$, respectively.
The complex parameter $H$ in Eq.~(\ref{Eq:Adler})
is assumed to be the same for both indices $j$ and $k$.
If $|\Delta_j| > |H|$ and $|\Delta_k| > |H|$, then
\begin{eqnarray*}
&&
\hspace{-7mm}
\langle e^{i \theta_j(t)} e^{-i \theta_k(t+\tau)} \rangle
- \langle e^{i \theta_j(t)} \rangle\: \langle e^{-i \theta_k(t)} \rangle
= \\[2mm]
&&
= \fr{4 \xi_j\: \mathrm{sgn}\:\Delta_j\:\xi_k\: \mathrm{sgn}\:\Delta_k}{|H|^2} \times\\[2mm]
&&
\times \sum\limits_{n,m=1}^\infty
\left( \fr{i \overline{H} e^{i \theta_j(0)} + \Delta_j - \xi_j\: \mathrm{sgn}\:\Delta_j}{i \overline{H} e^{i \theta_j(0)} + \Delta_j + \xi_j\: \mathrm{sgn}\:\Delta_j} \right)^n \times\\[2mm]
&&
\times\left( \fr{i H e^{-i \theta_k(0)} - \Delta_k + \xi_k\: \mathrm{sgn}\:\Delta_k}{i H e^{-i \theta_k(0)} - \Delta_k - \xi_k\: \mathrm{sgn}\:\Delta_k} \right)^m \times\\[2mm]
&&
\times e^{- i m \xi_k \tau\: \mathrm{sgn}\:\Delta_k}
\delta( n \xi_j\: \mathrm{sgn}\:\Delta_j - m \xi_k\: \mathrm{sgn}\:\Delta_k ),
\end{eqnarray*}
and
\begin{eqnarray*}
&&
\hspace{-7mm}
\langle e^{i \theta_j(t)} e^{i \theta_k(t+\tau)} \rangle
- \langle e^{i \theta_j(t)} \rangle\: \langle e^{i \theta_k(t)} \rangle = \\[2mm]
&&
= - \fr{4 \xi_j\: \mathrm{sgn}\:\Delta_j\:\xi_k\: \mathrm{sgn}\:\Delta_k}{\overline{H}^2} \times\\[2mm]
&&
\times \sum\limits_{n,m=1}^\infty
\left( \fr{i \overline{H} e^{i \theta_j(0)} + \Delta_j - \xi_j\: \mathrm{sgn}\:\Delta_j}{i \overline{H} e^{i \theta_j(0)} + \Delta_j + \xi_j\: \mathrm{sgn}\:\Delta_j} \right)^n \times\\[2mm]
&&
\times
\left( \fr{i \overline{H} e^{i \theta_k(0)} + \Delta_k - \xi_k\: \mathrm{sgn}\:\Delta_k}{i \overline{H} e^{i \theta_k(0)} + \Delta_k + \xi_k\: \mathrm{sgn}\:\Delta_k} \right)^m \times\\[2mm]
&&
\times e^{i m \xi_k \tau\: \mathrm{sgn}\:\Delta_k}
\delta( n \xi_j\: \mathrm{sgn}\:\Delta_j + m \xi_k\: \mathrm{sgn}\:\Delta_k ),
\end{eqnarray*}
where $\xi_j = \sqrt{\Delta_j^2 - |H|^2}$
and $\xi_k = \sqrt{\Delta_k^2 - |H|^2}$.
\label{Proposition:Covariances}
\end{proposition}

{\bf Proof:} The above formulas are straightforward consequences
of Proposition~\ref{Proposition:Expansions}
and the identity $\langle e^{i \nu t} \rangle = \delta(\nu)$.~\qed

\begin{remark}
If $\xi_j$ and $\xi_k$ in Proposition~\ref{Proposition:Covariances}
are incommensurable, then
$$
\langle e^{i \theta_j(t)} e^{- i \theta_k(t + \tau)} \rangle
=
\langle e^{i \theta_j(t)} \rangle\: \langle e^{- i \theta_k(t + \tau)} \rangle
$$
and
$$
\langle e^{i \theta_j(t)} e^{i \theta_k(t + \tau)} \rangle
=
\langle e^{i \theta_j(t)} \rangle\: \langle e^{i \theta_k(t + \tau)} \rangle.
$$
\end{remark}


\section{Daido's theory}
\label{app:daido}
In~\cite{Dai1990}, H. Daido proposed an analytical approach to approximate finite-size fluctuations
in the KS model~\eqref{Eq:KS:original} with $\lambda = 0$.
In particular, for the subcritical case he obtained explicit expressions
for the covariance function and the variance of the complex order parameter.
We follow here his approach and determine expressions for $\lambda\ne 0$
in the subcritical case, and show that unlike for a Lorentz distribution of the natural frequencies
this method is not well suited for a Gaussian frequency distribution.

Let us assume that the dynamics of the $i$th oscillator can be represented as
\begin{equation}
\psi_i(t) = \Theta_i(t) + \fr{1}{\sqrt{N}} \vartheta_i(t),
\label{Ansatz:Daido}
\end{equation}
where $\Theta_i, \vartheta_i = \mathcal{O}(1)$ for $N\to\infty$.
Upon substitution into the unnumbered equation
below (\ref{Eq:KS:original}) in the main text,
which we recall for convenience here, 
$$
\dot{\psi}_i = \omega_i + K\: \mathrm{Im}\left( Z(t) e^{-i \psi_i} e^{-i \lambda} \right),
$$
we obtain
$$
\dot{\Theta}_i + \fr{1}{\sqrt{N}} \dot{\vartheta}_i = \omega_i + K\: \mathrm{Im}\left( Z(t) e^{-i ( \Theta_i + \vartheta_i / \sqrt{N} )} e^{-i \lambda} \right).
$$
Next, assuming that $\langle Z(t)\rangle = 0$ and hence $Z(t) = \mathcal{O}(1/\sqrt{N})$,
and equating separately the terms of order $\mathcal{O}(1)$
and the terms of order $\mathcal{O}(1/\sqrt{N})$, we obtain
\begin{equation}
\dot{\Theta}_i = \omega_i
\label{Eq:Theta_i}
\end{equation}
and
\begin{equation}
\fr{1}{\sqrt{N}} \dot{\vartheta}_i = K\: \mathrm{Im}\left( Z(t) e^{-i \Theta_i} e^{-i \lambda} \right).
\label{Eq:vartheta_i}
\end{equation}
Eq.~\eqref{Eq:Theta_i} is solved by $\Theta_i(t) = \omega_i t + \Theta_i(0)$.
Substituting this into Eq.~\eqref{Eq:vartheta_i}, we obtain
$$
\fr{1}{\sqrt{N}} \dot{\vartheta}_i(t) = K \int_0^t \mathrm{Im}\left( Z(t') e^{-i ( \omega_i t' + \Theta_i(0) )} e^{-i \lambda} \right) dt'
$$
with $\vartheta_i(0) = 0$.
Finally, substituting the expressions for the solution to Eqs.~\eqref{Eq:Theta_i}
and~\eqref{Eq:vartheta_i} into~\eqref{Ansatz:Daido}
and then substituting $\theta_i(t) = \psi_i(t)$
into the definition of the complex order parameter~\eqref{Def:Z},
we obtain a self-consistency equation for $Z(t)$:
\begin{eqnarray}
Z(t) &=& \fr{1}{N} \sum\limits_{j=1}^N e^{i ( \omega_i t + \Theta_i(0) )} \times\\
&\times& \left( 1 + i K \int_0^t \mathrm{Im}\left( Z(t') e^{-i ( \omega_i t' + \Theta_i(0) )} e^{-i \lambda} \right) dt' \right) \nonumber\\
&=& \fr{1}{N} \sum\limits_{j=1}^N e^{i ( \omega_j t + \Theta_j(0) )}\nonumber\\
&+& \fr{K}{2} e^{-i \lambda} \int_0^t Z(t') \fr{1}{N} \sum\limits_{j=1}^N e^{i \omega_j (t - t')} dt' \nonumber\\
&-& \fr{K}{2} e^{i \lambda} \int_0^t \overline{Z}(t') \fr{1}{N} \sum\limits_{j=1}^N e^{i ( \omega_j (t + t') + 2 \Theta_j(0) )} dt',
\label{Eq:Z:Daido}
\end{eqnarray}
where we employed a Taylor expansion of the exponential function for $|Z(t)|\ll 1$.

In the next step, we will obtain an asymptotic formula for $Z(t)$ as $t\to\infty$.
For this, we first notice that the function
$$
\fr{1}{N} \sum\limits_{j=1}^N e^{i ( \omega_j (t + t') + 2 \Theta_j(0) )}
$$
is vanishing as $t\to\infty$ and hence can be neglected.
Furthermore, for $N\gg 1$ we have
$$
\fr{1}{N} \sum\limits_{j=1}^N e^{i \omega_j t} \mapsto \int_{-\infty}^\infty g(\omega) e^{i \omega t} d\omega
= \hat{g}(t).
$$
Thus, the complex order parameter $Z(t)$ given in~\eqref{Eq:Z:Daido}
is the solution of the following Volterra integral equation
\begin{eqnarray}
Z(t) &=& \fr{1}{N} \sum\limits_{j=1}^N e^{i ( \omega_j t + \Theta_j(0) )} \nonumber\\
&+& \fr{K}{2} e^{-i \lambda} \int_0^t \hat{g}(t - t') Z(t') dt'.
\label{Eq:Daido}
\end{eqnarray}
This equation can be solved using the Laplace transformation
$$
\mathcal{L}\::\: f(t) \mapsto F(s) = \int_0^\infty e^{- s t} f(t) dt.
$$
In particular, due to the convolution theorem, Eq.~\eqref{Eq:Daido} we write
$$
(\mathcal{L}Z)(s) = \fr{1}{N} \sum\limits_{j=1}^N \fr{e^{i \Theta_j(0)}}{s - i \omega_j} + \fr{K}{2} e^{-i \lambda} (\mathcal{L}\hat{g})(s) (\mathcal{L}Z)(s),
$$
and therefore obtain
$$
(\mathcal{L}Z)(s) = \fr{1}{N} \sum\limits_{j=1}^N \left( 1 - \fr{K}{2} e^{-i \lambda} (\mathcal{L}\hat{g})(s) \right)^{-1} \fr{e^{i \Theta_j(0)}}{s - i \omega_j}.
$$
Now, assuming that
$$
Q(s) := 1 - \fr{K}{2} e^{-i \lambda} (\mathcal{L}\hat{g})(s) \ne 0
\quad\mbox{for all}\quad \mathrm{Re}(s)\ge 0
$$
and that $1/Q(s)$ is a meromorphic function in the half-plane $\mathrm{Re}(s) < 0$ that satisfies
\begin{equation}
\lim\limits_{p\to\infty} \max\limits_{\pi/2\le |\phi| \le \pi} |1/Q( p e^{i \phi} )| = 0,
\label{Condition:Q}
\end{equation}
we apply the inverse Laplace transformation and find that the long-term asymptotics of $Z(t)$ is given by
$$
Z(t) = \fr{1}{N} \sum\limits_{j=1}^N \fr{e^{i ( \omega_j t + \Theta_j(0) )}}{Q(i \omega_j)}.
$$

The covariance function of the complex order parameter $Z(t)$ is now evaluated as
\begin{eqnarray*}
R_\mathrm{Daido}(\tau) &=& N \langle Z(t) \overline{Z(t+\tau)} \rangle \\
&=& \fr{1}{N} \sum\limits_{j,k=1}^N \fr{e^{-i ( \omega_k \tau + \Theta_k(0) - \Theta_j(0) )}}{Q(i \omega_j) \overline{Q(i \omega_k)}} \langle e^{i ( \omega_j - \omega_k ) t} \rangle \\
&=& \fr{1}{N} \sum\limits_{j=1}^N \fr{e^{-i \omega_j \tau}}{|Q(i \omega_j)|^2}
\approx \int_{-\infty}^\infty g(\omega) \fr{e^{-i \omega \tau}}{|Q(i \omega)|^2} d\omega,
\end{eqnarray*}
where we used $\langle e^{i ( \omega_j - \omega_k ) t} \rangle = \delta_{jk}$.
The variance of the order parameter $r(t) = |Z(t)|$ in the subcritical case is then given by
\begin{equation}
V_\mathrm{Daido} = N \langle |Z(t)|^2 \rangle - N \langle |Z(t)| \rangle^2 = R_\mathrm{Daido}(0) - \fr{\pi}{4},
\label{Formula:V:Daido}
\end{equation}
where we used our result~\eqref{Formula:incoh:r} for $\langle |Z(t)| \rangle$.
\begin{figure}
\begin{center}
\includegraphics[height=0.6\columnwidth,angle=270]{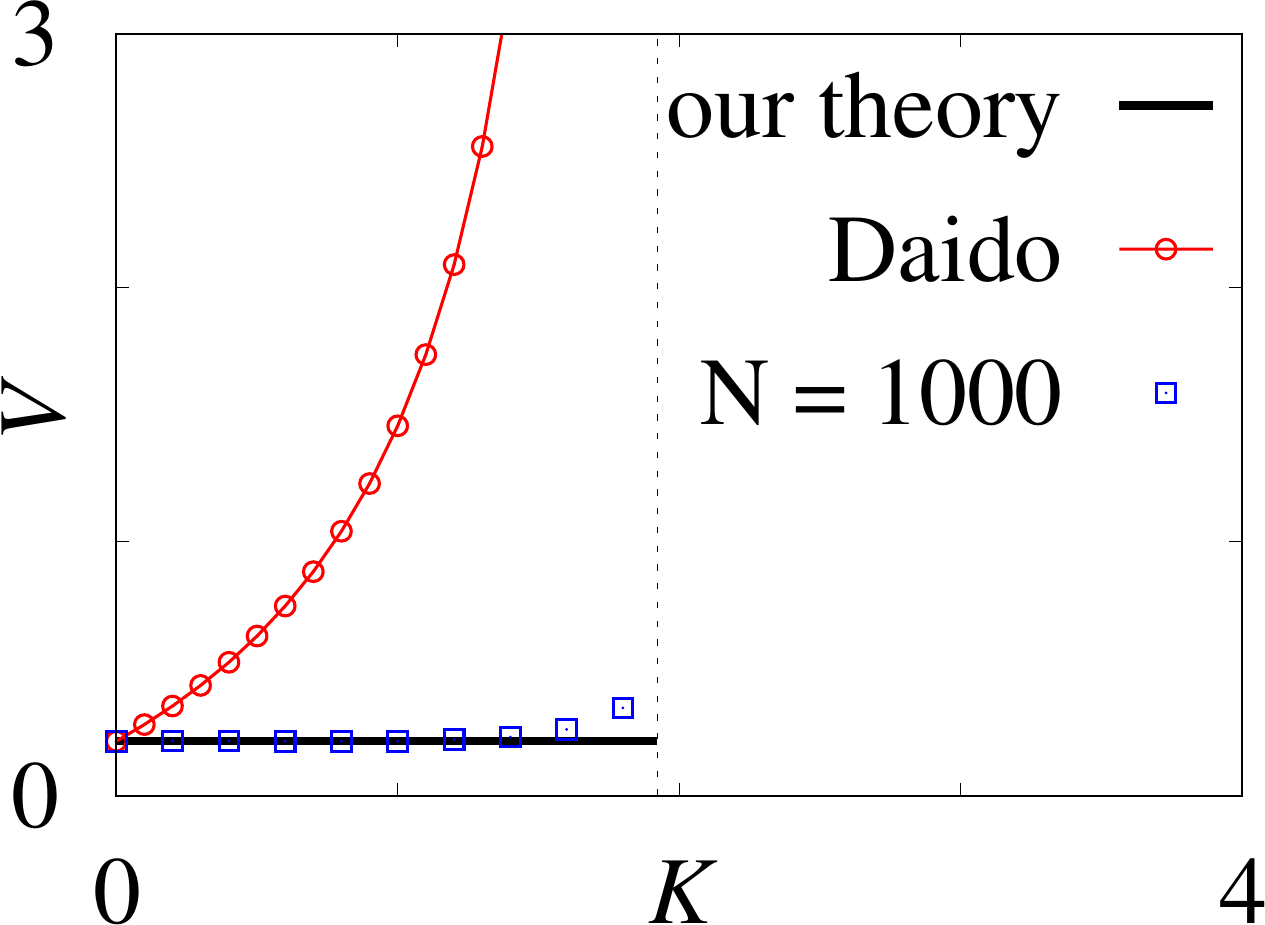}
\end{center}
\caption{Variance $V$ of the order parameter $r(t)$ versus coupling strength $K$
in the subcritical case of the KS model~\eqref{Eq:KS:original}
with $\lambda = \pi/4$ and a Gaussian distribution.
Theoretical predictions from our approach~\eqref{Formula:incoh:r} (thick line)
and from Daido's theory~\eqref{Formula:V:Daido} (circles)
are compared with the results of numerical simulations
with $N=1,000$ phase oscillators (squares).
The vertical dashed line indicates the critical coupling strength.
}
\label{Fig:OPVariance_Daido}
\end{figure}
Although the above formulas have been shown to be in good agreement with the numerical observations obtained from the full KS model~\eqref{Eq:KS:original} with a Lorentzian frequency distribution and $\lambda = 0$,
in the case of a Gaussian frequency distribution they fail to reproduce the observed behavior as seen in Fig.~\ref{Fig:OPVariance_Daido}.
In particular, they can be worse than our $(K,\lambda)$-independent formulas~\eqref{Formula:R:incoh} and~\eqref{Formula:incoh:r}.
Note that condition~\eqref{Condition:Q} is not satisfied for a Gaussian frequency distribution.


\bibliographystyle{apsrev4-2}
\bibliography{refs}

@article{Str2000,
 title = {From Kuramoto to Crawford: exploring the onset of synchronization in populations of coupled oscillators},
 author = {Strogatz, S.},
 year = {2000},
 journal = {Physica D},
 volume = {143},
 pages = {1--20}
}

@article{AceBVRS2005,
 title = {The Kuramoto model: A simple paradigm for synchronization phenomena},
 author = {Acebr\'on, J. and Bonilla, L. and P\'erez Vicente, C. and Ritort, F. and Spigler, R.},
 year = {2005},
 journal = {Rev. Modern Phys.},
 volume = {77},
 pages = {137--185}
}

@article{SakK1986,
 title = {A soluble active rotater model showing phase transitions via mutual entertainment},
 author = {Sakaguchi, H. and Kuramoto, Y.},
 year = {1986},
 journal = {Prog. Theor. Phys.},
 volume = {76},
 pages = {576--581}
}

@article{OmeW2012,
 title = {Nonuniversal transitions to synchrony in the Sakaguchi-Kuramoto model},
 author = {Omel'chenko, O. E. and Wolfrum, M.},
 year = {2012},
 journal = {Phys. Rev. Lett.},
 volume = {109},
 pages = {164101}
}

@article{OmeW2013,
 title = {Bifurcations in the Sakaguchi-Kuramoto model},
 author = {Omel'chenko, O. E. and Wolfrum, M.},
 year = {2013},
 journal = {Phys. D},
 volume = {263},
 pages = {74--85}
}

@article{KurN1987,
 title = {Statistical macrodynamics of large dynamical systems. Case of a phase transition in oscillator communities},
 author = {Kuramoto, Y. and Nishikawa, I.},
 year = {1987},
 journal = {J. Stat. Phys.},
 volume = {49},
 pages = {569--605}
}

@article{Dai1990,
 title = {Intrinsic fluctuations and a phase transition in a class of large populations of interacting oscillators},
 author = {Daido, H.},
 year = {1990},
 journal = {J. Stat. Phys.},
 volume = {60},
 pages = {753--800}
}

@article{Dai1994,
 title = {Generic scaling at the onset of macroscopic mutual entrainment in limit-cycle oscillators with uniform all-to-all coupling},
 author = {Daido, H.},
 year = {1994},
 journal = {Phys. Rev. Lett.},
 volume = {73},
 pages = {760}
}

@article{Dai1996,
 title = {Onset of cooperative entrainment in limit-cycle oscillators with uniform all-to-all interactions: bifurcations of the order function},
 author = {Daido, H.},
 year = {1994},
 journal = {Phys. D},
 volume = {91},
 pages = {24--66}
}

@article{CraD1999,
 title = {Synchronization of globally coupled phase oscillators: singularities and scaling for general couplings},
 author = {Crawford, J. D. and Davies, K. T. R.},
 year = {1999},
 journal = {Phys. D},
 volume = {125},
 pages = {1--46}
}

@article{NisTHA2012,
 title = {Long-term fluctuations in globally coupled phase oscillators
with general coupling: Finite size effects},
 author = {Nishikawa, I. and Tanaka, G. and Horita, T. and Aihara, K.},
 year = {2012},
 journal = {Chaos},
 volume = {22},
 pages = {013133}
}

@article{HonCTP2015,
 title = {Finite-size scaling, dynamic fluctuations, and hyperscaling relation in the Kuramoto model},
 author = {Hong, H. and Chat{\'e}, H. and Tang, L.-H. and Park, H.},
 year = {2015},
 journal = {Phys. Rev. E},
 volume = {92},
 pages = {022122}
}

@article{YueG2024,
 title = {A stochastic approximation for the finite-size {K}uramoto-{S}akaguchi model},
 author = {Yue, W. and Gottwald, G. A.},
 year = {2024},
 journal = {Phys. D},
 volume = {468},
 pages = {134292}
}

@article{PetP2018,
 title = {Transition to collective oscillations in finite Kuramoto ensembles},
 author = {Peter, F. and Pikovsky, A.},
 year = {2018},
 journal = {Phys. Rev. E},
 volume = {97},
 pages = {032310}
}

@article{PetGP2019,
 title = {Microscopic correlations in the finite-size Kuramoto model of coupled oscillators},
 author = {Peter, F. and Gong, C. C. and Pikovsky, A.},
 year = {2019},
 journal = {Phys. Rev. E},
 volume = {100},
 pages = {032210}
}

@article{SumJ2024,
 title = {Finite-size effect in Kuramoto oscillators with higher-order interactions},
 author = {Suman, A. and Jalan, S.},
 year = {2024},
 journal = {Chaos},
 volume = {34},
 pages = {101101}
}

@article{Luc2014,
 title = {Large time asymptotics for the fluctuation SPDE in the Kuramoto synchronization model},
 author = {Lu\c{c}on, E.},
 year = {2014},
 journal = {J. Funct. Anal.},
 volume = {266},
 pages = {6372--6417}
}

@article{HonOS2016,
 title = {Correlated disorder in the {K}uramoto model: Effects on
phase coherence, finite-size scaling, and dynamic
fluctuations},
 author = {Hong, H and O'Keeffe, K. P. and Strogatz, S. H.},
 year = {2016},
 journal = {Chaos},
 volume = {26},
 pages = {103105}
}

@article{PikDG2019,
 title = {Correlations of the states of non-entrained oscillators in the Kuramoto ensemble with noise in the mean field},
 author = {Pikovsky, A. S. and Dolmatova, A. V. and Goldobin, D. S.},
 year = {2019},
 journal = {Radiophys. Quantum Electron.},
 volume = {61},
 pages = {672--680}
}

@article{OttA2008,
 title = {Low dimensional behavior of large systems of globally coupled oscillators},
 author = {Ott, E. and Antonsen, T. M.},
 year = {2008},
 journal = {Chaos},
 volume = {18},
 pages = {037113}
}

@article{OttA2009,
 title = {Long time evolution of phase oscillator systems},
 author = {Ott, E. and Antonsen, T. M.},
 year = {2009},
 journal = {Chaos},
 volume = {19},
 pages = {023117}
}

@book{Kuramoto84,
   author = {Kuramoto, Y},
   title = {Chemical Oscillations, Waves, and Turbulence},
   publisher = {Springer},
   address = {Berlin},
   year = {1984}
}

@book{PikovskyRosenblumKurths01,
  author = {Pikovsky, A. and Rosenblum, M. and Kurths, J.},
  title = {Synchronization, a Universal Concept in Nonlinear Sciences},
  publisher = {Cambridge University Press},
  address = {Cambridge},
  year = {2001}
}

@article {Got2015,
	AUTHOR = {Gottwald, Georg A.},
	TITLE = {Model reduction for networks of coupled oscillators},
	JOURNAL = {Chaos},
	FJOURNAL = {Chaos. An Interdisciplinary Journal of Nonlinear Science},
	VOLUME = {25},
	YEAR = {2015},
	NUMBER = {5},
	PAGES = {053111, 12},
	DOI={10.1063/1.4921295}
}

@article{Gottwald17,
	author = {Georg A. Gottwald},
	title = {Finite-size effects in a stochastic {K}uramoto model},
	journal={Chaos},
	fjournal = {Chaos: An Interdisciplinary Journal of Nonlinear Science},
	volume = {27},
	number = {10},
	pages = {101103},
	year = {2017},
	doi = {10.1063/1.5004618}
}

@article{SmithGottwald20,
    author = {Smith, Lachlan D. and Gottwald, Georg A.},
    title = "{Model reduction for the collective dynamics of globally coupled oscillators: {F}rom finite networks to the thermodynamic limit}",
    journal = {Chaos},
    fjournal = {Chaos: An Interdisciplinary Journal of Nonlinear Science},
    volume = {30},
    number = {9},
    year = {2020},
    month = {09},
    issn = {1054-1500},
    doi = {10.1063/5.0009790},
    url = {https://doi.org/10.1063/5.0009790},
}

@InCollection{Lucon15,
  author    = {Lu\c{c}on, Eric},
  title     = {Large population asymptotics for interacting diffusions in a quenched random environment},
  booktitle = {From particle systems to partial differential equations. {II}},
  publisher = {Springer, Cham},
  year      = {2015},
  volume    = {129},
  series    = {Springer Proc. Math. Stat.},
  pages     = {231--251},
}

@article {GiacominPoquet15,
    AUTHOR = {Giacomin, Giambattista and Poquet, Christophe},
     TITLE = {Noise, interaction, nonlinear dynamics and the origin of
              rhythmic behaviors},
   JOURNAL = {Braz. J. Probab. Stat.},
  FJOURNAL = {Brazilian Journal of Probability and Statistics},
    VOLUME = {29},
      YEAR = {2015},
    NUMBER = {2},
     PAGES = {460--493},
}

@article{FialkowskiEtAl23,
  title = {Heterogeneous Nucleation in Finite-Size Adaptive Dynamical Networks},
  author = {Fialkowski, Jan and Yanchuk, Serhiy and Sokolov, Igor M. and Sch\"oll, Eckehard and Gottwald, Georg A. and Berner, Rico},
  journal = {Phys. Rev. Lett.},
  volume = {130},
  issue = {6},
  pages = {067402},
  numpages = {7},
  year = {2023},
  month = {Feb},
  publisher = {American Physical Society},
  doi = {10.1103/PhysRevLett.130.067402},
  url = {https://link.aps.org/doi/10.1103/PhysRevLett.130.067402}
}

@article{Givonetal04,
	Author = {Dror Givon and Raz Kupferman and Andrew Stuart},
	Journal = {Nonlinearity},
	Number = {6},
	Pages = {R55-127},
	Title = {Extracting macroscopic dynamics: Model problems and algorithms},
	Volume = {17},
	Year = {2004}}

@Article{ArenasEtAl08,
  author   = {Arenas, Alex and Diaz-Guilera, Albert and Kurths, J\"urgen and Moreno, Yamir and Zhou, Changsong},
  title    = {Synchronization in complex networks},
  journal  = {Phys. Rep.},
  year     = {2008},
  volume   = {469},
  number   = {3},
  pages    = {93--153},
  issn     = {0370-1573},
  doi      = {10.1016/j.physrep.2008.09.002},
  fjournal = {Physics Reports. A Review Section of Physics Letters},
  url      = {http://dx.doi.org/10.1016/j.physrep.2008.09.002},
}

@Article{RodriguesEtAl16,
  author  = {Francisco A. Rodrigues and Thomas K. DM. Peron and Peng Ji and J{\"u}rgen Kurths},
  title   = {The {K}uramoto model in complex networks},
  journal = {Phys. Rep.},
  year    = {2016},
  volume  = {610},
  pages   = {1 - 98},
}

@Article{DorflerBullo14,
  author  = {Florian D{\"o}rfler and Francesco Bullo},
  title   = {Synchronization in complex networks of phase oscillators: {A} survey},
  journal = {Automatica},
  year    = {2014},
  volume  = {50},
  number  = {6},
  pages   = {1539 - 1564},
}

@article {YamaguchiEtAl03,
	author = {Yamaguchi, Shun and Isejima, Hiromi and Matsuo, Takuya and Okura, Ryusuke and Yagita, Kazuhiro and Kobayashi, Masaki and Okamura, Hitoshi},
	title = {Synchronization of Cellular Clocks in the Suprachiasmatic Nucleus},
	volume = {302},
	number = {5649},
	pages = {1408--1412},
	year = {2003},
	doi = {10.1126/science.1089287},
	publisher = {American Association for the Advancement of Science},
	issn = {0036-8075},
	URL = {https://science.sciencemag.org/content/302/5649/1408},
	journal = {Science}
}

@InProceedings{BhowmikShanahan12,
  author    = {D. Bhowmik and M. Shanahan},
  title     = {How well do oscillator models capture the behaviour of biological neurons?},
  booktitle = {The 2012 International Joint Conference on Neural Networks (IJCNN)},
  year      = {2012},
  pages     = {1-8},
}

@article {KissEtAl02,
	author = {Kiss, Istv{\'a}n Z. and Zhai, Yumei and Hudson, John L.},
	title = {Emerging Coherence in a Population of Chemical Oscillators},
	volume = {296},
	number = {5573},
	pages = {1676--1678},
	year = {2002},
	doi = {10.1126/science.1070757},
	publisher = {American Association for the Advancement of Science},
	issn = {0036-8075},
	journal = {Science}
}

@article {TaylorEtAl09,
	author = {Taylor, Annette F. and Tinsley, Mark R. and Wang, Fang and Huang, Zhaoyang and Showalter, Kenneth},
	title = {Dynamical Quorum Sensing and Synchronization in Large Populations of Chemical Oscillators},
	volume = {323},
	number = {5914},
	pages = {614--617},
	year = {2009},
	doi = {10.1126/science.1166253},
	publisher = {American Association for the Advancement of Science},
	issn = {0036-8075},
	journal = {Science}
}

@article{FilatrellaEtAl08,
	Author = {Filatrella, G. and Nielsen, A. H. and Pedersen, N. F.},
	journal = {Eur. Phys J. B},
	FJournal = {The European Physical Journal B},
	Month = {Feb},
	Number = {4},
	Pages = {485--491},
	Title = {Analysis of a power grid using a {K}uramoto-like model},
	Volume = {61},
	Year = {2008},
	DOI={10.1140/epjb/e2008-00098-8}}

@article{BuiceChow07,
  title = {Correlations, fluctuations, and stability of a finite-size network of coupled oscillators},
  author = {Buice, Michael A. and Chow, Carson C.},
  journal = {Phys. Rev. E},
  volume = {76},
  issue = {3},
  pages = {031118},
  numpages = {25},
  year = {2007},
  month = {Sep},
  publisher = {American Physical Society},
  doi = {10.1103/PhysRevE.76.031118},
  url = {https://link.aps.org/doi/10.1103/PhysRevE.76.031118}
}

@article{YoonEtAl15,
  title = {Critical behavior of the relaxation rate, the susceptibility, and a pair correlation function in the {K}uramoto model on scale-free networks},
  author = {Yoon, S. and Sorbaro Sindaci, M. and Goltsev, A. V. and Mendes, J. F. F.},
  journal = {Phys. Rev. E},
  volume = {91},
  issue = {3},
  pages = {032814},
  numpages = {10},
  year = {2015},
  month = {Mar},
  publisher = {American Physical Society},
  doi = {10.1103/PhysRevE.91.032814},
  url = {https://link.aps.org/doi/10.1103/PhysRevE.91.032814}
}

@article{ParkPark24,
  title = {Finite-size scaling of the {K}uramoto model at criticality},
  author = {Park, Su-Chan and Park, Hyunggyu},
  journal = {Phys. Rev. E},
  volume = {110},
  issue = {3},
  pages = {034216},
  numpages = {12},
  year = {2024},
  month = {Sep},
  publisher = {American Physical Society},
  doi = {10.1103/PhysRevE.110.034216},
  url = {https://link.aps.org/doi/10.1103/PhysRevE.110.034216}
}

@article{HildebrandEtAl07,
  title = {Kinetic Theory of Coupled Oscillators},
  author = {Hildebrand, Eric J. and Buice, Michael A. and Chow, Carson C.},
  journal = {Phys. Rev. Lett.},
  volume = {98},
  issue = {5},
  pages = {054101},
  numpages = {4},
  year = {2007},
  month = {Jan},
  publisher = {American Physical Society},
  doi = {10.1103/PhysRevLett.98.054101},
  url = {https://link.aps.org/doi/10.1103/PhysRevLett.98.054101}
}

@article{SnyderEtAl21,
title = {Data-driven stochastic modeling of coarse-grained dynamics with finite-size effects using {L}angevin regression},
journal = {Physica D: Nonlinear Phenomena},
volume = {427},
pages = {133004},
year = {2021},
issn = {0167-2789},
doi = {https://doi.org/10.1016/j.physd.2021.133004},
url = {https://www.sciencedirect.com/science/article/pii/S0167278921001615},
author = {Jordan Snyder and Jared L. Callaham and Steven L. Brunton and J. Nathan Kutz},
}

@article{HongEtAl07,
  title = {Finite-Size Scaling in Complex Networks},
  author = {Hong, Hyunsuk and Ha, Meesoon and Park, Hyunggyu},
  journal = {Phys. Rev. Lett.},
  volume = {98},
  issue = {25},
  pages = {258701},
  numpages = {4},
  year = {2007},
  month = {Jun},
  publisher = {American Physical Society},
  doi = {10.1103/PhysRevLett.98.258701},
  url = {https://link.aps.org/doi/10.1103/PhysRevLett.98.258701}
}

@article{Daido89,
    author = {Daido, Hiroaki},
    title = {Intrinsic Fluctuation and Its Critical Scaling in a Class of Populations of Oscillators with Distributed Frequencies},
    journal = {Progress of Theoretical Physics},
    volume = {81},
    number = {4},
    pages = {727-731},
    year = {1989},
    month = {04},
    issn = {0033-068X},
    doi = {10.1143/PTP.81.727},
    url = {https://doi.org/10.1143/PTP.81.727},
    eprint = {https://academic.oup.com/ptp/article-pdf/81/4/727/5255143/81-4-727.pdf},
}

@article{CarGP2018,
  title = {Origin and scaling of chaos in weakly coupled phase oscillators},
  author = {Carlu, Mallory and Ginelli, Francesco and Politi, Antonio},
  journal = {Phys. Rev. E},
  volume = {97},
  issue = {1},
  pages = {012203},
  numpages = {11},
  year = {2018},
  month = {Jan},
  publisher = {American Physical Society},
  doi = {10.1103/PhysRevE.97.012203},
  url = {https://link.aps.org/doi/10.1103/PhysRevE.97.012203}
}

@article{MasKK2010,
doi = {10.1088/1367-2630/12/9/093007},
url = {https://doi.org/10.1088/1367-2630/12/9/093007},
year = {2010},
month = {sep},
publisher = {},
volume = {12},
number = {9},
pages = {093007},
author = {Masuda, Naoki and Kawamura, Yoji and Kori, Hiroshi},
title = {Collective fluctuations in networks of noisy components},
journal = {New Journal of Physics}
}

@article {DelgadinoEtAl21,
    AUTHOR = {Delgadino, Matias G. and Gvalani, Rishabh S. and Pavliotis,
              Grigorios A.},
     TITLE = {On the diffusive-mean field limit for weakly interacting
              diffusions exhibiting phase transitions},
   JOURNAL = {Arch. Ration. Mech. Anal.},
  FJOURNAL = {Archive for Rational Mechanics and Analysis},
    VOLUME = {241},
      YEAR = {2021},
    NUMBER = {1},
     PAGES = {91--148},
       DOI = {10.1007/s00205-021-01648-1},
       URL = {https://doi.org/10.1007/s00205-021-01648-1},
}

@article{ZagliEtAl23,
  title = {Dimension reduction of noisy interacting systems},
  author = {Zagli, Niccol\`o and Pavliotis, Grigorios A. and Lucarini, Valerio and Alecio, Alexander},
  journal = {Phys. Rev. Res.},
  volume = {5},
  issue = {1},
  pages = {013078},
  numpages = {13},
  year = {2023},
  month = {Feb},
  publisher = {American Physical Society},
  doi = {10.1103/PhysRevResearch.5.013078},
  url = {https://link.aps.org/doi/10.1103/PhysRevResearch.5.013078}
}

@article{PikR1999,
 title = {Finite-size effects in a population of interacting oscillators},
 author = {Pikovsky, A. and Ruffo, S.},
 year = {1999},
 journal = {Phys. Rev. E},
 volume = {59},
 pages = {1633}
}

@misc{PikBI2026,
 title = {Coherence properties of collective modes in ensembles of oscillators}, 
 author = {Pikovsky, A. and Bagnoli, F. and Iubini, S.},
 year = {2026},
 eprint = {2601.05963},
 archivePrefix = {arXiv},
 primaryClass = {nlin.CD},
 url = {https://arxiv.org/abs/2601.05963}
}

@article{Bue2025,
 title = {Mesoscopic Theory for Coupled Stochastic Oscillators},
 author = {Buendia, V.},
 year = {2025},
 journal = {Phys. Rev. Lett.},
 volume = {134},
 pages = {197201}
}

@article{TyuGKP2018,
 title = {Dynamics of Noisy Oscillator Populations beyond the {O}tt-{A}ntonsen Ansatz},
 author = {Tyulkina, I. V. and Goldobin, D. S. and Klimenko, L. S. and Pikovsky, A.},
 year = {2018},
 journal = {Phys. Rev. Lett.},
 volume = {120},
 pages = {264101}
}

@misc{IrvG2025,
 title = {Finte-size induced random switching of chimeras in a deterministic two-population {K}uramoto-{S}akaguchi model}, 
 author = {Henry Irvine and Georg A. Gottwald},
 year = {2025},
 eprint = {2510.12342},
 archivePrefix = {arXiv},
 primaryClass = {nlin.AO},
 url = {https://arxiv.org/abs/2510.12342}
}

@article{ThuSST2023,
  title = {Synchrony for Weak Coupling in the Complexified {K}uramoto Model},
  author = {Th\"umler, Moritz and Srinivas, Shesha G. M. and Schr\"oder, Malte and Timme, Marc},
  journal = {Phys. Rev. Lett.},
  volume = {130},
  issue = {18},
  pages = {187201},
  numpages = {6},
  year = {2023},
  month = {May},
  publisher = {American Physical Society},
  doi = {10.1103/PhysRevLett.130.187201},
  url = {https://link.aps.org/doi/10.1103/PhysRevLett.130.187201}
}

@article{LeeBBSTT2024,
  title = {Complexified synchrony},
  author = {Lee, Seungjae and Braun, Lucas and B\"onisch, Frieder and Schr\"oder, Malte and Th\"umler, Moritz and Timme, Marc},
  journal = {Chaos},
  volume = {34},
  pages = {053141},
  year = {2024}
}

\end{document}